\documentstyle[12pt]{article}

\newcounter{eq}
\newcounter{sc}

\def\overleftrightarrow#1{\vbox{\ialign{##\crcr
 $\leftrightarrow$\crcr\noalign{\kern-1pt\nointerlineskip}
 $\hfil\displaystyle{#1}\hfil$\crcr}}}

\newlength{\minitwocolumn}
\begin{document}

\begin{flushright}
DPUR/TH/82\\
May, 2025\\
\end{flushright}
\vspace{20pt}

\pagestyle{empty}
\baselineskip15pt

\begin{center}
{\large\bf  Manifestly Covariant Canonical Formalism of Quadratic Gravity 
\vskip 1mm }

\vspace{20mm}

Ichiro Oda\footnote{
           E-mail address:\ ioda@cs.u-ryukyu.ac.jp
                  }

\vspace{10mm}
           Department of Physics, Faculty of Science, University of the 
           Ryukyus,\\
           Nishihara, Okinawa 903-0213, Japan\\

\end{center}


\vspace{10mm}
\begin{abstract}

We present the manifestly covariant quantization of quadratic gravity or higher-derivative gravity in the de Donder gauge condition 
(or harmonic gauge condition) for general coordinate invariance on the basis of the BRST transformation. We explicitly calculate various 
equal-time commutation relations (ETCRs), in particlular, the ETCRs between the metric tensor and its time derivatives in detail, 
and show that they are identically vanishing. We also clarify global symmetries, the physical content of quadratic gravity, and clearly show 
that this theory is not unitary and has a massive scalar, massive ghost and massless graviton as physical modes. Finally, we comment on 
confinement of the massive ghost, thereby recovering the unitarity of the physical S-matrix in quadratic gravity.

\end{abstract}

\newpage
\pagestyle{plain}
\pagenumbering{arabic}


\section{Introduction}

The construction of a quantum field theory of gravity is one of the most important and fascinating problems 
in modern particle physics. In recent years, we watch a revival of a considerable interest in the perturbatively renormalizable quantum 
gravity \cite{Luca}, which is nowdays called $\it{quadratic \, gravity}$, where quadratic curvature terms are added to 
the Einstein-Hilbert action \cite{Stelle1}. 

This revival mainly arises from the following two reasons: First, quadratic gravity is a renormalizable theory so it not only provides us 
with an example of ultraviolet (UV)-complete theory but also is highly predictive in the sense that the results obtained from this theory 
can be tested in the near future. For instance, quadratic gravity contains the Starobinsky model of inflation \cite{Starobinsky}, 
which is currently in perfect agreement with the cosmic microwave background observations and will be checked in future's 
cosmological observations.

Second, many of new ideas have been recently put forward to solve an infamous problem in quadratic gravity, which is the unitarity 
violation owing to the existence of physical massive ghost \cite{Anselmi}-\cite{Luca2}. It is worthwhile to point out that if the problem 
of the physical ghost were solved, we could obtain the renormalizable quantum gravity for the first time within the framework 
of quantum field theories. 

However, these new ideas are almost based on Feynman's path integral approach and toy models. As is well known, quantum field theories 
can be formulated in two different but mathematically equivalent formalisms, one of which is path integral formalism and the other is the
canonical operator formalism \cite{Kugo-Ojima, Nakanishi, N-O-text}. Moreover, the toy models describe only some aspects of quadratic 
gravity. It is therefore timely to attempt to construct the manifestly covariant canonical operator formalism of quadratic gravity 
and investigate the problem of the unitarity violation in quadratic gravity.  
In this article, we present such a manifestly covariant operator formalism of quadratic gravity in the de Donder gauge condition, or 
equivalently harmonic gauge condition for general coordinate invariance on the basis of the BRST transformation.
We take the metric tensor field $g_{\mu\nu}$ as a fundamental field and quantize it in the canonical formalism without recourse to 
perturbation theory. To put it differently, our formalism is essentially in nonperturbative nature and independent of a background space-time 
at the action level in the sense that we do not make use of any particular space-time metric such as a flat Minkowski metric $\eta_{\mu\nu}$. 
Since the metric tensor field $g_{\mu\nu}$ is treated as a q-number, the concept of the causality might become dubious at the operator level.  

The canonical formalism of quadratic gravity has been already considered in the series of papers \cite{Kimura1, Kimura2, Kimura3}.
However, it is a pity that these papers neither present the detail of calculations nor give us a complete proof for physical states on the
basis of the BRST formalism, which makes it difficult for us to grasp the full content of the theory, so we wish to fill the gaps in this article.
In addition to it, we wish to show a remakable feature of quadratic gravity that the equal-time commutation relations (ETCRs) between 
the metric and its derivatives are identically vanishing despite the nonvanishing ETCRs in conformal gravity \cite{Oda-Ohta, Oda-Conf}
and $f(R)$ gravity \cite{Oda-f} having higher-derivative curvature terms (See also related works \cite{Oda-Q, Oda-W, Oda-Saake, Oda-Corfu}).
This peculiar feature of quadratic gravity might imply that the metric tensor field $g_{\mu\nu}$ could lose its dynamical role at high energies 
in quadratic gravity, and be related to its renormalizability since this feature emerges only under the situation that an $R^2$ term coexists with
a $C_{\mu\nu\rho\sigma}^2$ term in the action.  
  
We close this section with an overview of this article. In Section 2, we consider a classical quadratic gravity and rewrite it into 
the first-order form by introducing auxiliary fields \cite{Kimura2}. In Section 3, using the de Donder gauge condition we construct 
a BRST-invariant action, derive field equations and argue its global symmetries. In Section 4, we derive the canonical conjugate momenta 
and set up the canonical commutation relations (CCRs). In Section 5, we calculate various ETCRs where a special attention is paid for 
the ETCRs between the metric and its time derivatives. In Section 6, by expanding the metric field around a flat Minkowski metric $\eta_{\mu\nu}$, 
we construct a linearized theory of quadratic gravity. In Section 7, we derive the four-dimensional commutation relations (4D CRs), 
analyze the physical content by imposing the Kugo-Ojima condition \cite{Kugo-Ojima} and show that the physical S-matrix is not unitary owing
to the presence of massive ghost. The final section is devoted to the conclusion. 

Five appendices are put for technical details. In Appendix A, a derivation of an equal-time commutation relation,
$[ \dot K_{\rho\sigma}, K_{\mu\nu}^\prime ]$ is given at the linearized level. In Appendix B we give useful ETCRs in a flat 
Minkowski background without the detailed proofs. In Appendix C, we present a derivation of a 4D CR, $[ h_{\mu\nu} (x), h_{\rho\sigma} (y) ]$.
In Appendix D, we illustrate a derivation of the Fourier transform of a 4D CR, $[ h_{\mu\nu} (p), \tilde A_\rho^\dagger (q) ]$ as an example. 
A proof of Eq. (\ref{Simple-field2}) is given in the final Appendix E.

\section{Classical quadratic gravity}

In this section, we consider quadratic gravity which is invariant under general coordinate transformation (GCT) 
in four space-time dimensions. Our classical Lagrangian consists of the Einstein-Hilbert term, an $R^2$ term and a squared term of 
conformal tensor\footnote{We follow the notation and conventions of Misner-Thorne-Wheeler (MTW) textbook \cite{MTW}. 
Lowercase Greek letters $\mu, \nu, \dots$ and Latin ones $i, j, \dots$ are used for space-time and spatial indices, respectively; 
for instance, $\mu= 0, 1, 2, 3$ and $i = 1, 2, 3$. The Riemann curvature tensor and the Ricci tensor are, respectively, defined by 
$R^\rho{}_{\sigma\mu\nu} = \partial_\mu \Gamma^\rho_{\sigma\nu} 
- \partial_\nu \Gamma^\rho_{\sigma\mu} + \Gamma^\rho_{\lambda\mu} \Gamma^\lambda_{\sigma\nu} 
- \Gamma^\rho_{\lambda\nu} \Gamma^\lambda_{\sigma\mu}$ and $R_{\mu\nu} = R^\rho{}_{\mu\rho\nu}$. 
The Minkowski metric tensor is denoted by $\eta_{\mu\nu}$; $\eta_{00} = - \eta_{11} = - \eta_{22} 
= - \eta_{33} = -1$ and $\eta_{\mu\nu} = 0$ for $\mu \neq \nu$. Moreover, $x^\mu = ( x^0, x^i ) = ( t, x^i )$
and $p^\mu = ( p^0, p^i ) = ( E, p^i )$.} 
\begin{eqnarray}
{\cal L}_0 = {\cal L}_{EH} + {\cal L}_{R^2} + {\cal L}_{C^2},
\label{Class-Lag0}  
\end{eqnarray}
where each term on the right-hand side (RHS) is defined as
\begin{eqnarray}
{\cal L}_{EH} =  \sqrt{-g} \, \frac{1}{2 \kappa^2} R,   \qquad
{\cal L}_{R^2} =  \sqrt{-g} \, \alpha_R R^2,   \qquad
{\cal L}_{C^2} = - \sqrt{-g} \, \alpha_C C_{\mu\nu\rho\sigma} C^{\mu\nu\rho\sigma},
\label{Class-Lag1}  
\end{eqnarray}
where $\kappa$ is defined as $\kappa^2 = 8 \pi G = \frac{1}{M_{Pl}^2}$ through the Newton constant $G$ and the
reduced Planck mass $M_{Pl}$, $R$ the scalar curvature, $\alpha_R, \alpha_C$ dimensionless positive coupling constants, 
and $C_{\mu\nu\rho\sigma}$ is conformal tensor defined as
\begin{eqnarray}
C_{\mu\nu\rho\sigma} &=& R_{\mu\nu\rho\sigma} - \frac{1}{2} ( g_{\mu\rho} R_{\nu\sigma}
- g_{\mu\sigma} R_{\nu\rho} - g_{\nu\rho} R_{\mu\sigma} + g_{\nu\sigma} R_{\mu\rho} )
\nonumber\\
&+& \frac{1}{6} ( g_{\mu\rho} g_{\nu\sigma} - g_{\mu\sigma} g_{\nu\rho} ) R.
\label{C-tensor}  
\end{eqnarray}

Since the Lagrangians ${\cal L}_{R^2}$ and ${\cal L}_{C^2}$ involve higher-derivative terms, it is necessary to rewrite them
into the first-order form for the canonical quantization procedure by introducing an auxiliary, symmetric tensor field
$K_{\mu\nu} = K_{\nu\mu}$. Furthermore, we have to introduce a St\"{u}ckelberg-like vector field $A_\mu$ 
in order to avoid the second-class constraint \cite{Kimura1, Kimura2, Kimura3}.  Then, the Lagrangians ${\cal L}_{R^2}$ 
and ${\cal L}_{C^2}$ can be cast to the form
\begin{eqnarray}
{\cal L}_{HD} \equiv {\cal L}_{R^2} + {\cal L}_{C^2} = \sqrt{-g} \, \Bigl[ \gamma G_{\mu\nu} K^{\mu\nu} 
+ \beta_1 ( K_{\mu\nu} - \nabla_\mu A_\nu - \nabla_\nu A_\mu )^2 
+ \beta_2 ( K - 2 \nabla_\rho A^\rho )^2 \Bigr],
\label{Mod-Lag}  
\end{eqnarray}
where $G_{\mu\nu} \equiv R_{\mu\nu} - \frac{1}{2} g_{\mu\nu} R$ denotes the Einstein tensor,
and $\gamma, \beta_1$ and $\beta_2$ are dimensionless coupling constants which obey a relation:
\begin{eqnarray}
\alpha_C = \frac{\gamma^2}{8 \beta_1},  \qquad
\alpha_R = - \frac{( \beta_1 + \beta_2 ) \gamma^2}{12 \beta_1 ( \beta_1 + 4 \beta_2 )}.
\label{Couplings}  
\end{eqnarray}
Here we assume that  $\beta_1 > 0, \beta_1 \neq - \beta_2, \beta_1 \neq 0, \beta_1 + 4 \beta_2 \neq 0$. 
In case of $\beta_1 = - \beta_2$, we have the well-known conformal gravity \cite{Oda-Ohta, Oda-Conf}.
It is easy to see that carrying out the path integral over $K_{\mu\nu}$ in ${\cal L}_{HD}$ produces 
the Lagrangian of ${\cal L}_{R^2}$ plus ${\cal L}_{C^2}$.

Actually, taking the variation of $K_{\mu\nu}$ leads to the equation:
\begin{eqnarray}
K_{\mu\nu} - \nabla_\mu A_\nu - \nabla_\nu A_\mu + \frac{\beta_2}{\beta_1} g_{\mu\nu} ( K - 2 \nabla_\rho A^\rho )
= - \frac{\gamma}{2 \beta_1} G_{\mu\nu}.
\label{K-var}  
\end{eqnarray}
Moreover, taking the trace of this equation yields
\begin{eqnarray}
K - 2 \nabla_\rho A^\rho = \frac{\gamma}{2 ( \beta_1 + 4 \beta_2 )} R.
\label{K-trace}  
\end{eqnarray}
Inserting (\ref{K-trace}) to (\ref{K-var}) gives us the expression of $K_{\mu\nu}$:
\begin{eqnarray}
K_{\mu\nu} = \nabla_\mu A_\nu + \nabla_\nu A_\mu - \frac{\gamma}{2 \beta_1} 
\left[ R_{\mu\nu} - \frac{\beta_1 + 2 \beta_2}{2 ( \beta_1 + 4 \beta_2 )} g_{\mu\nu} R \right].
\label{K-form}  
\end{eqnarray}
Finally, substituting Eqs. (\ref{K-trace}) and (\ref{K-form}) into ${\cal L}_{HD}$ in (\ref{Mod-Lag})
and using the relation (\ref{Couplings}), we can arrive at the Lagrangian ${\cal L}_{R^2}$
plus ${\cal L}_{C^2}$ in (\ref{Class-Lag1}) up to surface terms. This can be achieved by use of the identity
\begin{eqnarray}
C_{\mu\nu\rho\sigma}^2  = I + 2 R_{\mu\nu}^2 - \frac{2}{3} R^2,
\label{C-ident}  
\end{eqnarray}
where $I$ is defined as
\begin{eqnarray}
I = R_{\mu\nu\rho\sigma}^2 - 4 R_{\mu\nu}^2 + R^2,
\label{E-def}  
\end{eqnarray}
which is locally a total derivative in four dimensions.
As a classical Lagrangian ${\cal L}_c$ we shall take a linear combination of the Einstein-Hilbert term ${\cal L}_{EH}$
and ${\cal L}_{HD}$:
\begin{eqnarray}
{\cal L}_c &\equiv& {\cal L}_{EH} + {\cal L}_{HD}
\nonumber\\
&=& \sqrt{-g} \Bigl[ \frac{1}{2 \kappa^2} R + \gamma G_{\mu\nu} K^{\mu\nu} 
+ \beta_1 ( K_{\mu\nu} - \nabla_\mu A_\nu - \nabla_\nu A_\mu )^2 + \beta_2 ( K - 2 \nabla_\rho A^\rho )^2 \Bigr].
\label{Class-Lag}  
\end{eqnarray}
   
The classical Lagrangian ${\cal L}_c$ is invariant under both general coordinate transformation (GCT)
and the St\"{u}ckelberg transformation. The infinitesimal GCT is described as
\begin{eqnarray}
&{}& \delta^{(1)} g_{\mu\nu} = - ( \nabla_\mu \xi_\nu + \nabla_\nu \xi_\mu )
= - ( \xi^\alpha \partial_\alpha g_{\mu\nu} + \partial_\mu \xi^\alpha g_{\alpha\nu} 
+ \partial_\nu \xi^\alpha g_{\alpha\mu} ), 
\nonumber\\
&{}& \delta^{(1)} K_{\mu\nu} = - \xi^\alpha \nabla_\alpha K_{\mu\nu} - \nabla_\mu \xi^\alpha K_{\alpha\nu}
- \nabla_\nu \xi^\alpha K_{\mu\alpha},
\nonumber\\
&{}& \delta^{(1)} A_\mu = - \xi^\alpha \nabla_\alpha A_\mu - \nabla_\mu \xi^\alpha A_\alpha.
\label{GCT}  
\end{eqnarray}
while the St\"{u}ckelberg transformation takes the form:
\begin{eqnarray}
\delta^{(2)} g_{\mu\nu} =  0, \qquad
\delta^{(2)} K_{\mu\nu} = \nabla_\mu \varepsilon_\nu + \nabla_\nu \varepsilon_\mu, 
\qquad
\delta^{(2)} A_\mu = \varepsilon_\mu.
\label{Stuckel}  
\end{eqnarray}
In the above, $\xi_\mu$ and $\varepsilon_\mu$ are infinitesimal transformation parameters.

To close this section, let us count the number of phyical degrees of freedom since it is known that 
this counting is more subtle in higher derivative theories than in conventional second-order derivative
theories. In the formalism at hand, however, the introduction of the auxiliary field
$K_{\mu\nu}$ makes it possible to rewrite the Lagrangians with fourth-order derivatives to a second-order
derivative theory, so we can apply the usual counting method. The fields $g_{\mu\nu}, K_{\mu\nu}$ and $A_\mu$
have 10, 10 and 4 degrees of freedom, respectively. We have two kinds of local symmetries, those are,
the GCT and St\"{u}ckelberg symmetries with 4 and 4 degrees of freedom, respectively. Thus,
we have totally $(10 + 10 + 4) - (4 + 4) \times 2 = 8$ physical degrees of freedom, which will turn out to
be a massive scalar field of 1 physical degree and the massless graviton of 2 degrees, both of which have positive-definite 
norm, and the spin 2 massive ghost of 5 degrees with negative norm. In considering the construction of 
a renormalizable and unitary quantum gravity, it is worthwhile to highlight that both the massive scalar and
the massless graviton, which have positive norm, comes from the Einstein-Hilbert term ${\cal L}_{EH}$ and
the $R^2$ term ${\cal L}_{R^2}$ whereas the massive ghost with negative norm does from the Lagrangian of
conformal gravity, ${\cal L}_{C^2}$. Thus, without ${\cal L}_{C^2}$, the theory becomes unitary but the renormalization
forces this term to appear in the action, thereby violating the unitarity.

\section{Quantum quadratic gravity}

Since we have established our classical theory in the previous section, in this section we turn our attention to 
its quantum aspects. The classical Lagrangian (\ref{Class-Lag}) is invariant under general coordinate transformation (GCT) in (\ref{GCT})
and St\"{u}ckelberg transformation in (\ref{Stuckel}). In quantum field theories, these local gauge symmetries must be fixed by
gauge conditions. The appropriate gauge fixing condition for the GCT, which preserves the maximal global symmetry, i.e. the general
linear transformation $GL(4)$, is provided by the de Donder gauge condition (or the harmonic gauge condition):
\begin{eqnarray}
\partial_\mu \tilde g^{\mu\nu} = 0,
\label{Donder}  
\end{eqnarray}
where we have defined $\tilde g^{\mu\nu} \equiv \sqrt{-g} g^{\mu\nu}$.  It seems that there could be a higher-derivative
extension of the de Donder gauge condition since quadratic gravity includes fourth-order derivatives. However, such 
the extension does not make sense owing to the metric condition $\nabla_\rho g_{\mu\nu} = 0$. Note that
the requirement of the $GL(4)$ symmetry forbids the flat Minkowski metric $\eta_{\mu\nu}$ to appear in the
gauge fixing condition. For the St\"{u}ckelberg transformation, we will impose the gauge condition on $K_{\mu\nu}$:
\begin{eqnarray}
\nabla_\mu K^{\mu\nu} = 0,
\label{K-gauge}  
\end{eqnarray}
which is also invariant under the $GL(4)$.  

After gauge fixing, the Lagrangian is not invariant under the local gauge transformations any longer, but it is invariant
under the corresponding BRST transformations, which are obtained from classical transformations by replacing the infinitesimal transformation 
parameters $\xi_\mu$ and $\varepsilon_\mu$ with Faddeev-Popov (FP) ghosts $c_\mu$ and $\zeta_\mu$, respectively. 
The BRST transformation for the GCT, which we call GCT BRST transformation, is then given by
\begin{eqnarray}
&{}& \delta^{(1)}_B g_{\mu\nu} = - ( \nabla_\mu c_\nu + \nabla_\nu c_\mu )
= - ( c^\alpha \partial_\alpha g_{\mu\nu} + \partial_\mu c^\alpha g_{\alpha\nu} 
+ \partial_\nu c^\alpha g_{\alpha\mu} ), 
\nonumber\\
&{}& \delta^{(1)}_B K_{\mu\nu} = - c^\alpha \nabla_\alpha K_{\mu\nu} - \nabla_\mu c^\alpha K_{\alpha\nu}
- \nabla_\nu c^\alpha K_{\mu\alpha}, 
\nonumber\\
&{}& \delta^{(1)}_B A_\mu = - c^\alpha \nabla_\alpha A_\mu - \nabla_\mu c^\alpha A_\alpha, \qquad
\delta^{(1)}_B c^\mu = - c^\alpha \partial_\alpha c^\mu,
\nonumber\\
&{}& \delta^{(1)}_B \bar c_\mu = i B_\mu, \qquad
\delta^{(1)}_B B_\mu = 0, \qquad
\delta^{(1)}_B b_\mu = - c^\alpha \partial_\alpha b_\mu,
\label{GCT-BRST}  
\end{eqnarray}
where $\bar c_\mu$ and $B_\mu$ are respectively an antighost and a Nakanishi-Lautrup (NL) field, and
a new NL field $b_\mu$ is defined as
\begin{eqnarray}
b_\mu = B_\mu - i c^\alpha \partial_\alpha \bar c_\mu,
\label{new-b}  
\end{eqnarray}
which will be used in place of $B_\mu$ in what follows.
The BRST transformation for the St\"{u}ckelberg transformation, which we call ST BRST transformation, 
is of form:
\begin{eqnarray}
&{}& \delta^{(2)}_B g_{\mu\nu} =  0, \qquad
\delta^{(2)}_B K_{\mu\nu} = \nabla_\mu \zeta_\nu + \nabla_\nu \zeta_\mu, 
\nonumber\\
&{}& \delta^{(2)}_B A_\mu = \zeta_\mu,  \qquad
\delta^{(2)}_B \bar \zeta_\mu = i \beta_\mu,  \qquad
\delta^{(2)}_B \zeta_\mu = \delta^{(2)}_B \beta_\mu = 0,
\label{ST-BRST}  
\end{eqnarray}
where $\bar \zeta_\mu$ and $\beta_\mu$ are respectively an antighost and a Nakanishi-Lautrup (NL) field. 

It is obvious that the two BRST transformations are nilpotent, $(\delta^{(1)}_B)^2 = (\delta^{(2)}_B)^2 = 0$.  
In order to make the two BRST transformations be anticommuting with each other, i.e.
$\{ \delta^{(1)}_B, \delta^{(2)}_B \} = 0$, the remaining BRST transformations are fixed to
\begin{eqnarray}
&{}& \delta^{(1)}_B \zeta_\mu = - c^\alpha \nabla_\alpha \zeta_\mu - \nabla_\mu c^\alpha \zeta_\alpha, \qquad
\delta^{(1)}_B \bar \zeta_\mu = - c^\alpha \nabla_\alpha \bar \zeta_\mu - \nabla_\mu c^\alpha \bar \zeta_\alpha,
\nonumber\\
&{}& \delta^{(1)}_B \beta_\mu = - c^\alpha \nabla_\alpha \beta_\mu - \nabla_\mu c^\alpha \beta_\alpha, \qquad
\delta^{(2)}_B b_\mu = \delta^{(2)}_B c^\mu = \delta^{(2)}_B \bar c_\mu = 0.
\label{Remain-BRST}  
\end{eqnarray}
 
Now that we have selected suitable gauge fixing conditions and established all the BRST transformations, 
it is straightforward to make a gauge fixed, BRST invariant quantum Lagrangian by following the standard recipe:
\begin{eqnarray}
{\cal L}_q &\equiv& {\cal L}_c + i \delta_B^{(1)} ( \tilde g^{\mu\nu} \partial_\mu \bar c_\nu ) 
+ i \delta_B^{(2)} ( \sqrt{-g} \bar \zeta_\nu \nabla_\mu K^{\mu\nu} )
\nonumber\\
&=& \sqrt{-g} \Bigl[ \frac{1}{2 \kappa^2} R + \gamma G_{\mu\nu} K^{\mu\nu} 
+ \beta_1 ( K_{\mu\nu} - \nabla_\mu A_\nu - \nabla_\nu A_\mu )^2 + \beta_2 ( K - 2 \nabla_\rho A^\rho )^2 \Bigr]
\nonumber\\
&-& \tilde g^{\mu\nu} \partial_\mu b_\nu - i \tilde g^{\mu\nu} \partial_\mu \bar c_\rho \partial_\nu c^\rho
+ \sqrt{-g} [ - \nabla_\mu K^{\mu\nu} \beta_\nu + i \nabla^\mu \bar  \zeta^\nu ( \nabla_\mu \zeta_\nu 
+ \nabla_\nu \zeta_\mu ) ],
\label{Quant-Lag}  
\end{eqnarray}
where surface terms are dropped.

Starting with the quantum Lagrangian (\ref{Quant-Lag}), it is straightforward to derive field equations by taking the variation
of each fundamental field in order whose result is given by
\begin{eqnarray}
&{}& \frac{1}{2 \kappa^2} G_{\mu\nu} - \frac{1}{2} g_{\mu\nu} ( \gamma G_{\rho\sigma} K^{\rho\sigma}
+ \beta_1 \hat K_{\rho\sigma}^2 + \beta_2 \hat K^2 )
+ \gamma \Big[ 2 G_{\rho(\mu} K_{\nu)}{}^\rho + \frac{1}{2} \Box K_{\mu\nu} 
+ \frac{1}{2} K_{\mu\nu} R 
\nonumber\\
&{}& - \frac{1}{2} ( R_{\mu\nu} + g_{\mu\nu} \Box - \nabla_\mu \nabla_\nu ) K \Big]
+ 2 \beta_1 [ \hat K_{\rho(\mu} \hat K_{\nu)}{}^\rho  + 2 \nabla_\rho ( A_{(\mu} \hat K_{\nu)}{}^\rho )
- \nabla_\rho ( \hat K_{\mu\nu} A^\rho ) ]
\nonumber\\
&{}& + 2 \beta_2 [ \hat K \hat K_{\mu\nu} + 2 \nabla_{(\mu} ( A_{\nu)} \hat K ) 
- g_{\mu\nu} \nabla_\rho ( A^\rho \hat K ) \Big] 
- \frac{1}{2} \Big( E_{\mu\nu} - \frac{1}{2} g_{\mu\nu} E ) + L_{\mu\nu} = 0.
\label{Field-eq1}
\\
&{}& \hat K_{\mu\nu} + \frac{\beta_2}{\beta_1} g_{\mu\nu} \hat K = - \frac{\gamma}{2 \beta_1} G_{\mu\nu} 
- \frac{1}{2 \beta_1} \nabla_{(\mu} \beta_{\nu)}. 
\label{Field-eq2}
\\
&{}& \nabla^\mu ( \hat K_{\mu\nu} + \frac{\beta_2}{\beta_1} g_{\mu\nu} \hat K ) = 0. 
\label{Field-eq3}
\\
&{}& \partial_\mu \tilde g^{\mu\nu} = 0.
\label{Field-eq4}
\\
&{}& g^{\mu\nu} \partial_\mu \partial_\nu c^\rho = g^{\mu\nu} \partial_\mu \partial_\nu \bar c_\rho = 0.
\label{Field-eq5}
\\
&{}& \nabla_\mu K^{\mu\nu} = 0.
\label{Field-eq6}
\\
&{}& \nabla^\mu ( \nabla_\mu \zeta_\nu + \nabla_\nu \zeta_\mu ) = \nabla^\mu ( \nabla_\mu \bar \zeta_\nu 
+ \nabla_\nu \bar \zeta_\mu ) = 0.
\label{Field-eq7}
\end{eqnarray}
In the above, we have defined the following quantities:
\begin{eqnarray}
&{}& \Box = g^{\mu\nu} \nabla_\mu \nabla_\nu, \qquad
E_{\mu\nu} = \partial_\mu b_\nu+ i \partial_\mu\bar c_\rho\partial_\nu c^\rho+ ( \mu\leftrightarrow \nu),  \qquad
E = g^{\mu\nu} E_{\mu\nu},
\nonumber\\
&{}& \hat K_{\mu\nu} = K_{\mu\nu} - \nabla_\mu A_\nu - \nabla_\nu A_\mu,  \qquad
\hat K = g^{\mu\nu} \hat K_{\mu\nu} = K - 2 \nabla_\rho A^\rho,
\nonumber\\
&{}& L_{\mu\nu} = K_{\rho(\mu} \nabla_{\nu)} \beta^\rho - \frac{1}{2} g_{\mu\nu} K^{\rho\sigma} \nabla_\rho \beta_\sigma
+ \frac{1}{2} \nabla_\rho ( K_{\mu\nu} \beta^\rho )
\nonumber\\
&{}& + i [ ( \nabla_{(\mu} \bar \zeta^\rho + \nabla^\rho \bar \zeta_{(\mu} ) ( \nabla_{\nu)} \zeta_\rho
+ \nabla_{|\rho|} \zeta_{\nu)} ) + \nabla^\rho \bar \zeta_{(\mu} ( \nabla_{|\rho|} \zeta_{\nu)} + \nabla_{\nu)} \zeta_\rho )
- \frac{1}{2} g_{\mu\nu} \nabla^\rho \bar \zeta^\sigma ( \nabla_\rho \zeta_\sigma + \nabla_\sigma \zeta_\rho ) ]
\nonumber\\
&{}& - i \nabla^\rho [ \bar \zeta_{(\mu} ( \nabla_{\nu)} \zeta_\rho + \nabla_{|\rho|} \zeta_{\nu)} )
- \bar \zeta_\rho \nabla_{(\mu} \zeta_{\nu)} + \nabla_{(\mu} \bar \zeta_{|\rho|} \zeta_{\nu)}
+ \nabla_\rho \bar \zeta_{(\mu} \zeta_{\nu)} - \nabla_{(\mu} \bar \zeta_{\nu)} \zeta_\rho ].
\label{Various-defs}
\end{eqnarray}
We have also introduced symmetrization with weight one by round bracket, e.g. $A_{(\mu} B_{\nu)} \equiv
\frac{1}{2} ( A_\mu B_\nu + A_\nu B_\mu )$. 

The GCT BRST transformation of the field equation for $\bar c_\rho$ in (\ref{Field-eq5}) 
enables us to derive the field equation for $b_\rho$ \cite{Oda-Saake}:
\begin{eqnarray}
g^{\mu\nu} \partial_\mu \partial_\nu b_\rho = 0.
\label{b-rho-eq}  
\end{eqnarray}
We can check that this field equation is also obtained by acting the covariant derivative $\nabla^\mu$ on the Einstein equation 
(\ref{Field-eq1}).

In other words, setting $X^M = \{ x^\mu, b_\mu, c^\mu, \bar c_\mu \}$,\footnote{The fact that the space-time coordinates $x^\mu$
belong to a set of fields might suggest that $x^\mu$ would be promoted to quantum fields in quantum gravity.}$ X^M$ turns out to obey 
the very simple equation: 
\begin{eqnarray}
g^{\mu\nu} \partial_\mu \partial_\nu X^M = 0.
\label{Alembert-eq}  
\end{eqnarray}
With the help of the de Donder gauge condition (\ref{Donder}), this equation provides us with two kinds of 
conserved currents
\begin{eqnarray}
{\cal{P}}^{\mu M} &=& g^{\mu\nu} \partial_\nu X^M = g^{\mu\nu} ( 1 \overleftrightarrow{\partial}_\nu X^M ),
\nonumber\\
{\cal{M}}^{\mu M N} &=& g^{\mu\nu} ( X^M \overleftrightarrow{\partial}_\nu Y^N ),
\label{IOSp-current}  
\end{eqnarray}
and the corresponding charges
\begin{eqnarray}
{\cal{P}}^{M} = \int d^3 x \, {\cal{P}}^{0 M}, \qquad 
{\cal{M}}^{M N} = \int d^3 x \, {\cal{M}}^{0 M N},
\label{IOSp-charge}  
\end{eqnarray}
where $X^M \overleftrightarrow{\partial}_\mu Y^N \equiv X^M \partial_\mu Y^N -
(\partial_\mu X^M) Y^N$. Using these charges, we can show that there is a Poincare-like symmetry $IOSp (8|8)$ 
as in general relativity \cite{Nakanishi, N-O-text}. Note that ${\cal{P}}^{M}$ and ${\cal{M}}^{M N}$ represent
$16$ and $128$ symmetry generators, respectively, so we have totally $144$ generators. 

Here it is worth commenting on this global symmetry in more detail since in quantum gravity the meaning 
of global symmetries is more subtle than that in elementary particle physics. It is well known that
global symmetries except for those relevant to the geometry such as the angular momentum are broken 
by the no-hair theorem of black holes \cite{MTW}. Moreover, in string theory, which is a strong candidate 
of quantum gravity, we never get any additive conservation laws and at least in known string vacua, 
the additive global symmetries turn out to be either gauge symmetries or explicitly violated \cite{Banks}.
Even in such a situation, in quantum gravity we have important $\it{quantum}$ global symmetries such as 
the BRST symmetry and the conservation law of the ghost number. The BRST symmetry, which is a residual global symmetry
emerging after the gauge-fixing procedure of local gauge symmetries, plays a role in proving the unitarity of the theory 
and deducing the Ward identities among the Green functions \cite{Kugo-Ojima}. On the other hand, the conservation 
of the ghost number is violated in a two dimensional quantum theory, and leads to the ghost number anomaly,
which is closely related to the Riemann-Roch theorem on the closed Riemann surfaces \cite{Peskin}. 
From this viewpoint, it is valuable to derive global symmetries as many as possible, which include
the BRST symmetry and the symmetry of the FP ghost number, in quantum gravity. If such the symmetries are 
purely built from quantum fields such as the FP (anti)ghosts and the Nakanishi-Lautrup auxiliary fields, 
they could escape from the no-hair theorem of black holes, and might have some important applications.
The Poincare-like symmetry $IOSp (8|8)$ is such a quantum symmetry and includes the BRST symmetry and
ghost number conservation law as a subgroup, so it might shed some light on the future development of 
quantum gravity.

\section{Canonical commutation relations}

In this section, we derive the concrete expressions of canonical conjugate momenta and set up the canonical 
(anti)commutation relations (CCRs), which will be used in evaluating various equal-time (anti)commutation relations (ETCRs) 
among fundamental variables in the next section. To simplify various expressions, we obey the following abbreviations 
adopted in the textbook of Nakanishi and Ojima \cite{N-O-text}:
\begin{eqnarray}
[ A, B^\prime ] &=& [ A(x), B(x^\prime) ] |_{x^0 = x^{\prime 0}},
\qquad \delta^3 = \delta(\vec{x} - \vec{x}^\prime), 
\nonumber\\
\tilde f &=& \frac{1}{\tilde g^{00}} = \frac{1}{\sqrt{-g} g^{00}},
\label{abbreviation}  
\end{eqnarray}
where we assume that $\tilde g^{00}$ is invertible. 

To remove second-order derivatives of the metric tensor involved in $R$ and $G_{\mu\nu}$, and regard $b_\mu$
and $\beta_\mu$ as non-canonical variables, we perform the integration by parts and rewrite the Lagrangian 
(\ref{Quant-Lag}) as
\begin{eqnarray}
{\cal L}_q &=& - \frac{1}{2 \kappa^2} \tilde g^{\mu\nu} ( \Gamma^\alpha_{\sigma\alpha} \Gamma^\sigma_{\mu\nu} 
- \Gamma^\alpha_{\sigma\nu} \Gamma^\sigma_{\mu\alpha} ) 
- \gamma \sqrt{-g} ( \Gamma^\alpha_{\mu\nu} \partial_\alpha - \Gamma^\alpha_{\mu\alpha} \partial_\nu
+ \Gamma^\beta_{\mu\alpha} \Gamma^\alpha_{\beta\nu} - \Gamma^\alpha_{\mu\alpha} \Gamma^\beta_{\nu\beta} )
\bar K^{\mu\nu} 
\nonumber\\
&+& \beta_1 \sqrt{-g} ( K_{\mu\nu} - \nabla_\mu A_\nu - \nabla_\nu A_\mu )^2 + \beta_2 \sqrt{-g} ( K - 2 \nabla_\mu A^\mu )^2  
+ \partial_\mu \tilde g^{\mu\nu} b_\nu
\nonumber\\
&-&  i \tilde g^{\mu\nu} \partial_\mu \bar c_\rho \partial_\nu c^\rho - \sqrt{-g} \, \nabla_\mu K^{\mu\nu} \cdot \beta_\nu 
+ i \sqrt{-g} \, \nabla^\mu \bar \zeta^\nu ( \nabla_\mu \zeta_\nu + \nabla_\nu \zeta_\mu ) + \partial_\mu {\cal{V}}^\mu,
\label{Can-Quant-Lag}  
\end{eqnarray}
where $\bar K_{\mu\nu}$ is defined as
\begin{eqnarray}
\bar K_{\mu\nu} \equiv K_{\mu\nu} - \frac{1}{2} g_{\mu\nu} K,  \qquad
\bar K \equiv g^{\mu\nu} \bar K_{\mu\nu} = - K,
\label{bar-K}  
\end{eqnarray}
and a surface term ${\cal{V}}^\mu$ is given by
\begin{eqnarray}
{\cal{V}}^\mu =  \frac{1}{2 \kappa^2} ( \tilde g^{\alpha\beta} \Gamma^\mu_{\alpha\beta} 
- \tilde g^{\mu\nu} \Gamma^\alpha_{\nu\alpha} ) 
+ \gamma \sqrt{-g} ( \Gamma^\mu_{\alpha\beta} \bar K^{\alpha\beta} - \Gamma^\alpha_{\alpha\nu} \bar K^{\mu\nu} )
- \tilde g^{\mu\nu} b_\nu.
\label{surface}  
\end{eqnarray}
Since the NL fields $b_\mu$ and $\beta_\mu$ have no derivatives in ${\cal L}_q$, we can regard them as non-canonical 
variables.

Since the Lagrangian (\ref{Can-Quant-Lag}) is now written in the first-order formalism, it is straightforward to derive 
the concrete expressions of canonical conjugate momenta. The result is given by
\begin{eqnarray}
\pi_g^{\mu\nu} &=& \frac{\partial {\cal L}_q}{\partial \dot g_{\mu\nu}} 
\nonumber\\
&=& - \frac{1}{4 \kappa^2} \sqrt{-g} \, \Bigl[ - g^{0 \lambda} g^{\mu\nu} g^{\sigma\tau} 
- g^{0 \tau} g^{\mu\lambda} g^{\nu\sigma} - g^{0 \sigma} g^{\mu\tau} g^{\nu\lambda} 
+ g^{0 \lambda} g^{\mu\tau} g^{\nu\sigma} 
+ g^{0 \tau} g^{\mu\nu} g^{\lambda\sigma}
\nonumber\\
&+& \frac{1}{2} ( g^{0 \mu} g^{\nu\lambda} + g^{0 \nu} g^{\mu\lambda} ) g^{\sigma\tau} \Biggr] \partial_\lambda g_{\sigma\tau}
- \frac{1}{2} \sqrt{-g} \, ( g^{0 \mu} g^{\nu\rho} + g^{0 \nu} g^{\mu\rho} - g^{0 \rho} g^{\mu\nu} )  b_\rho
\nonumber\\
&-& \gamma \sqrt{-g} \, \Biggl\{ \nabla^{(\mu} \bar K^{\nu)0} - \frac{1}{2} \nabla^0 \bar K^{\mu\nu} 
- \frac{1}{2} g^{\mu\nu} \partial_\rho \bar K^{0\rho} - g^{\mu\nu} \Gamma^\rho_{\rho\sigma} \bar K^{0\sigma}
- \Gamma^0_{\rho\sigma} \Big[ 2 g^{\rho(\mu} \bar K^{\nu)\sigma}
\nonumber\\
&+& \frac{1}{2} ( g^{\rho(\mu} g^{\nu)\sigma} K - g^{\rho\sigma} K^{\mu\nu} ) \Big]  
+ \Gamma^\sigma_{\rho\sigma} \Big[ g^{\rho(\mu} \bar K^{\nu)0} + g^{0(\mu} \bar K^{\nu)\rho} 
+ \frac{1}{2} ( g^{\rho(\mu} g^{\nu)0} K - g^{0\rho} K^{\mu\nu} ) \Big] \Biggr\}
\nonumber\\
&+& 2 \beta_1 \sqrt{-g} \, ( 2 \hat K^{0(\mu} A^{\nu)}  - \hat K^{\mu\nu} A^0 ) 
+ 2 \beta_2 \sqrt{-g} \, \hat K ( 2 g^{0(\mu} A^{\nu)} - g^{\mu\nu} A^0 )
\nonumber\\
&+& \frac{1}{2} \sqrt{-g} \, ( 2 g^{0 (\mu} K^{\nu)\rho} + g^{0\rho} K^{\mu\nu} - g^{\mu\nu} K^{0 \rho} ) \beta_\rho
\nonumber\\
&-& i \sqrt{-g} \, ( \bar \zeta^{(\mu} \nabla^{\nu)} \zeta^0 + \bar \zeta^{(\mu} \nabla^{|0|} \zeta^{\nu)}
- \bar \zeta^0 \nabla^{(\mu} \zeta^{\nu)} + \nabla^{(\mu} \bar \zeta^{|0|} \zeta^{\nu)} + \nabla^0 \bar \zeta^{(\mu} \zeta^{\nu)} 
- \nabla^{(\mu} \bar \zeta^{\nu)} \zeta^0 ),
\nonumber\\
\pi_K^{\mu\nu} &=& \frac{\partial {\cal L}_q}{\partial \dot{K}_{\mu\nu}} 
= - \gamma \sqrt{-g} \, \Bigl[ ( g^{\mu\rho} g^{\nu\sigma} - \frac{1}{2} g^{\mu\nu} g^{\rho\sigma} ) 
\Gamma^0_{\rho\sigma}
- \frac{1}{2} ( g^{0\mu} g^{\nu\rho} + g^{0\nu} g^{\mu\rho} - g^{\mu\nu} g^{0\rho} ) \Gamma^\sigma_{\rho\sigma} \Bigr]
\nonumber\\ 
&-& \frac{1}{2} \sqrt{-g} \, ( g^{0\mu} \beta^\nu + g^{0\nu} \beta^\mu ),
\nonumber\\ 
\pi_A^\mu &=& \frac{\partial {\cal L}_q}{\partial \dot{A}_\mu} = - 4 \sqrt{-g} \, ( \beta_1 \hat K^{0\mu} + \beta_2 g^{0\mu} \hat K ),
\nonumber\\
\pi_{c \mu} &=& \frac{\partial {\cal L}_q}{\partial \dot c^\mu} = - i \tilde g^{0 \nu} \partial_\nu \bar c_\mu,  \qquad
\pi_{\bar c}^\mu = \frac{\partial {\cal L}_q}{\partial \dot {\bar c}_\mu} = i \tilde g^{0 \nu} \partial_\nu c^\mu,
\nonumber\\
\pi_\zeta^\mu &=& \frac{\partial {\cal L}_q}{\partial \dot \zeta_\mu} = i \sqrt{-g} \, ( \nabla^\mu \bar \zeta^0
+ \nabla^0 \bar \zeta^\mu ),  \qquad
\pi_{\bar \zeta}^\mu = \frac{\partial {\cal L}_q}{\partial \dot {\bar \zeta}_\mu} = - i \sqrt{-g} \, ( \nabla^\mu \zeta^0
+ \nabla^0 \zeta^\mu ),
\label{CCM}  
\end{eqnarray}
where the dot stands for the derivative with respect to time, e.g. $\dot g_{\mu\nu} \equiv \frac{\partial g_{\mu\nu}}{\partial t}
\equiv \frac{\partial g_{\mu\nu}}{\partial x^0} \equiv \partial_0 g_{\mu\nu}$, and the differentiation of ghosts is taken 
from the right.  

Next let us set up the canonical (anti)commutation relations (CCRs): 
\begin{eqnarray}
&{}& [ g_{\mu\nu}, \pi_g^{\prime\rho\lambda} ] = [ K_{\mu\nu}, \pi_K^{\prime\rho\lambda} ] 
= i \frac{1}{2} ( \delta_\mu^\rho\delta_\nu^\lambda 
+ \delta_\mu^\lambda\delta_\nu^\rho) \delta^3,   \qquad
[ A_\mu, \pi_A^{\prime\nu} ] = i \delta_\mu^\nu \delta^3, 
\nonumber\\
&{}& \{ c^\mu, \pi_{c \nu}^\prime \} = \{ \bar c_\nu, \pi_{\bar c}^{\prime\mu} \}
= \{ \bar{\zeta}_\nu, \pi_{\bar \zeta}^{\prime\mu} \} = \{ \zeta_\nu, \pi_\zeta^{\prime\mu} \} = i \delta_\nu^\mu \delta^3,      
\label{CCRs}  
\end{eqnarray}
where the other (anti)commutation relations vanish. 
In setting up these CCRs, it is valuable to distinguish non-canonical variables from canonical ones. 
Recall again that in our formalism, the NL fields $b_\mu$ and $\beta_\mu$ are not canonical variables.

Finally, for later convenience, let us consider the de Donder gauge condition (\ref{Donder}), 
from which we have identities:
\begin{eqnarray}
g^{\mu\nu} \Gamma^\lambda_{\mu\nu} = 0, \qquad  g^{\lambda\mu} \partial_\lambda g_{\mu\nu} 
= \Gamma^\lambda_{\lambda\nu}.
\label{Donder-iden}  
\end{eqnarray}
These identities and the de Donder gauge condition will be often utilized to obtain the equal-time commutation
relations later. Moreover, since the equation $g^{\mu\nu} \Gamma^\lambda_{\mu\nu} = 0$ reads 
\begin{equation}
( 2 g^{\lambda\mu} g^{\nu\rho} - g^{\mu\nu} g^{\lambda\rho} )\partial_\rho g_{\mu\nu}=0,
\label{Donder-iden2}
\end{equation}
it is possible to express the time derivative of the metric field in terms of its spacial one as
\begin{eqnarray}
{\cal{D}}^{\lambda\mu\nu} \dot g_{\mu\nu} = ( 2 g^{\lambda\mu} g^{\nu k} - g^{\mu\nu} g^{\lambda k} )
\partial_k g_{\mu\nu}, 
\label{D-eq}  
\end{eqnarray}
where the operator ${\cal{D}}^{\lambda\mu\nu}$ is defined by
\begin{eqnarray}
{\cal{D}}^{\lambda\mu\nu} = g^{0 \lambda} g^{\mu\nu} - 2 g^{\lambda\mu} g^{0 \nu}.
\label{D-op}  
\end{eqnarray}

\section{Equal-time commutation relations}

Since we have presented the canonical (anti)commutation relations (CCRs) in the previous section, 
in this section we will calculate various nontrivial equal-time (anti)commutation relations (ETCRs), 
which are necessary for the algebra of symmetries and computations in later sections,   
by using the CCRs, field equations and BRST transformations. 
In deriving ETCRs, we often use an identity for generic fields $\Phi$ and $\Psi$
\begin{eqnarray}
[ \Phi, \dot \Psi^\prime] = \partial_0 [ \Phi, \Psi^\prime] - [ \dot \Phi, \Psi^\prime],
\label{identity}  
\end{eqnarray}
which holds for the anticommutation relation as well. 

To begin with, we wish to deduce the ETCR between $g_{\mu\nu}$ and $b_\mu$, which is one of 
the important ETCRs and plays a role in proving the algebra of symmetries. For this purpose,
let us first consider the antiCCR, $\{ c^\mu, \pi_{c\nu}^\prime \} = i \delta_\nu^\mu \delta^3$,
which gives us 
\begin{eqnarray}
\{ c^\mu, \dot{\bar c}_\nu^\prime \} = - \tilde f \delta_\nu^\mu \delta^3.
\label{c-cbar-GCT}  
\end{eqnarray}
Next, we find that the CCR, $[ g_{\mu\nu}, \pi_{c\rho}^\prime ] = 0$ leads to
\begin{eqnarray}
[ g_{\mu\nu}, \dot{\bar c}_\rho^\prime ] = [ \dot g_{\mu\nu}, \bar c_\rho^\prime ] = 0,
\label{g-cbar-GCT}  
\end{eqnarray}
where we have used the CCR, $[ g_{\mu\nu}, \bar c_\rho^\prime ] = 0$ and the formula (\ref{identity}).
It then turns out that the GCT BRST transformation (\ref{GCT-BRST}) of the CCR, $[ g_{\mu\nu}, 
\bar c_\rho^\prime ] = 0$ yields
\begin{eqnarray}
[ g_{\mu\nu}, b_\rho^\prime ] = - i \tilde f ( \delta_\mu^0 g_{\rho\nu} + \delta_\nu^0 g_{\rho\mu} ) \delta^3,
\label{g-b}  
\end{eqnarray}
where we have used Eqs. (\ref{new-b}), (\ref{c-cbar-GCT}) and (\ref{g-cbar-GCT}).
From this ETCR, we can easily derive the following ETCRs:
\begin{eqnarray}
&{}& [ g^{\mu\nu}, b_\rho^\prime ] = i \tilde f ( g^{\mu0} \delta_\rho^\nu+ g^{\nu0} \delta_\rho^\mu) \delta^3,
\nonumber\\
&{}& [ \tilde g^{\mu\nu}, b_\rho^\prime ] = i \tilde f ( \tilde g^{\mu0} \delta_\rho^\nu+ \tilde g^{\nu0} \delta_\rho^\mu 
- \tilde g^{\mu\nu} \delta_\rho^0 ) \delta^3.
\label{3-g-b}  
\end{eqnarray}

Now we are ready to evaluate a very important ETCR, $[ \dot g_{\rho\sigma}, g_{\mu\nu}^\prime ]$.
To this aim, let us focus on the canonical conjugate momentum $\pi_K^{\mu\nu}$, from which
we can describe $\dot g_{ij}$ as
\begin{eqnarray}
\dot g_{ij} = \frac{2}{\gamma} \tilde f \Big[ ( g_{i\mu} g_{j\nu} - \frac{1}{2} g_{ij} g_{\mu\nu} ) 
\pi_K^{\mu\nu} - \frac{1}{2} \tilde g_{ij} \beta^0 \Big]
+ \tilde f [ \tilde g^{0\rho} ( \partial_i g_{j\rho} + \partial_j g_{i\rho} ) 
- \tilde g^{0k} \partial_k g_{ij} ].
\label{dot-g-piK}  
\end{eqnarray}
This expression immediately gives us the ETCR:
\begin{eqnarray}
[ \dot g_{ij}, g_{\mu\nu}^\prime ] = 0.
\label{g-dot-g}  
\end{eqnarray}
Here we have used the ETCR:
\begin{eqnarray}
[ \beta_\rho, g_{\mu\nu}^\prime ] = 0,
\label{beta-g}  
\end{eqnarray}
which can be easily shown by taking the ST BRST transformation (\ref{ST-BRST})
of the CCR, $[ \bar \zeta_\rho, g_{\mu\nu}^\prime ] = 0$. 

For the purpose of calculating the remaining ETCR, $[ \dot g_{0\rho}, g_{\mu\nu}^\prime ]$ we utilize 
the de Donder gauge condition (\ref{Donder-iden}).
The former identity in (\ref{Donder-iden}) makes it possible to express $\dot g_{0\rho}$ in terms of $\dot g_{ij}$ 
as follows:
\begin{eqnarray}
\dot g_{00} &=& \tilde f ( \tilde g^{ij} \dot g_{ij} - 2 \tilde g^{i\rho} \partial_i g_{0\rho} ),
\nonumber\\
\dot g_{0i} &=& \tilde f \Big( - \tilde g^{0j} \dot g_{ij} + \frac{1}{2} \tilde g^{\rho\sigma} \partial_i g_{\rho\sigma} 
- \tilde g^{j\rho} \partial_j g_{\rho i} \Big).
\label{dot-g-0rho}  
\end{eqnarray}
Then, using Eq. (\ref{g-dot-g}), we find that $[ \dot g_{0\rho}, g_{\mu\nu}^\prime ] = 0$, so we have a surprisingly 
simple ETCR: 
\begin{eqnarray}
[ \dot g_{\rho\sigma}, g_{\mu\nu}^\prime ] = 0.
\label{Zero ETCR}  
\end{eqnarray}

Since the ETCR (\ref{Zero ETCR}) is one of the most important ETCRs in the theory at hand, 
let us present an alternative proof based on its symmetry. 
First, let us note that $[ \dot g_{\rho\sigma}, g_{\mu\nu}^\prime ]$ has a symmetry under the simultaneous exchange 
of $(\mu\nu) \leftrightarrow (\rho\sigma)$ and primed $\leftrightarrow$ unprimed as well as the usual symmetry
$\mu\leftrightarrow \nu$ and $\rho\leftrightarrow \sigma$, because the time derivative $\partial_0$ of 
$[g_{\rho\sigma}, g_{\mu\nu}^\prime ]=0$ leads to $[\dot g_{\rho\sigma}, g_{\mu\nu}^\prime ]=[\dot g_{\mu\nu}^\prime, g_{\rho\sigma} ]$.
Then, we can write down its generic expression:
\begin{eqnarray}
[ \dot g_{\rho\sigma}, g_{\mu\nu}^\prime ] &=& \Big\{ c_1 g_{\rho\sigma} g_{\mu\nu} + c_2 ( g_{\rho\mu} g_{\sigma\nu}
+ g_{\rho\nu} g_{\sigma\mu} )
\nonumber\\
&+& \sqrt{-g} \tilde f [ c_3 ( \delta_\rho^0 \delta_\sigma^0 g_{\mu\nu} + \delta_\mu^0 \delta_\nu^0 g_{\rho\sigma} )
+ c_4 ( \delta_\rho^0 \delta_\mu^0 g_{\sigma\nu} + \delta_\rho^0 \delta_\nu^0 g_{\sigma\mu} 
\nonumber\\
&+& \delta_\sigma^0 \delta_\mu^0 g_{\rho\nu} + \delta_\sigma^0 \delta_\nu^0 g_{\rho\mu} ) ] 
+ ( \sqrt{-g} \tilde f )^2 c_5 \delta_\rho^0 \delta_\sigma^0 \delta_\mu^0 \delta_\nu^0 \Big\} \delta^3,  
\label{dot-g-g}  
\end{eqnarray}
where $c_i ( i = 1, \cdots, 5)$ are suitable coefficients to be fixed shortly. Here we have assumed that this ETCR
is proportional to $\delta^3$, but our proof can be generalized to the case where it is proportional to
$\partial_k \delta^3$.

Second, we find that the ${\cal{D}}$-equation (\ref{D-eq}) stemming from the de Donder gauge condition 
gives rise to the equation
\begin{eqnarray}
[ {\cal{D}}^{\lambda\rho\sigma} \dot g_{\rho\sigma}, g_{\mu\nu}^\prime ] = 0, 
\label{D-eq2}  
\end{eqnarray}
from which we can derive relations among the coefficients:
\begin{eqnarray}
c_3 = 2 ( c_1 + c_2 ),  \qquad
c_4 = - c_2, \qquad
c_5 = 2 c_3.
\label{C3-C5}  
\end{eqnarray}

Finally, we further take account of the CCR, $[ \pi_K^{\alpha\beta}, g_{\mu\nu}^\prime ] = 0$
which can be cast to the form:
\begin{eqnarray}
[ g^{0\alpha} ( g^{\beta\sigma} g^{0\rho} - g^{0\beta} g^{\rho\sigma} ) 
+ g^{\alpha\rho} ( g^{0\beta} g^{0\sigma} - g^{00} g^{\beta\sigma} ) 
+ g^{\alpha\beta} ( g^{00} g^{\rho\sigma} - g^{0\rho} g^{0\sigma} ) ]
[ \dot g_{\rho\sigma}, g_{\mu\nu}^\prime ] = 0,
\label{pi-K-g}  
\end{eqnarray}
where Eq. (\ref{beta-g}) was used. This equation gives us 
\begin{eqnarray}
c_1 = c_2 = c_3 = 0.
\label{C1-C3}  
\end{eqnarray}
Thus, together with Eq. (\ref{C3-C5}), all the coefficients $c_i$ must identically vanish, which
provides an alternative proof of Eq. (\ref{Zero ETCR}). 

This vanishing ETCR is a peculiar feature of quadratic gravity, so it is valuable to investigate its
consequence further. First, we wish to calculate the ETCR, $[ \dot g_{\rho\sigma}, \dot g_{\mu\nu}^\prime ]$,
which is antisymmetric under the exchange of $(\mu\nu) \leftrightarrow (\rho\sigma)$ and primed $\leftrightarrow$ unprimed,
so it must have a generic expression:
\begin{eqnarray}
[ \dot g_{\rho\sigma}, \dot g_{\mu\nu}^\prime ] &=& [ c_1 ( \dot g_{\rho\sigma} g_{\mu\nu} 
- \dot g_{\mu\nu} g_{\rho\sigma} ) + c_2 \sqrt{-g} \tilde f ( \delta_\rho^0 \delta_\sigma^0 \dot g_{\mu\nu} 
- \delta_\mu^0 \delta_\nu^0 \dot g_{\rho\sigma}  ) ] \delta^3
\nonumber\\
&+& c_3 ( \sqrt{-g} \tilde f \delta_\rho^0 \delta_\sigma^0 g_{\mu\nu} g^{0k}
+ \sqrt{-g^\prime} \tilde f^\prime \delta_\mu^0 \delta_\nu^0 g_{\rho\sigma}^\prime g^{\prime 0k} )
\partial_k \delta^3
\nonumber\\
&+& \Big\{ [ c_4 g_{\rho\sigma} g_{\mu\nu} + c_5 ( g_{\rho\mu} g_{\sigma\nu} + g_{\rho\nu} g_{\sigma\mu}
+ g_{\sigma\mu} g_{\rho\nu} + g_{\sigma\nu} g_{\rho\mu} )
\nonumber\\
&+& c_6 \sqrt{-g} \tilde f ( \delta_\rho^0 \delta_\mu^0 g_{\sigma\nu} + \delta_\rho^0 \delta_\nu^0 g_{\sigma\mu}
+ \delta_\sigma^0 \delta_\mu^0 g_{\rho\nu} + \delta_\sigma^0 \delta_\nu^0 g_{\rho\mu} )
\nonumber\\
&+& c_7 (\sqrt{-g} \tilde f)^2 \delta_\rho^0 \delta_\sigma^0 \delta_\mu^0 \delta_\nu^0 ] g^{0k} 
+ ({\rm{unprimed}} \rightarrow {\rm{primed}}) \Big\} \partial_k \delta^3,  
\label{dot-g-dot-g}  
\end{eqnarray}
where $c_i ( i = 1, \cdots, 7)$ are some coefficients. From Eq. (\ref{D-eq}) coming from the Donder gauge condition
(\ref{Donder}), we can impose the operator ${\cal{D}}^{\lambda\rho\sigma}$ on (\ref{dot-g-dot-g}):
\begin{eqnarray}
[ {\cal{D}}^{\lambda\rho\sigma} \dot g_{\rho\sigma}, \dot g_{\mu\nu}^\prime ] = 0,
\label{D-eq3}  
\end{eqnarray}
where Eq. (\ref{Zero ETCR}) was used. This equation leads to $c_i = 0$, which means that
\begin{eqnarray}
[ \dot g_{\rho\sigma}, \dot g_{\mu\nu}^\prime ] = 0.
\label{dot-g-dot-g-result}  
\end{eqnarray}
Of course, we can also verify that this ECTR is correct by a direct calculation based on Eqs. (\ref{dot-g-piK}) and  
(\ref{dot-g-0rho}).\footnote{For instance, from Eq. (\ref{dot-g-piK}), we find that $[ \dot g_{ij}, \dot g_{kl}^\prime ] = 0$ 
requires us to prove $[ \beta_\mu, \beta_\nu^\prime ] = 0$ and $[ \beta_\mu, \pi_K^{\prime \rho\sigma} ] = 0$. 
The former equation can be shown by taking the ST BRST transformation of $[ \beta_\mu, \bar \zeta_\nu^\prime ] = 0$,
which is derived from $[ \pi_K^{\prime \rho\sigma}, \bar \zeta_\nu^\prime ] = 0$ and 
$[ \dot g_{\rho\sigma}, \bar \zeta_\nu^\prime ] = 0$ in Eq. (\ref{dot-g-Phi}). The latter equation can be obtained 
via $[ \dot g_{\rho\sigma}, \beta_\mu^\prime ] = 0$ in Eq. (\ref{dot-g-Phi}) and $[ \beta_\mu, \beta_\nu^\prime ] = 0$
as well as $[ g_{\rho\sigma}, \beta_\mu^\prime ] = 0$.} 
  
We cannot resist from deducing the more general ETCRs involving an arbitrary number of derivatives of the metric tensor.
For this aim, we find that the ETCRs (\ref{Zero ETCR}) and (\ref{dot-g-dot-g-result}) directly produce
\begin{eqnarray}
[ \ddot g_{\rho\sigma}, g_{\mu\nu}^\prime ] = 0.
\label{2dot-g-g}  
\end{eqnarray}
With the help of Eq. (\ref{D-eq}), this equation gives us a new equation:
\begin{eqnarray}
[ {\cal{D}}^{\lambda\rho\sigma} \dot g_{\rho\sigma}, \ddot g_{\mu\nu}^\prime ] = 0.
\label{D-dot-g-g}  
\end{eqnarray}
By using the symmetry $[ \dot g_{\rho\sigma}, \ddot g_{\mu\nu}^\prime ] = [ \dot g_{\mu\nu}^\prime, \ddot g_{\rho\sigma} ]$
and the similar argument to Eqs. (\ref{dot-g-g}) and (\ref{dot-g-dot-g}), we can also show that
\begin{eqnarray}
[ \dot g_{\rho\sigma}, \ddot g_{\mu\nu}^\prime ] = 0.
\label{dot-g-2dot-g}  
\end{eqnarray}
Proceeding along almost the same line of argument repeatedly, we can prove remarkably surprising ETCRs:
\begin{eqnarray}
\left[ \frac{\partial^m g_{\rho\sigma}}{\partial t^m}, \frac{\partial^n g_{\mu\nu}^\prime}{\partial t^n} \right] = 0,
\label{general-dot-g-dot-g}  
\end{eqnarray}
where $m, n = 0, 1, 2, \cdots$. These vanishing ETCRs enable us to do calculations involving the metric tensor 
and its time derivatives very easily, and indicate that the metric field itself has the trivial canonical structure 
in quadratic gravity. 

It is worthwhile to stress a few points relevant to the ETCRs (\ref{general-dot-g-dot-g}).
First, let us note that we can derive such remarkable ETCRs only in quadratic gravity where
$C_{\mu\nu\rho\sigma}^2$ and $R^2$ terms coexist in the action whereas both in conformal gravity with only
$C_{\mu\nu\rho\sigma}^2$ term \cite{Oda-Ohta} and in $f(R)$ gravity having $R^2$ and its higher
polynomials \cite{Oda-f} we do have the nonvanishing ETCRs. From this observation, it is tempting to imagine that
Eq. (\ref{general-dot-g-dot-g}) might be connected with the renormalizability of quadratic gravity.

Second, it is of interest to see whether such vanishing ETCRs between a dynamical variable and its time derivative 
occur in the other theories. As the simplest example, let us take account of quantum electrodynamics (QED).     
The gauge-fixing Lagrangian of QED in a flat space-time is usually given by\footnote{In order to take the
Landau gauge $\alpha = 0$ into consideration, one has to start with this Lagrangian instead of the more
familiar one, ${\cal{L}}_{gf} =  - \frac{1}{2 \alpha} (\partial^\mu A_\mu)^2$.}
\begin{eqnarray}
{\cal{L}}_{gf} = B \partial^\mu A_\mu + \frac{1}{2} \alpha B^2,
\label{GF-QED}  
\end{eqnarray}
where $A_\mu$ is the $U(1)$ gauge field, B the Nakanishi-Lautrup auxiliary field and $\alpha$ the
gauge parameter. Ignoring the kinetic term $- \frac{1}{4} F_{\mu\nu}^2$ for the time being, 
let us carry out the canonical quantization for ${\cal{L}}_{gf}$. This can be also regarded as the canonical
quantization for topological field theories where the classical action is trivially zero or some topological
invariant though Eq. (\ref{GF-QED}) is only the partial gauge fixing of topological symmetry,
$\delta A_\mu (x) = \varepsilon_\mu (x)$. The canonical conjugate momentum $\pi_0$ for $A_0$ is given by
\begin{eqnarray}
\pi_0 = \frac{\partial {\cal{L}}_{gf}}{\partial \dot A_0} = - B.
\label{pi0-QED}  
\end{eqnarray}
Then the canonical commutation relation, $[ A_0, \pi_0^\prime ] = i \delta^3$ reads
\begin{eqnarray}
[ A_0, B^\prime ] = - i \delta^3.
\label{AB-QED}  
\end{eqnarray}
By virtue of the field equation, $B = - \frac{1}{\alpha} \partial^\mu A_\mu$, we can obtain
\begin{eqnarray}
[ A_0, \dot A_0^\prime ] = i \alpha \delta^3.
\label{AA0-QED}  
\end{eqnarray}
With the Landau gauge $\alpha =0$, we have the vanishing ETCR:
\begin{eqnarray}
[ A_0, \dot A_0^\prime ] = 0.
\label{AA0-QED2}  
\end{eqnarray}
Someone might be anxious that the similar situation occurs in quadratic gravity as well. However,
there is a critical difference between the above topological QED and quadratic gravity.  The vanishing ETCR
in (\ref{AA0-QED}) depends on the gauge parameter $\alpha$, so it is not a physical result. This gauge-dependent 
result is very natural since the ETCR in (\ref{AA0-QED}) is the one between the gauge degree of freedom, $A_0$. 
If we add the kinetic term for the gauge field, $- \frac{1}{4} F_{\mu\nu}^2$ to the gauge-fixing 
Lagrangian in (\ref{GF-QED}), it turns out that the ETCR between the dynamical degrees of freedom
associated with $A_i ( i = 1, 2, 3)$ and its derivatives takes the nonvanishing value:
\begin{eqnarray}
[ A_i, \dot A_\mu^\prime ] = i \delta_{i\mu} \delta^3.
\label{AAi-QED}  
\end{eqnarray}
Since we have kinetic terms $R, R^2$ and $C_{\mu\nu\rho\sigma}^2$ for the metric tensor in quadratic gravity,
the ETCR between the all components of the metric tensor and its derivatives in Eq. (\ref{Zero ETCR})
corresponds to Eq. (\ref{AAi-QED}) in QED. Thus, the vanishing ETCRs (\ref{general-dot-g-dot-g}) as well as
(\ref{Zero ETCR}) are a very nontrivial and peculiar feature in quadratic gravity. 

Finally, we wish to spell out in detail what physical implications the vanishing ETCRs (\ref{general-dot-g-dot-g})
have. To this end, let us note that Eq. (\ref{general-dot-g-dot-g}) has the metric tensor field satisfy the commutation relation:
\begin{eqnarray}
[ g_{\rho\sigma} (x), g_{\mu\nu} (x^\prime) ] = 0,
\label{4D-g-g-CR}  
\end{eqnarray}
for $x^\mu$ and $x^{\prime \mu}$ being timelike separated. To derive Eq. (\ref{4D-g-g-CR}), let us set
$x^\mu \equiv ( t, \vec{x} )$ and $x^{\prime \mu} \equiv ( t + \delta t, \vec{x}^\prime )$. Expanding
the LHS on Eq. (\ref{4D-g-g-CR}) around $x^0 = t$, we have
\begin{eqnarray}
[ g_{\rho\sigma} (x), g_{\mu\nu} (x^\prime) ] = [ g_{\rho\sigma}, g_{\mu\nu}^\prime]
+ \left[ g_{\rho\sigma}, \frac{\partial g_{\mu\nu}^\prime}{\partial t} \right] \delta t
+ \frac{1}{2} \left[ g_{\rho\sigma}, \frac{\partial^2 g_{\mu\nu}^\prime}{\partial t^2} \right] (\delta t)^2 + \cdots.
\label{4D-g-g-CR2}  
\end{eqnarray}
Then, we find that Eq. (\ref{general-dot-g-dot-g}) leads to Eq. (\ref{4D-g-g-CR}). The key point is that Eq. (\ref{4D-g-g-CR}) is
valid when $x^\mu$ and $x^{\prime \mu}$ are timelike separated.\footnote{When $x^\mu$ and $x^{\prime \mu}$ are spacelike separated,
Eq. (\ref{4D-g-g-CR}) is nothing but the law of microcausality, i.e. a measurement of $ g_{\mu\nu}$ at $x^{\prime \mu}$ cannot influence
the value of $ g_{\mu\nu}$ at $x^\mu$. Precisely speaking, Eq. (\ref{4D-g-g-CR}) does not make sense since we do not
know if $x^\mu$ and $x^{\prime \mu}$ are timelike, null or spacelike separated until we know the metric tensor.
This is the chicken-or-the-egg controversy in quantum gravity, which we ignore in this article.} This fact means that in quadratic gravity the metric
tensor behaves as if it were not a quantum operator but a classical object at least when we consider correlation functions 
between the metric tensors. 

This situation can be understood from a different context as well.  
In case of general relativity, it is known \cite{Nakanishi, N-O-text} that
\begin{eqnarray}
[ \dot g_{\rho\sigma}, g_{\mu\nu}^\prime ] = 2 i \kappa^2 \tilde f \Biggl[ 2 g_{\rho(\mu} g_{\nu)\sigma}
- g_{\rho\sigma} g_{\mu\nu} - \frac{2}{g^{00}} ( \delta_\rho^0 \delta_{(\mu}^0 g_{\nu)\sigma}
+ \delta_\sigma^0 \delta_{(\mu}^0 g_{\nu)\rho} ) \Biggr] \delta^3,
\label{GR-gg-CR}  
\end{eqnarray}
from which we can show that $g_{\mu\nu}$ has the graviton as the physical mode. On the other hand, in quadratic
gravity we have $[ \dot g_{\rho\sigma}, g_{\mu\nu}^\prime ] = 0$, so it is expected that it is not $g_{\mu\nu}$ but 
a different operator that carries information on the graviton.   
In this sense, the vanishing ETCRs (\ref{general-dot-g-dot-g}) would exhibit that the metric tensor itself loses
the role as dynamical variables, and the whole information on dynamical degrees of freedom for gravity is encoded 
in the ETCRs among the metric tensor, auxiliary fields and FP ghosts. 
Actually, as will be seen later in Eqs. (\ref{Graviton}) and (\ref{Graviton-eq}), the degree of the graviton is included in not
$\varphi_{\mu\nu}$ but $h_{\mu\nu}$ at the linearized level.   
  
At this point, it is of interest to recall that an origin of the difficulty for constructing a consistent quantum gravity can be
traced to the dual role played by the metric tensor field. In any field theory one needs a metric to contract indices in the action 
and to define the volume element; in this role the metric provides us with the geometrical standard according 
to which lengths and angles are measured. On the other hand, the metric also plays the role of dynamical variables. 
This duality is, however, the source of  the difficulties, which are encountered in the quantization of gravity \cite{Wald}.  
Namely, in order to quantize the dynamical degrees of freedom of the gravitational field, we have to give a quantum
mechanical description of space-time structure. This problem is a charactristic feature of quantum gravity and
has no analog for the other quantum field theories which are formulated on a fixed background space-time. 
In the present theory, even if the metric tensor gives us the geometrical standard as in Einstein's general relativity 
it loses the role of dynamical variables owing to its vanishing ETCRs and instead a suitable linear combination of
the metric, auxiliary fields and FP ghosts plays the role of dynamical variables. In this sense,
there is no dual role played by the metric field in the theory at hand.
  
Since we have established the ETCRs among the metric tensor and its time derivatives, we would like to derive 
a type of ETCRs, $[ \dot g_{\rho\sigma}, \Phi^\prime ]$ where $\Phi$ denotes a set of fields, 
$\Phi = \{ g_{\mu\nu}, K_{\mu\nu}, A_\mu, c^\mu, \bar c_\mu, \zeta_\mu, \bar \zeta_\mu, \beta_\mu \}$.  
For $g_{\mu\nu}$ and $\bar c_\mu$, we have already obtained vanishing results as seen in Eqs. (\ref{g-cbar-GCT}) 
and (\ref{Zero ETCR}).  In a similar way, it is easy to show that
\begin{eqnarray}
[ \dot g_{\rho\sigma}, c^{\prime \mu} ] = [ \dot g_{\rho\sigma}, \zeta_\mu^\prime ]
= [ \dot g_{\rho\sigma}, \bar \zeta_\mu^\prime ] = [ \dot g_{\rho\sigma}, \beta_\mu^\prime ] = 0.
\label{dot-g-Phi}  
\end{eqnarray}
Here the last ETCR can be shown by taking the ST BRST transformation of 
$[ \dot g_{\rho\sigma}, \bar \zeta_\mu^\prime ] = 0$.  As for $\Phi = A_\mu$, we use Eqs. (\ref{dot-g-piK}), (\ref{dot-g-0rho})
and $[ \beta_\rho, A_\mu^\prime ] = 0$, which can be derived from the ST BRST transformation of CCR,
$[ \bar \zeta_\rho, A_\mu^\prime ] = 0$, and the result is given by
\begin{eqnarray}
[ \dot g_{\rho\sigma}, A_\mu^\prime ] = 0.
\label{dot-g-A}  
\end{eqnarray}

In order to derive the remaining ETCR, $[ \dot g_{\rho\sigma}, K_{\mu\nu}^\prime ]$, we need some calculations. 
First, let us note that the CCR, $\{ \bar \zeta_\mu, \pi_{\bar \zeta}^{\prime \nu} \} = i \delta_\mu^\nu \delta^3$ yields
the ETCR:
\begin{eqnarray}
\{ \bar \zeta_\mu, \dot \zeta_\nu^\prime \} = - \tilde f \left( g_{\mu\nu} 
- \frac{1}{2 g^{00}} \delta_\mu^0 \delta_\nu^0 \right) \delta^3.
\label{d-bar-zeta-zeta}  
\end{eqnarray}
Next, we find that the ST BRST transformation of the CCR, $[ \bar \zeta_\rho, K_{\mu\nu}^\prime ] = 0$ produces
\begin{eqnarray}
[ \beta_\rho, K_{\mu\nu}^\prime ] = - i \{ \bar \zeta_\rho, \nabla_\mu \zeta_\nu^\prime 
+ \nabla_\nu \zeta_\mu^\prime \}.
\label{beta-K}  
\end{eqnarray}
From Eqs. (\ref{dot-g-Phi}) and (\ref{d-bar-zeta-zeta}), this equation can be cast to the form:
\begin{eqnarray}
[ \beta_\rho, K_{\mu\nu}^\prime ] = i \tilde f \left( \delta_\mu^0 g_{\rho\nu} + \delta_\nu^0 g_{\rho\mu} 
- \frac{1}{g^{00}} \delta_\rho^0 \delta_\mu^0 \delta_\nu^0 \right) \delta^3.
\label{beta-K2}  
\end{eqnarray}
Using this equation, Eqs. (\ref{dot-g-piK}) and  (\ref{dot-g-0rho}), it is straightforward to derive 
\begin{eqnarray}
[ \dot g_{\rho\sigma}, K_{\mu\nu}^\prime ] &=& i \frac{1}{\gamma} \tilde f \Big[ g_{\rho\sigma} g_{\mu\nu} - 2 g_{\rho(\mu} g_{\nu)\sigma} 
- \frac{1}{g^{00}} g_{\rho\sigma} \delta_\mu^0 \delta_\nu^0 +  \frac{2}{g^{00}} ( g_{\rho(\mu} \delta_{\nu)}^0 \delta_\sigma^0
+ g_{\sigma(\mu} \delta_{\nu)}^0 \delta_\rho^0 )
\nonumber\\
&-& \frac{2}{(g^{00})^2} \delta_\rho^0 \delta_\sigma^0 \delta_\mu^0 \delta_\nu^0 \Big] \delta^3.
\label{d-g-K-ETCR}  
\end{eqnarray}
 
Now we wish to calculate the ETCRs involving the $b_\mu$ field, $[ \Phi, b_\rho^\prime ]$ since the $b_\mu$ field plays a role 
as the generator of global translation \cite{Oda-Q}.  Since we have already given the ETCR, 
$[ g_{\mu\nu}, b_\rho^\prime ]$ in Eq. (\ref{g-b}), we first evaluate the ETCR, $[ K_{\mu\nu}, b_\rho^\prime ]$.  
From the CCRs, $[ K_{\mu\nu}, \bar c_\rho^\prime ] = [ K_{\mu\nu}, \pi_{c\rho}^\prime ] = 0$, we can obtain that
\begin{eqnarray}
[ K_{\mu\nu}, \dot{\bar c}_\rho^\prime ] = [ \dot K_{\mu\nu}, \bar c_\rho^\prime ] = 0.
\label{K-bar-c}  
\end{eqnarray}
Taking the GCT BRST transformation of $[ K_{\mu\nu}, \bar c_\rho^\prime ] = 0$ yields
\begin{eqnarray}
- \{ c^\alpha \nabla_\alpha K_{\mu\nu} + \nabla_\mu c^\alpha K_{\alpha\nu} 
+ \nabla_\nu c^\alpha K_{\mu\alpha}, \bar c_\rho^\prime \}
+ [ K_{\mu\nu}, i B_\rho^\prime ] = 0.
\label{K&bar-c}  
\end{eqnarray}
By means of Eqs. (\ref{new-b}), (\ref{c-cbar-GCT}), (\ref{g-cbar-GCT}), and (\ref{K-bar-c}),  
we reach the ETCR between $K_{\mu\nu}$ and $b_\rho$:
\begin{eqnarray}
[ K_{\mu\nu}, b_\rho^\prime ] = - i \tilde f ( \delta_\mu^0 K_{\rho\nu} + \delta_\nu^0 K_{\rho\mu} ) \delta^3.
\label{K-b-rho}  
\end{eqnarray}
Along the same line of argument, it is easy to show that  
\begin{eqnarray}
[ A_\mu, b_\rho^\prime ] = - i \tilde f \delta_\mu^0 A_\rho \delta^3,  \qquad
[ \zeta_\mu, b_\rho^\prime ] = - i \tilde f \delta_\mu^0 \zeta_\rho \delta^3,  \qquad
[ \bar \zeta_\mu, b_\rho^\prime ] = - i \tilde f \delta_\mu^0 \bar \zeta_\rho \delta^3.
\label{Var-b-rho}  
\end{eqnarray}
The remaining ETCRs for $\Phi = \{ \beta_\mu, c^\mu, \bar c_\mu\}$ can be obtained as follows:
The ST BRST transformation of the last ETCR in (\ref{Var-b-rho}) directly produces 
\begin{eqnarray}
[ \beta_\mu, b_\rho^\prime ] = - i \tilde f \delta_\mu^0 \beta_\rho \delta^3.
\label{Var-b-rho2}  
\end{eqnarray}
Similarly, the GCT BRST transformation of the CCR, $\{ c^\mu, \bar c_\rho^\prime \} = 0$ produces
\begin{eqnarray}
[ c^\mu, b_\rho^\prime ] = 0.
\label{Var-b-rho3}  
\end{eqnarray}
Finally, the ETCR, $[ \bar c_\mu, b_\rho^\prime ]$ can be calculated from the CCR, 
$[ \bar c_\mu, \pi_g^{\prime 0\nu} ] = 0$\footnote{$\pi_g^{0\nu}$ does not depend on both $\dot g_{\mu\nu}$ 
and $\dot K_{\mu\nu}$.} by using the ETCRs, $[ \bar c_\mu, \dot A_\nu^\prime ]
= [ \bar c_\mu, \dot \zeta_\nu^\prime ] = [ \bar c_\mu, \dot{\bar \zeta^\prime}_\nu ]
= [ \bar c_\mu, \beta_\nu^\prime ] = 0$, whose validity is easily shown.  The result is 
\begin{eqnarray}
[ \bar c_\mu, b_\rho^\prime ] = 0.
\label{Var-b-rho4}  
\end{eqnarray}

Next we wish to comment on the type of ETCRs, $[ \dot \Phi, b_\rho^\prime ]$. 
First of all, the ETCR, $[ \dot g_{\mu\nu}, b_\rho^\prime ]$ has been already calculated by using the method 
developed in our previous article \cite{Oda-Q}. Only the result is written out as
\begin{eqnarray}
[ \dot g_{\mu\nu}, b_\rho^\prime ] &=& - i \Bigl\{ \tilde f ( \partial_\rho g_{\mu\nu} 
+ \delta_\mu^0 \dot g_{\rho\nu} + \delta_\nu^0 \dot g_{\rho\mu} ) \delta^3 
\nonumber\\
&+& [ ( \delta_\mu^k - 2 \delta_\mu^0 \tilde f \tilde g^{0 k} ) g_{\rho\nu}
+ (\mu\leftrightarrow \nu) ] \partial_k ( \tilde f \delta^3 ) \Bigr\},
\label{dot g-b}  
\end{eqnarray}
or equivalently
\begin{eqnarray}
[ g_{\mu\nu}, \dot b_\rho^\prime ] &=& i \Bigl\{ [ \tilde f \partial_\rho g_{\mu\nu} 
- \partial_0 \tilde f ( \delta_\mu^0 g_{\rho\nu} + \delta_\nu^0 g_{\rho\mu} ) ] \delta^3 
\nonumber\\
&+& [ ( \delta_\mu^k - 2 \delta_\mu^0 \tilde f \tilde g^{0k} ) g_{\rho\nu} + (\mu \leftrightarrow \nu) ]
\partial_k (\tilde f \delta^3) \Bigr\}.
\label{g-dot b}  
\end{eqnarray}
Furthermore, following the derivation in Appendix A in Ref. \cite{Oda-f}, we have 
\begin{eqnarray}
[ \dot A_\mu, b_\rho^\prime ] = i ( 2 \tilde f \tilde g^{0i} \delta_\mu^0 - \delta_\mu^i ) 
A_\rho \partial_i ( \tilde f \delta^3 ) 
- i \tilde f ( \partial_\rho A_\mu + \delta_\mu^0 \partial_0 A_\rho ) \delta^3.  
\label{dot-Amu-b}  
\end{eqnarray}
In a similar manner, we can also show that
\begin{eqnarray}
&{}& [ \dot K_{\mu\nu}, b_\rho^\prime ] = [ i ( 2 \tilde f \tilde g^{0i} \delta_\mu^0 - \delta_\mu^i ) 
K_{\rho\nu} + (\mu \leftrightarrow \nu) ] \partial_i ( \tilde f \delta^3 ) 
\nonumber\\
&{}& - i \tilde f ( \partial_\rho K_{\mu\nu} + \delta_\mu^0 \partial_0 K_{\rho\nu} 
+ \delta_\nu^0 \partial_0 K_{\rho\mu} ) \delta^3,
\nonumber\\
&{}& [ \dot c^\mu, b_\rho^\prime ] =  - i \tilde f \partial_\rho c^\mu \delta^3, \quad
[ \dot{\bar c}_\mu, b_\rho^\prime ] =  - i \tilde f \partial_\rho \bar c_\mu \delta^3,
\nonumber\\
&{}& [ \dot \beta_\mu, b_\rho^\prime ] = i ( 2 \tilde f \tilde g^{0i} \delta_\mu^0 - \delta_\mu^i ) 
\beta_\rho \partial_i ( \tilde f \delta^3 ) 
- i \tilde f ( \partial_\rho \beta_\mu + \delta_\mu^0 \partial_0 \beta_\rho ) \delta^3,
\nonumber\\
&{}& [ \dot \zeta_\mu, b_\rho^\prime ] = i ( 2 \tilde f \tilde g^{0i} \delta_\mu^0 - \delta_\mu^i ) 
\zeta_\rho \partial_i ( \tilde f \delta^3 ) 
- i \tilde f ( \partial_\rho \zeta_\mu + \delta_\mu^0 \partial_0 \zeta_\rho ) \delta^3,
\nonumber\\
&{}& [ \dot {\bar \zeta}_\mu, b_\rho^\prime ] = i ( 2 \tilde f \tilde g^{0i} \delta_\mu^0 - \delta_\mu^i ) 
\bar \zeta_\rho \partial_i ( \tilde f \delta^3 ) 
- i \tilde f ( \partial_\rho \bar \zeta_\mu + \delta_\mu^0 \partial_0 \bar \zeta_\rho ) \delta^3.  
\label{dot-vec1-b}  
\end{eqnarray}
Finally, from our previous calculation \cite{Oda-Q}, we can prove
\begin{eqnarray}
[ b_\mu, b_\nu^\prime ] = 0,   \qquad
[ b_\mu, \dot b_\nu^\prime ] = i \tilde f ( \partial_\mu b_\nu + \partial_\nu b_\mu ) \delta^3. 
\label{b-b}  
\end{eqnarray}

Before closing this section, let us mention the remaining ETCRs. Let us first consider a type of ETCRs,
$[ \dot A_\rho, \Phi^\prime ]$. To evaluate such the ETCRs, we must express $\dot A_\rho$ in terms of 
the other canonical variables. It is easy to do so by solving $\pi_A^\mu$ in (\ref{CCM}) with respect to 
$\dot A_\rho$, and the result is of form:
\begin{eqnarray}
\dot A_0 &=& \frac{1}{4 \beta_1} \tilde f \left[ g_{0\mu} - \frac{\beta_1 + 2 \beta_2}{2 (\beta_1 + \beta_2) g^{00}} \delta_\mu^0
\right] \pi_A^\mu  
- \frac{1}{2 (\beta_1 + \beta_2) g^{00}} \left( \frac{\beta_1 + 2 \beta_2}{g^{00}} g^{0i} g^{0j} - \beta_2 g^{ij} \right) \hat K_{ij}
\nonumber\\
&+& \frac{1}{2} K_{00} + \Gamma_{00}^\mu A_\mu,
\nonumber\\
\dot A_i &=& \frac{1}{g^{00}} g^{0j} \hat K_{ij} + \frac{1}{4 \beta_1} \tilde f g_{i\mu} \pi_A^\mu + K_{0i} 
+ \Gamma_{0i}^\mu A_\mu - \nabla_i A_0. 
\label{dot-A}  
\end{eqnarray}
For instance, using this equation, the ETCRs, $[ \dot A_\rho, A_\mu^\prime ]$ and 
$[ \dot A_\rho, K_{\mu\nu}^\prime ]$ can be calculated to be
\begin{eqnarray}
[ \dot A_\rho, A_\mu^\prime ] = - i \frac{1}{4 \beta_1} \tilde f \left[ g_{\rho\mu} 
- \frac{\beta_1 + 2 \beta_2}{2 (\beta_1 + \beta_2) g^{00}} \delta_\rho^0 \delta_\mu^0 \right] \delta^3,
\label{dot-A-A}  
\end{eqnarray}
and
\begin{eqnarray}
[ \dot A_\rho, K_{\mu\nu}^\prime ] &=& i \frac{1}{\gamma} \tilde f \Bigg[ \Big( g_{\mu\nu} - \frac{1}{g^{00}} 
\delta_\mu^0 \delta_\nu^0 \Big) A_\rho - 2 g_{\rho(\mu} A_{\nu)} 
+ \frac{2}{g^{00}} g_{\rho(\mu} \delta_{\nu)}^0 A^0 
\nonumber\\
&-& \frac{2 \beta_2}{(\beta_1 + \beta_2) g^{00}} g_{0(\mu} \delta_{\nu)}^0 \delta_\rho^0 A^0
+ \frac{2}{g^{00}} \delta_{(\mu}^0 A_{\nu)} \delta_\rho^0 - \frac{\beta_1 + 2 \beta_2}{2 (\beta_1 + \beta_2) g^{00}} g_{\mu\nu} \delta_\rho^0 A^0
\nonumber\\
&-& \frac{3 \beta_1 + 2 \beta_2}{2 (\beta_1 + \beta_2) (g^{00})^2} \delta_\mu^0 \delta_\nu^0 \delta_\rho^0 A^0
+ \frac{\beta_2}{\beta_1 + \beta_2} g_{0\mu} g_{0\nu} \delta_\rho^0 A^0 \Bigg] \delta^3.
\label{dot-A-K}  
\end{eqnarray}

In a similar manner, we can also evaluate a type of ETCRs, $[ \dot K_{\rho\sigma}, \Phi^\prime ]$ by expressing 
$\dot K_{\rho\sigma}$ in terms of the other canonical variables. However, the result is in general quite complicated 
since $\pi_g^{\mu\nu}$ in (\ref{CCM}) and the ETCR in (\ref{d-g-K-ETCR}) have a long expression. Luckily enough,
we do not need such the ETCRs in what follows so we omit to write down their complicated ETCRs. However,  
it is worthwhile to stress that any ETCRs can be in principle calculated through the CCRs, field equations, 
the BRST transformations although we have not given all the ETCRs explicitly in this article.

\section{Linearized theory}

The main goal of this section is to analyze asymptotic fields under the assumption that all elementary fields have their own
asymptotic fields and there is no bound state. We also assume that all asymptotic fields are
governed by the quadratic part of the quantum Lagrangian apart from possible renormalization.

Let us expand the gravitational field $g_{\mu\nu}$ around a flat Minkowski metric $\eta_{\mu\nu}$ 
as
\begin{eqnarray}
g_{\mu\nu} = \eta_{\mu\nu} + \varphi_{\mu\nu},
\label{Background}  
\end{eqnarray}
where $\varphi_{\mu\nu}$ denotes fluctuations. In general relativity without higher derivative terms, instead of 
(\ref{Background}) it is conventional to make the expansion like $g_{\mu\nu} = \eta_{\mu\nu} + \kappa \varphi_{\mu\nu}$. 
However, this may not be the best choice for quadratic gravity since there are the other coupling constants 
$\alpha_R$ and $\alpha_C$ \cite{Buchbinder}. Actually, we have two options to caninically normalize the kinetic terms: 
One option is to normalize the kinetic term with two derivatives and the other one is to do so with four derivatives.
The expansion (\ref{Background}) corresponds to the latter option where $\varphi_{\mu\nu}$ is dimensionless while
the conventional expansion does to the former one where $\varphi_{\mu\nu}$ has a mass dimension. 

For the sake of simplicity, we use the same notation for the other asymptotic fields as that for the
interacting fields. Then, up to surface terms the quadratic part of the quantum Lagrangian (\ref{Quant-Lag}) reads:
\begin{eqnarray}
&{}& {\cal L}_q = \frac{1}{2 \kappa^2} \left( \frac{1}{4} \varphi_{\mu\nu} \Box \varphi^{\mu\nu} 
- \frac{1}{4} \varphi \Box \varphi - \frac{1}{2} \varphi^{\mu\nu} \partial_\mu \partial_\rho \varphi_\nu{}^\rho
+ \frac{1}{2} \varphi^{\mu\nu} \partial_\mu \partial_\nu \varphi \right)
\nonumber\\
&{}& + \gamma \left[ \left(\partial_\mu \partial_\rho \varphi_\nu{}^\rho - \frac{1}{2} \Box \varphi_{\mu\nu}
- \frac{1}{2} \partial_\mu \partial_\nu \varphi \right) K^{\mu\nu} + \frac{1}{2} ( \Box \varphi 
- \partial_\mu \partial_\nu \varphi^{\mu\nu} ) K \right]
\nonumber\\
&{}& + \beta_1 ( K_{\mu\nu} - \partial_\mu A_\nu - \partial_\nu A_\mu )^2
+ \beta_2 ( K - 2 \partial_\rho A^\rho )^2 + \left( \varphi^{\mu\nu} - \frac{1}{2} \eta^{\mu\nu} \varphi \right)
\partial_\mu b_\nu
\nonumber\\
&{}& - i \partial_\mu \bar c_\rho \partial^\mu c^\rho - \partial_\mu K^{\mu\nu} \beta_\nu
+ i \partial^\mu \bar \zeta^\nu ( \partial_\mu \zeta_\nu + \partial_\nu \zeta_\mu). 
\label{Free-Lag}  
\end{eqnarray}
In this and next sections, the spacetime indices $\mu, \nu, \dots$ are raised or lowered by the Minkowski metric $\eta^{\mu\nu}
= \eta_{\mu\nu} = \rm{diag} ( -1, 1, 1, 1)$, and we define $\Box \equiv \eta^{\mu\nu} \partial_\mu \partial_\nu$
and $\varphi \equiv \eta^{\mu\nu} \varphi_{\mu\nu}$.

From this Lagrangian, it is straightforward to derive the following linearized field equations: 
\begin{eqnarray}
&{}& \frac{1}{2 \kappa^2} \biggl[ \frac{1}{2} \Box \varphi_{\mu\nu} - \partial_\rho \partial_{(\mu} \varphi_{\nu)}{}^\rho 
+ \frac{1}{2} \partial_\mu \partial_\nu \varphi - \frac{1}{2} \eta_{\mu\nu} ( \Box \varphi 
- \partial_\rho \partial_\sigma \varphi^{\rho\sigma} ) \biggr]
\nonumber\\
&{}& - \frac{\gamma}{2} ( \Box K_{\mu\nu} - \eta_{\mu\nu} \Box K + \partial_\mu \partial_\nu K )
+ \partial_{(\mu} b_{\nu)} - \frac{1}{2} \eta_{\mu\nu} \partial_\rho b^\rho = 0.
\label{Linear-Eq1}
\\
&{}& - \gamma \biggl[ \frac{1}{2} \Box \varphi_{\mu\nu} - \partial_\rho \partial_{(\mu} \varphi_{\nu)}{}^\rho 
+ \frac{1}{2} \partial_\mu \partial_\nu \varphi - \frac{1}{2} \eta_{\mu\nu} ( \Box \varphi 
- \partial_\rho \partial_\sigma \varphi^{\rho\sigma} ) \biggr]
\nonumber\\
&{}& + 2 \beta_1 ( K_{\mu\nu} - \partial_\mu A_\nu - \partial_\nu A_\mu )
+ 2 \beta_2 \eta_{\mu\nu} ( K - 2 \partial_\rho A^\rho )
+ \partial_{(\mu} \beta_{\nu)} = 0.
\label{Linear-Eq2}
\\
&{}& \Box A_\mu + \frac{\beta_1 + 2 \beta_2}{\beta_1} \partial_\mu \partial_\nu A^\nu 
- \frac{\beta_2}{\beta_1} \partial_\mu K = 0. 
\label{Linear-Eq3}  
\\
&{}& \partial^\nu \varphi_{\mu\nu} - \frac{1}{2} \partial_\mu \varphi = 0. 
\label{Linear-Eq4}  
\\
&{}& \partial_\mu K^{\mu\nu} = 0. 
\label{Linear-Eq5}  
\\
&{}& \Box c^\rho = \Box \bar c_\rho = 0. 
\label{Linear-Eq6}  
\\
&{}& \Box \zeta_\mu + \partial_\mu \partial^\nu \zeta_\nu 
= \Box \bar \zeta_\mu + \partial_\mu \partial^\nu \bar \zeta_\nu= 0. 
\label{Linear-Eq7}  
\end{eqnarray}

Now we are ready to simplify the field equations obtained above. 
First, operating $\partial^\mu$ on the linearized Einstein equation (\ref{Linear-Eq1}), we have
\begin{eqnarray}
\Box b_\mu = 0,
\label{L-b-rho-eq}  
\end{eqnarray}
which is a linearized analog of Eq. (\ref{b-rho-eq}). This equation can be also obtained by taking 
the linearized GCT BRST transformation $\delta_B^{(L)} \bar c_\mu = i b_\mu$
of $\Box \bar c_\mu = 0$ in Eq. (\ref{Linear-Eq6}).
Next, the trace part of the linearized Einstein equation (\ref{Linear-Eq1}) yields
\begin{eqnarray}
\Box \varphi + 4 \kappa^2 ( \partial_\rho b^\rho - \gamma \Box K ) = 0,
\label{Trace-L-Ein-eq}  
\end{eqnarray}
where Eq. (\ref{Linear-Eq4}) was used.

Similarly, operating $\partial^\mu$ on Eq. (\ref{Linear-Eq2}) leads to
\begin{eqnarray}
\Box A_\mu + \frac{\beta_1 + 2 \beta_2}{\beta_1} \partial_\mu \partial_\nu A^\nu 
- \frac{\beta_2}{\beta_1} \partial_\mu K - \frac{1}{4 \beta_1} ( \Box \beta_\mu 
+ \partial_\mu \partial_\nu \beta^\nu ) = 0.
\label{d-L-eq2}  
\end{eqnarray}
Comparing this equation with Eq. (\ref{Linear-Eq3}), we find the field equation for $\beta_\mu$:
\begin{eqnarray}
\Box \beta_\mu + \partial_\mu \partial_\nu \beta^\nu = 0.
\label{Lbeta-field-eq}  
\end{eqnarray}
Note that this field equation is also derived by taking the ST BRST transformation of (\ref{Linear-Eq7}).
Acting $\partial^\mu$ on Eq. (\ref{Lbeta-field-eq}) produces
\begin{eqnarray}
\Box \partial_\mu \beta^\mu = 0.
\label{Lbeta-field-eq2}  
\end{eqnarray}
Then, together with Eq. (\ref{Lbeta-field-eq2}), Eq. (\ref{Lbeta-field-eq}) leads to
\begin{eqnarray}
\Box^2 \beta_\mu = 0,
\label{Lbeta-field-eq3}  
\end{eqnarray}
which means that $\beta_\mu$ is a massless dipole ghost field.
The trace part of Eq. (\ref{Linear-Eq2}) gives us
\begin{eqnarray}
\Box \varphi + \frac{4 ( \beta_1 + 4 \beta_2)}{\gamma} ( K - 2 \partial_\rho A^\rho )
+ \frac{2}{\gamma} \partial_\mu \beta^\mu = 0.
\label{Trace-L-eq2}  
\end{eqnarray}
Furthermore, a comparison between Eq. (\ref{Linear-Eq1}) and (\ref{Linear-Eq2}) exhibits
\begin{eqnarray}
&{}& \Box K_{\mu\nu} - \eta_{\mu\nu} \Box K + \partial_\mu \partial_\nu K
- \frac{2}{\gamma} \partial_{(\mu} b_{\nu)} + \frac{1}{\gamma} \eta_{\mu\nu} \partial_\rho b^\rho
\nonumber\\
&{}& - \frac{1}{\kappa^2 \gamma^2} [ 2 \beta_1 ( K_{\mu\nu} - \partial_\mu A_\nu - \partial_\nu A_\mu )
+ 2 \beta_2 \eta_{\mu\nu} ( K - 2 \partial_\rho A^\rho ) + \partial_{(\mu} \beta_{\nu)} ] = 0.
\label{K-rel}
\end{eqnarray}
In particular, the trace part of this equation reads
\begin{eqnarray}
\Box K = - \frac{1}{\kappa^2 \gamma^2} \left[ ( \beta_1 + 4 \beta_2 ) ( K - 2 \partial_\rho A^\rho ) 
+ \frac{1}{2} \partial_\rho \beta^\rho \right] + \frac{1}{\gamma} \partial_\rho b^\rho.
\label{K-rel-trace}
\end{eqnarray}

Next, let us take account of Eq. (\ref{Linear-Eq3}). Operating $\partial^\mu$ on this equation, we have
\begin{eqnarray}
\Box K = \frac{2 (\beta_1 + \beta_2)}{\beta_2} \Box \partial_\mu A^\mu.
\label{d-Lfield-eq3}  
\end{eqnarray}
Using this equation together with (\ref{Lbeta-field-eq2}) and (\ref{Trace-L-eq2}), we can obtain 
\begin{eqnarray}
\Box^2 \varphi + \frac{4 \beta_1 ( \beta_1 + 4 \beta_2)}{( \beta_1 + \beta_2) \gamma}  \Box K = 0.
\label{Trace-L-eq3}  
\end{eqnarray}
Operating the d'Alembertian operator $\Box$ on Eq. (\ref{Trace-L-Ein-eq}) and using the field equation for $b_\mu$ in 
(\ref{L-b-rho-eq}), we have
\begin{eqnarray}
\Box^2 \varphi = 4 \kappa^2 \gamma \Box^2 K.
\label{Box-K-phi}  
\end{eqnarray}
Eqs. (\ref{Trace-L-eq3}) and (\ref{Box-K-phi}) lead to 
\begin{eqnarray}
\Box ( \Box - m^2 ) K = 0,
\label{K-eq-final}  
\end{eqnarray}
where we have defined a mass squared by
\begin{eqnarray}
m^2 = - \frac{\beta_1 ( \beta_1 + 4 \beta_2 )}{( \beta_1 + \beta_2 ) \kappa^2 \gamma^2}.
\label{mass-square}  
\end{eqnarray}
Let us recall that $\frac{1}{\kappa}$ has mass dimension $1$ while the coupling constants $\beta_1, \beta_2$ 
and $\gamma$ are dimensionless.  Moreover, acting the d'Alembertian operator $\Box$ on Eq. (\ref{Trace-L-eq3}) again 
and using Eq. (\ref{Box-K-phi}), we can obtain
\begin{eqnarray}
\Box^2 ( \Box - m^2 ) \varphi = 0.
\label{phi-eq-final}  
\end{eqnarray}

Now let us turn our attention to the linearized Einstein equation (\ref{Linear-Eq1}) itself. 
Using Eqs. (\ref{Linear-Eq4}) and (\ref{Trace-L-Ein-eq}), the linearized Einstein equation (\ref{Linear-Eq1}) can be 
simplified to be
\begin{eqnarray}
\Box \varphi_{\mu\nu} - 2 \kappa^2 \gamma ( \Box K_{\mu\nu} + \partial_\mu \partial_\nu K ) 
+ 4 \kappa^2 \partial_{(\mu} b_{\nu)} = 0.
\label{LEin-Eq}  
\end{eqnarray}
Operating an operator $\Box ( \Box - m^2 )$ on this equation and using Eqs. (\ref{L-b-rho-eq}) and (\ref{K-eq-final}), 
we arrive at
\begin{eqnarray}
\Box^2 ( \Box - m^2 ) \tilde \varphi_{\mu\nu} = 0,
\label{L-varphi-Eq}  
\end{eqnarray}
where $\tilde \varphi_{\mu\nu}$ is defined as
\begin{eqnarray}
\tilde \varphi_{\mu\nu} = \varphi_{\mu\nu} - 2 \kappa^2 \gamma K_{\mu\nu}.
\label{tilde-varphi}  
\end{eqnarray}

Eqs. (\ref{K-eq-final}) and (\ref{L-varphi-Eq}) suggest that the fields $K$ and $\tilde \varphi_{\mu\nu}$ involve
massless and massive modes. To separate a massive scalar mode from these modes in a consistent way, it is useful
to consider the following linear combination:
\begin{eqnarray}
\phi = \hat K - \frac{\kappa^2 \gamma}{\beta_1 + 4 \beta_2} \partial_\rho b^\rho
+ \frac{1}{2 (\beta_1 + 4 \beta_2)} \partial_\rho \beta^\rho.
\label{Mass-scalar}  
\end{eqnarray}
From (\ref{K-rel-trace}), we find that this can be rewritten as 
\begin{eqnarray}
\phi = - \frac{\kappa^2 \gamma^2}{\beta_1 + 4 \beta_2} \Box K.
\label{Mass-scalar2}  
\end{eqnarray}
Then, Eq. (\ref{K-eq-final}) shows that the new field $\phi$ obeys a massive Klein-Gordon equation:
\begin{eqnarray}
( \Box - m^2 ) \phi = 0.
\label{KG-scalar}  
\end{eqnarray}

Before extracting the massless graviton mode from $\varphi_{\mu\nu}$, it is more convenient to consider 
the massive ghost which is essentially included in the auxiliary field $K_{\mu\nu}$ by taking the following linear 
combination of fields:
\begin{eqnarray}
\psi_{\mu\nu} &=& \hat K_{\mu\nu} - \frac{\beta_1 + \beta_2}{3 \beta_1} \left( \eta_{\mu\nu} 
+ \frac{\kappa^2 \gamma^2}{\beta_1} \partial_\mu \partial_\nu \right) \phi
+ \frac{\kappa^2 \gamma}{\beta_1} \Bigg[ \partial_{(\mu} b_{\nu)} 
- \frac{\beta_1 + 2 \beta_2}{2 (\beta_1 + 4 \beta_2)} \eta_{\mu\nu} \partial_\rho b^\rho
\nonumber\\
&-& \frac{(\beta_1 + 2 \beta_2) \kappa^2 \gamma^2}{2 \beta_1 (\beta_1 + 4 \beta_2)} 
\partial_\mu \partial_\nu \partial_\rho b^\rho \Bigg]
+ \frac{1}{2 \beta_1} \Bigg[ \partial_{(\mu} \beta_{\nu)} 
- \frac{\beta_2}{\beta_1 + 4 \beta_2} \eta_{\mu\nu} \partial_\rho \beta^\rho
\nonumber\\
&-& \frac{\beta_2 \kappa^2 \gamma^2}{\beta_1 (\beta_1 + 4 \beta_2)} 
\partial_\mu \partial_\nu \partial_\rho \beta^\rho \Bigg].
\label{Massive-ghost}  
\end{eqnarray}
This expression is chosen in a such way that $\psi_{\mu\nu}$ satisfies the equations describing the massive ghost
\begin{eqnarray}
( \Box - M^2 ) \psi_{\mu\nu} = \partial^\mu \psi_{\mu\nu} = \eta^{\mu\nu} \psi_{\mu\nu} = 0,
\label{Massive-ghost-eq}  
\end{eqnarray}
where the squared mass of the ghost is defined as
\begin{eqnarray}
M^2 = \frac{2 \beta_1}{\kappa^2 \gamma^2}.
\label{Mass-ghost}  
\end{eqnarray}
Then, given the new field\footnote{To tell the truth, the logic is reverse. We have looked for $h_{\mu\nu}$ such that
it obeys Eq. (\ref{Graviton-eq}). Then, it turns out that a general expression takes the form:
\begin{eqnarray}
h_{\mu\nu} &=& \varphi_{\mu\nu} - 2 \kappa^2 \gamma \psi_{\mu\nu} 
- \frac{2 ( \beta_1 + \beta_2 ) \kappa^2 \gamma}{3 \beta_1} \Bigg( \eta_{\mu\nu} 
+ \frac{2}{m^2} \partial_\mu \partial_\nu \Bigg) \phi + a_1 \partial_{(\mu} b_{\nu)}
\nonumber\\
&+& a_2 \partial_\mu \partial_\nu \partial_\rho b^\rho + a_3 \Big( \partial_{(\mu} \beta_{\nu)} 
- \frac{1}{2} \eta_{\mu\nu} \partial_\rho \beta^\rho \Big) + a_4 \partial_\mu \partial_\nu \partial_\rho \beta^\rho,
\label{Graviton0}  
\end{eqnarray}
where $a_i ( i = 1, \cdots, 4 )$ are arbitrary constants.  We can verify that $h_{\mu\nu}$ in (\ref{Graviton}) is 
the simplest choice since the terms depending on $a_i$ do not make a contribution to the 
four-dimensional commutation relations in the next section.} 
\begin{eqnarray}
h_{\mu\nu} = \varphi_{\mu\nu} - 2 \kappa^2 \gamma \psi_{\mu\nu} 
- \frac{2 ( \beta_1 + \beta_2 ) \kappa^2 \gamma}{3 \beta_1} \Bigg( \eta_{\mu\nu} 
+ \frac{2}{m^2} \partial_\mu \partial_\nu \Bigg) \phi,
\label{Graviton}  
\end{eqnarray}
we find that it obeys the field equations:   
\begin{eqnarray}
\Box^2 h_{\mu\nu} = \partial^\mu h_{\mu\nu} - \frac{1}{2} \partial_\nu h = 0.
\label{Graviton-eq}  
\end{eqnarray}
It turns out later that $h_{\mu\nu}$ describes the massless graviton. 

Finally, we are willing to consider the St\"{u}ckelberg-like vector field $A_\mu$. As can be seen in Eq. (\ref{Linear-Eq3}),
this field is neither a massless simple pole field nor a dipole ghost one. To make it be a massless dipole ghost field,
it is sufficient to take account of the following combination:
\begin{eqnarray}
\tilde A_\mu = A_\mu - \frac{\beta_2}{2 \beta_1 m^2} \partial_\mu \phi + c \, b_\mu,
\label{new-Stuckel}  
\end{eqnarray}
where $c$ is an arbitrary constant. The following results do not depend on this constant $c$, so we shall select
$c = 0$ for the sake of simplicity.\footnote{This statement holds except for the 4D CR,
$[ h_{\mu\nu}(x), \tilde A_\rho(y) ]$ where an additional term $- 2 i c \eta_{\rho(\mu} \partial_{\nu)}
D(x-y)$ must be added.}  It is easy to see that the new field $\tilde A_\mu$ is a dipole ghost field satisfying
the equations:
\begin{eqnarray}
\Box \tilde A_\mu &=& - \frac{\beta_1 + 2 \beta_2}{\beta_1} \partial_\mu \partial_\nu A^\nu 
+ \frac{\beta_2}{\beta_1} \partial_\mu K - \frac{\beta_2}{2 \beta_1} \partial_\mu \phi,
\nonumber\\
\Box^2 \tilde A_\mu &=& 0.
\label{new-Stuckel2}  
\end{eqnarray}

\section{Analysis of physical states}

In this section, on the basis of the BRST formalism we would like to clearly show that physical states of quadratic gravity
are constituted of a massive scalar, the massless graviton and the spin $2$ massive ghost, and the massive ghost has 
negative norm, by which the unitarity is violated.    

First, by following the standard technique, we calculate the four-dimensional (anti)commutation 
relations (4D CRs) between asymptotic fields. In order to account for the method of the calculation, as an example, 
let us take the Nakanishi-Lautrup field $b_\mu (x)$ obeying the simple pole field equation, $\Box b_\mu = 0$ which
can be represented by the invariant delta function $D(x)$ as
\begin{eqnarray}
b_\mu (x) = - \int d^3 z \, D(x-z) \overleftrightarrow{\partial}_0^z b_\mu (z).
\label{D-b1}  
\end{eqnarray}
Here the invariant delta function $D(x)$ for massless simple pole fields and its properties
are described as
\begin{eqnarray}
&{}& D(x) = - \frac{i}{(2 \pi)^3} \int d^4 k \, \epsilon (k^0) \delta (k^2) e^{i k x}, \qquad
\Box D(x) = 0,
\nonumber\\
&{}& D(-x) = - D(x), \qquad D(0, \vec{x}) = 0, \qquad 
\partial_0 D(0, \vec{x}) = - \delta^3 (x), 
\label{D-function}  
\end{eqnarray}
where $\epsilon (k^0) \equiv \frac{k^0}{|k^0|}$. With these properties, it is easy to verify that
the right-hand side (RHS) of Eq. (\ref{D-b1}) is independent of $z^0$, which will be crucially
utilized in calculating the 4D CRs through the ETCRs.

To illustrate the detail of the calculation, let us evaluate a 4D CR, $[ h_{\mu\nu} (x), b_\rho (y) ]$ explicitly.
Using Eq. (\ref{D-b1}), it can be described as
\begin{eqnarray}
[ h_{\mu\nu} (x), b_\rho (y) ] 
&=& - \int d^3 z \, D(y-z) \overleftrightarrow{\partial}_0^z [ h_{\mu\nu} (x), b_\rho (z) ]
\nonumber\\
&=& - \int d^3 z \Bigl( D(y-z) [ h_{\mu\nu} (x), \dot b_\rho (z) ] 
- \partial_0^z D(y-z) [ h_{\mu\nu} (x), b_\rho (z) ] \Bigr).
\label{4D-h&b}  
\end{eqnarray}
As mentioned above, since the RHS of Eq. (\ref{4D-h&b}) is independent of $z^0$, we can put $z^0 = x^0$ 
and use relevant ETCRs:
\begin{eqnarray}
&{}& [ h_{\mu\nu} (x), b_\rho (z) ] = i ( \delta_\mu^0 \eta_{\rho\nu}
+ \delta_\nu^0 \eta_{\rho\mu} ) \delta^3 (x-z),
\nonumber\\
&{}& [ h_{\mu\nu} (x), \dot b_\rho (z) ] = - i ( \delta_\mu^k \eta_{\rho\nu}
+ \delta_\nu^k \eta_{\rho\mu} ) \partial_k \delta^3 (x-z).
\label{4D-h&b2}  
\end{eqnarray}
Substituting Eq. (\ref{4D-h&b2}) into Eq. (\ref{4D-h&b}), we can easily obtain
\begin{eqnarray}
[ h_{\mu\nu} (x), b_\rho (y) ] 
= - i ( \eta_{\mu\rho} \partial_\nu + \eta_{\nu\rho} \partial_\mu ) D(x-y).
\label{4D-h&b3}  
\end{eqnarray}

In a similar manner, we can calculate the four-dimensional (anti)commutation relations among 
$\phi, \psi_{\mu\nu}, h_{\mu\nu}, b_\mu, \beta_\mu, c^\mu$ and $\bar c_\mu$. For instance, since $\phi$ obeys 
the massive Klein-Gordon equation (\ref{KG-scalar}), it can be expressed in terms of the invariant delta function 
$\Delta(x; m^2)$ for massive simple pole fields as
\begin{eqnarray}
\phi (x) = - \int d^3 z \, \Delta (x-z; m^2) \overleftrightarrow{\partial}_0^z \phi (z),
\label{phi-Delta}  
\end{eqnarray}
where $\Delta(x; m^2)$ is defined as
\begin{eqnarray}
&{}& \Delta(x; m^2) = - \frac{i}{(2 \pi)^3} \int d^4 k \, \epsilon (k^0) \delta (k^2 + m^2) e^{i k x}, \quad
(\Box - m^2) \Delta(x; m^2) = 0,
\nonumber\\
&{}& \Delta(-x; m^2) = - \Delta(x; m^2), \quad \Delta(0, \vec{x}; m^2) = 0, 
\nonumber\\
&{}& \partial_0 \Delta(0, \vec{x}; m^2) = - \delta^3 (x),  \qquad
\Delta(x; 0) = D(x). 
\label{Delta-function}  
\end{eqnarray}
In a perfectly similar way, $\psi_{\mu\nu}$ satisfies the massive Klein-Gordon equation (\ref{Massive-ghost-eq})
with the mass $M$ instead of the mass $m$, so that we can represent that
\begin{eqnarray}
\psi_{\mu\nu} (x) = - \int d^3 z \, \Delta (x-z; M^2) \overleftrightarrow{\partial}_0^z \psi_{\mu\nu} (z).
\label{psi-Delta}  
\end{eqnarray}

As for $h_{\mu\nu}$, since $h_{\mu\nu}$ is a massless dipole ghost field as can be seen 
in Eq. (\ref{Graviton-eq}), it can be described as
\begin{eqnarray}
&{}& h_{\mu\nu} (x) = - \int d^3 z \left[ D(x-z) \overleftrightarrow{\partial}_0^z h_{\mu\nu} (z)
+ E(x-z) \overleftrightarrow{\partial}_0^z \Box h_{\mu\nu} (z) \right]
\nonumber\\
&{}& = - \int d^3 z \Biggl( D(x-z) \overleftrightarrow{\partial}_0^z h_{\mu\nu} (z)
+ E(x-z) \overleftrightarrow{\partial}_0^z \Biggl\{ - 4 \kappa^2 \Biggl[ \partial_{(\mu} b_{\nu)}
\nonumber\\
&{}& - \frac{(\beta_1 + 2 \beta_2) \kappa^2 \gamma^2}{2 \beta_1 (\beta_1 + 4 \beta_2)}
\partial_\mu \partial_\nu \partial_\rho b^\rho \Biggr]
+ \frac{2 \beta_2 \kappa^2 \gamma}{\beta_1 (\beta_1 + 4 \beta_2)}
\partial_\mu \partial_\nu \partial_\rho \beta^\rho \Biggr\} \Biggr) (z),
\label{E-h-exp}  
\end{eqnarray}
where we have used the equation
\begin{eqnarray}
\Box h_{\mu\nu} = - 4 \kappa^2 \Biggl[ \partial_{(\mu} b_{\nu)}
- \frac{(\beta_1 + 2 \beta_2) \kappa^2 \gamma^2}{2 \beta_1 (\beta_1 + 4 \beta_2)}
\partial_\mu \partial_\nu \partial_\rho b^\rho \Biggr]
+ \frac{2 \beta_2 \kappa^2 \gamma}{\beta_1 (\beta_1 + 4 \beta_2)}
\partial_\mu \partial_\nu \partial_\rho \beta^\rho, 
\label{Box-h}  
\end{eqnarray}
which can be derived from Eqs. (\ref{K-rel}), (\ref{LEin-Eq}), (\ref{Mass-scalar}), (\ref{Mass-scalar2}), 
(\ref{KG-scalar}), (\ref{Massive-ghost-eq}) 
and (\ref{Graviton}), and we have introduced the invariant delta function $E(x)$ for massless 
dipole ghost fields and its properties are given by
\begin{eqnarray}
&{}& E(x) = - \frac{i}{(2 \pi)^3} \int d^4 k \, \epsilon (k^0) \delta^\prime (k^2) e^{i k x}, \qquad  
\Box E(x) = D(x),
\nonumber\\
&{}& E(-x) = - E(x), \qquad 
E(0, \vec{x}) = \partial_0 E(0, \vec{x}) = \partial_0^2 E(0, \vec{x}) = 0, 
\nonumber\\ 
&{}& \partial_0^3 E(0, \vec{x}) = \delta^3 (x), \qquad
E(x) = \frac{\partial}{\partial m^2} \Delta(x; m^2)|_{m^2 = 0}. 
\label{E-function}  
\end{eqnarray}
As in Eq. (\ref{D-b1}), we can also check that the RHS of (\ref{phi-Delta}), (\ref{psi-Delta}) and (\ref{E-h-exp}) is independent of $z^0$. 

After a straightforward but a bit lengthy calculation, we find the following 4D CRs:\footnote{As an illustration
of the calculation, in Appendix C we present a derivation of Eq. (\ref{4D-CR5}), which is the most complicated derivation
in all the 4D CRs.}
\begin{eqnarray}
&{}& [ \phi (x), \phi (y) ] = i \frac{3}{2}  \left[ \frac{\beta_1}{(\beta_1 + \beta_2) \kappa \gamma} \right]^2 
\Delta (x-y; m^2).
\label{4D-CR1}
\\
&{}& [ \phi (x), \psi_{\mu\nu} (y) ] = [ \phi (x), h_{\mu\nu} (y) ] = [ \phi (x), \tilde A_\mu (y) ] = 0.
\label{4D-CR2}
\\
&{}& [ \psi_{\mu\nu} (x), \psi_{\rho\sigma} (y) ] = i \Biggl[ \frac{1}{\kappa^2 \gamma^2} \Biggl( - \eta_{\mu(\rho} \eta_{\sigma)\nu}
+ \frac{1}{3} \eta_{\mu\nu} \eta_{\rho\sigma} \Biggr) - \frac{1}{6 \beta_1} ( \eta_{\mu\nu} \partial_\rho \partial_\sigma 
+ \eta_{\rho\sigma} \partial_\mu \partial_\nu )
\nonumber\\
&{}& + \frac{1}{2 \beta_1} ( \eta_{\mu(\rho} \partial_{\sigma)} \partial_\nu + \eta_{\nu(\rho} \partial_{\sigma)} \partial_\mu )
- \frac{\kappa^2 \gamma^2}{6 \beta_1^2} \partial_\mu \partial_\nu \partial_\rho \partial_\sigma \Biggr] \Delta (x-y; M^2).
\label{4D-CR3}
\\
&{}& [ \psi_{\mu\nu} (x), h_{\rho\sigma} (y) ] = [ \psi_{\mu\nu} (x), \tilde A_\rho (y) ] = 0.
\label{4D-CR4}
\\
&{}& [ h_{\mu\nu} (x), h_{\rho\sigma} (y) ] = i \Biggl[ 2 \kappa^2 ( 2 \eta_{\mu(\rho} \eta_{\sigma)\nu} - \eta_{\mu\nu} \eta_{\rho\sigma} )
+ \frac{2 (\beta_1 + 2 \beta_2) \kappa^4 \gamma^2}{\beta_1 (\beta_1 + 4 \beta_2)} ( \eta_{\mu\nu} \partial_\rho \partial_\sigma 
+ \eta_{\rho\sigma} \partial_\mu \partial_\nu )
\nonumber\\
&{}& - \frac{2 \kappa^4 \gamma^2}{\beta_1} ( \eta_{\mu(\rho} \partial_{\sigma)} \partial_\nu + \eta_{\nu(\rho} \partial_{\sigma)} \partial_\mu )
- \frac{2 (\beta_1 + 2 \beta_2) (\beta_1 - 2 \beta_2) \kappa^6 \gamma^4}{\beta_1^2 (\beta_1 + 4 \beta_2)^2} 
\partial_\mu \partial_\nu \partial_\rho \partial_\sigma \Biggr] D (x-y)
\nonumber\\
&{}& + i \Biggl[ - 4 \kappa^2 ( \eta_{\mu(\rho} \partial_{\sigma)} \partial_\nu + \eta_{\nu(\rho} \partial_{\sigma)} \partial_\mu )
+ \frac{4 (\beta_1 + 2 \beta_2) \kappa^4 \gamma^2}{\beta_1 (\beta_1 + 4 \beta_2)} 
\partial_\mu \partial_\nu \partial_\rho \partial_\sigma \Biggr] E (x-y).
\label{4D-CR5}
\\
&{}& [ h_{\mu\nu} (x), \tilde A_\rho (y) ] = i \frac{\beta_2}{2 (\beta_1 + \beta_2) \gamma m^2} 
\Biggl( \eta_{\mu\nu} + \frac{2}{m^2} \partial_\mu \partial_\nu \Biggr) \partial_\rho D(x-y)
\nonumber\\
&{}& + i \frac{\beta_2}{(\beta_1 + \beta_2) \gamma m^2} \partial_\mu \partial_\nu \partial_\rho E(x-y).
\label{4D-CR6}
\\
&{}& [ \tilde A_\mu (x), \tilde A_\nu (y) ] = i \Biggl\{ - \frac{1}{4 \beta_1} \eta_{\mu\nu} 
+ \frac{3}{2} \Biggl[ \frac{\beta_2}{2 (\beta_1 + \beta_2) \kappa \gamma m^2} \Biggr]^2 \partial_\mu \partial_\nu \Biggr\} D(x-y)
\nonumber\\
&{}& + i \frac{\beta_1 + 5 \beta_2}{8 \beta_1 (\beta_1 + 4 \beta_2)} \partial_\mu \partial_\nu E(x-y).
\label{4D-CR7}
\\
&{}& [ \phi (x), b_\mu (y) ] = [ \phi (x), \beta_\mu (y) ] = 0.
\label{4D-CR8}
\\
&{}& [ \psi_{\mu\nu} (x), b_\rho (y) ] = [ \psi_{\mu\nu} (x), \beta_\rho (y) ] = 0.
\label{4D-CR9}
\\
&{}& [ h_{\mu\nu} (x), b_\rho (y) ] = - i ( \eta_{\mu\rho} \partial_\nu + \eta_{\nu\rho} \partial_\mu ) D(x-y),
\qquad [ h_{\mu\nu} (x), \beta_\rho (y) ] = 0.
\label{4D-CR10}
\\
&{}& [ \tilde A_\mu (x), b_\nu (y) ] = 0, \qquad 
[ \tilde A_\mu (x), \beta_\nu (y) ] = - i \eta_{\mu\nu} D(x-y) 
+ i \frac{1}{2} \partial_\mu \partial_\nu E(x-y).
\label{4D-CR11}
\\
&{}& [ b_\mu (x), b_\nu (y) ] = [ b_\mu (x), \beta_\nu (y) ] = [ \beta_\mu (x), \beta_\nu (y) ] = 0.
\label{4D-CR12}
\\
&{}& \{ c^\mu (x), \bar c_\nu (y) \} = \delta_\nu^\mu D(x-y). 
\label{4D-CR13}
\\
&{}& \{ \zeta_\mu (x), \bar \zeta_\nu (y) \} = - \eta_{\mu\nu} D(x-y) + \frac{1}{2} \partial_\mu \partial_\nu E(x-y). 
\label{4D-CR14}
\end{eqnarray}

Here we make remarks on two points. The one point is that Eqs. (\ref{4D-CR2}) and (\ref{4D-CR4}) show that three fields 
$\psi_{\mu\nu}, h_{\mu\nu}$ and $\phi$ are independent fields. The other point is that at the lowest order ${\cal{O}} (\kappa^2)$,
the 4D CR in (\ref{4D-CR5}) reduces to
\begin{eqnarray}
[ h_{\mu\nu} (x), h_{\rho\sigma} (y) ] &=& 2 i \kappa^2 ( 2 \eta_{\mu(\rho} \eta_{\sigma)\nu} - \eta_{\mu\nu} \eta_{\rho\sigma} )
D (x-y) - 4 i \kappa^2 ( \eta_{\mu(\rho} \partial_{\sigma)} \partial_\nu 
\nonumber\\
&+& \eta_{\nu(\rho} \partial_{\sigma)} \partial_\mu )
E (x-y).
\label{4D-CR-hh}
\end{eqnarray}
This expression agrees with that of general relativity up to an overall constant which stems from the difference of the 
conventions \cite{N-O-text}. Thus, it is not $\varphi_{\mu\nu} (x)$ but $h_{\mu\nu} (x)$ that plays the role of the graviton
in quadratic gravity.

As is well known nowadays, the physical Hilbert space $|\rm{phys} \rangle$ is defined by the Kugo-Ojima subsidiary 
conditions \cite{Kugo-Ojima}
\begin{eqnarray}
\rm{Q_B^{(1)}} |\rm{phys} \rangle = \rm{Q_B^{(2)}} |\rm{phys} \rangle = 0,
\label{Phys-Hilbert}  
\end{eqnarray}
where $\rm{Q_B^{(1)}}$ and $\rm{Q_B^{(2)}}$ are the BRST charges associated with the general coordinate transformation  
and the St\"{u}ckelberg transformation, respectively. Note that this equation is consistently imposed because of $\{ \rm{Q_B^{(1)}},
\rm{Q_B^{(2)}} \} = 0$. 

Now we would like to clarify physical modes and discuss the issue of the unitarity of the theory at hand. To do that, it is
convenient to perform the Fourier transformation of Eqs. (\ref{4D-CR1})-(\ref{4D-CR14}).
However, for a dipole field we cannot use the three-dimensional Fourier expansion to define 
the creation and annihilation operators. We therefore make use of the four-dimensional 
Fourier expansion \cite{N-O-text}:\footnote{With an abuse of notation we denote the Fourier-transformed field 
by the same symbol, but now it is a function of not the coordinates $x^\mu$ but the momenta $p^\mu$.}
\begin{eqnarray}
\Phi (x) = \frac{1}{(2 \pi)^{\frac{3}{2}}} \int d^4 p \, \theta (p^0) [ \Phi (p) e^{i p x}
+ \Phi^\dagger (p) e^{- i p x} ],
\label{4D-FT}  
\end{eqnarray}
where $\Phi (x)$ is a generic field which includes a simple pole field as well as a dipole field. 

Here let us recall that for a massless simple pole field $\Phi_1 (x)$, the three-dimensional Fourier expansion is defined as
\begin{eqnarray}
\Phi_1 (x) = \frac{1}{(2 \pi)^{\frac{3}{2}}} \int d^3 p \, \frac{1}{\sqrt{2 |\vec{p}|}}  
[ \Phi_1 (\vec{p}) e^{- i |\vec{p}| x^0 + i \vec{p} \cdot \vec{x} } 
+ \Phi_1^\dagger (\vec{p}) e^{ i |\vec{p}| x^0 - i \vec{p} \cdot \vec{x} } ].
\label{3D-FT}  
\end{eqnarray}
Thus, the annihilation operator $\Phi_1 (p)$ in the four-dimensional Fourier expansion has a connection with 
the annihilation operator $\Phi_1 (\vec{p})$ in the three-dimensional Fourier expansion like\footnote{For a massive
field of the mass $m$, $\delta(p^2)$ and $|\vec{p}|$ are simply replaced with $\delta(p^2 + m^2)$ and $\sqrt{\vec{p}^2 + m^2}$,
respectively.}
\begin{eqnarray}
\Phi_1 (p) = \theta (p^0) \delta (p^2) \sqrt{2 |\vec{p}|} \Phi_1 (\vec{p}).
\label{3D-4D}  
\end{eqnarray}
Note that for a massless simple pole field the Fourier transform $\Phi_1 (p)$ takes the form
\begin{eqnarray}
\Phi_1 (p) = \frac{i}{(2 \pi)^{\frac{3}{2}}} \theta (p^0) \delta(p^2) \int d^3 z \, e^{-ipz}
\overleftrightarrow{\partial}_0^z \Phi_1 (z),
\label{Phi-1}  
\end{eqnarray}
whereas for a massless dipole field its Fourier transform $\Phi_2 (p)$ does
\begin{eqnarray}
\Phi_2 (p) = \frac{i}{2 (2 \pi)^{\frac{3}{2}}} \theta (p^0) \int d^3 z \, e^{-ipz}
\overleftrightarrow{\partial}_0^z [ \delta(p^2) \Phi_2 (z) + \delta^\prime(p^2) \Box \Phi_2 (z) ].
\label{Phi-2}  
\end{eqnarray}

Based on these Fourier expansions, we can calculate the nontrivial Fourier transform of Eqs. 
(\ref{4D-CR1})-(\ref{4D-CR14}):\footnote{For instance, a derivation of Eq. (\ref{FT-4D-CR4}) is 
given in Appendix D.}
\begin{eqnarray}
&{}& [ \phi (p), \phi^\dagger (q) ] = \frac{3}{2} \Biggl[ \frac{\beta_1}{(\beta_1 + \beta_2) \kappa \gamma} \Biggr]^2
\theta (p^0) \delta(p^2 + m^2) \delta^4 (p-q).
\label{FT-4D-CR1}
\\
&{}& [ \psi_{\mu\nu} (p), \psi_{\rho\sigma}^\dagger (q) ] = \theta (p^0) \delta(p^2 + M^2) \delta^4 (p-q)
\Bigg[ \frac{1}{\kappa^2 \gamma^2} \Big( - \eta_{\mu(\rho} \eta_{\sigma)\nu} 
+ \frac{1}{3} \eta_{\mu\nu} \eta_{\rho\sigma} \Big) 
\nonumber\\
&{}& + \frac{1}{6 \beta_1} ( \eta_{\mu\nu} p_\rho p_\sigma + \eta_{\rho\sigma} p_\mu p_\nu ) 
- \frac{1}{2 \beta_1} ( \eta_{\mu(\rho} p_{\sigma)} p_\nu + \eta_{\nu(\rho} p_{\sigma)} p_\mu )
- \frac{\kappa^2 \gamma^2}{6 \beta_1^2} p_\mu p_\nu p_\rho p_\sigma \Bigg].
\label{FT-4D-CR2}
\\
&{}& [ h_{\mu\nu} (p), h_{\rho\sigma}^\dagger (q) ] = \frac{\kappa^2}{2} \theta (p^0) \delta^4 (p-q)
\Bigg\{ \delta(p^2) \Bigg[ 2 \eta_{\mu(\rho} \eta_{\sigma)\nu} - \eta_{\mu\nu} \eta_{\rho\sigma} 
\nonumber\\
&{}& - \frac{(\beta_1 + 2 \beta_2) \kappa^2 \gamma^2}{\beta_1 ( \beta_1 + 4 \beta_2 )} 
( \eta_{\mu\nu} p_\rho p_\sigma + \eta_{\rho\sigma} p_\mu p_\nu )
+ \frac{\kappa^2 \gamma^2}{\beta_1} ( \eta_{\mu(\rho} p_{\sigma)} p_\nu + \eta_{\nu(\rho} p_{\sigma)} p_\mu )
\nonumber\\
&{}& - \frac{(\beta_1 + 2 \beta_2) (\beta_1 - 2 \beta_2) \kappa^4 \gamma^4}{\beta_1^2 ( \beta_1 + 4 \beta_2 )^2}
p_\mu p_\nu p_\rho p_\sigma \Bigg] + 6 \delta^\prime (p^2) \Bigg[ \eta_{\mu(\rho} p_{\sigma)} p_\nu 
+ \eta_{\nu(\rho} p_{\sigma)} p_\mu
\nonumber\\
&{}&  + \frac{(\beta_1 + 2 \beta_2) \kappa^2 \gamma^2}{\beta_1 ( \beta_1 + 4 \beta_2 )} 
p_\mu p_\nu p_\rho p_\sigma \Bigg] \Bigg\}.
\label{FT-4D-CR3}
\\
&{}& [ h_{\mu\nu} (p), \tilde A_\rho^\dagger (q) ] = i \frac{\beta_2}{8 (\beta_1 + \beta_2) \gamma m^2} 
\theta (p^0) \delta^4 (p-q) \Biggl[ \delta(p^2) \Biggl( \eta_{\mu\nu} - \frac{2}{m^2} p_\mu p_\nu \Biggr)
\nonumber\\
&{}& - 6 \delta^\prime(p^2)  p_\mu p_\nu \Biggr] p_\rho. 
\label{FT-4D-CR4}
\\
&{}& [ \tilde A_\mu (p), \tilde A_\nu^\dagger (q) ] = - \frac{1}{4} \theta (p^0) \delta^4 (p-q) 
\Biggl( \delta(p^2) \Biggl\{ \frac{1}{4 \beta_1} \eta_{\mu\nu} 
+ \frac{3}{2} \Biggl[ \frac{\beta_2}{2 (\beta_1 + \beta_2) \kappa \gamma m^2} \Biggr]^2 p_\mu p_\nu
\nonumber\\
&{}& + \frac{3 (\beta_1 + 5 \beta_2)}{8 \beta_1 (\beta_1 + 4 \beta_2)} \delta^\prime(p^2)  p_\mu p_\nu \Biggr). 
\label{FT-4D-CR5}
\\
&{}& [ h_{\mu\nu} (p), b_\rho^\dagger (q) ] = - i \frac{1}{2} ( \eta_{\mu\rho} p_\nu + \eta_{\nu\rho} p_\mu ) 
\theta (p^0) \delta(p^2) \delta^4 (p-q). 
\label{FT-4D-CR6}
\\
&{}& [ \tilde A_\mu (p), \beta_\nu^\dagger (q) ] =  - \frac{1}{4} \theta (p^0) \delta^4 (p-q) 
\Biggl[ \delta(p^2) \eta_{\mu\nu} + \frac{3}{2} \delta^\prime (p^2) p_\mu p_\nu \Biggr]. 
\label{FT-4D-CR7}
\\
&{}& \{ c^\mu (p), \bar c_\nu^\dagger (q) \} = - i \delta^\mu_\nu \theta (p^0) \delta(p^2) \delta^4 (p-q). 
\label{FT-4D-CR8}  
\\
&{}& \{ \zeta_\mu (p), \bar \zeta_\nu^\dagger (q) \} = i \frac{1}{4} \theta (p^0) \delta^4 (p-q)
\Bigg[ \delta(p^2) \eta_{\mu\nu} + \frac{3}{2} \delta^\prime (p^2) p_\mu p_\nu \Bigg]. 
\label{FT-4D-CR9}  
\end{eqnarray}

Next, let us turn our attention to the linearized field equations. In the Fourier transformation, 
the de Donder condition in Eq. (\ref{Graviton-eq}) takes the form:
\begin{eqnarray}
p^\nu h_{\mu\nu} - \frac{1}{2} p_\mu h = 0.
\label{FT-Linear-Eq3}
\end{eqnarray}
Since this equation gives us four independent relations in ten components of $h_{\mu\nu} (p)$,
the number of the independent components of $h_{\mu\nu} (p)$ is six. To deal with
the six independent components, it is convenient to take a specific Lorentz 
frame which is defined by $p^\mu = ( p, 0, 0, p )$ with $p > 0$, and choose the six components as follows:
\begin{eqnarray}
&{}& h_1 (p) = h_{11} (p),  \qquad
h_2 (p) = h_{12} (p),  \qquad
\omega_0 (p) = \frac{1}{2 p} [ h_{00} (p) - h_{11} (p) ],   
\nonumber\\
&{}& \omega_I (p) = \frac{1}{p} h_{0I} (p),  \qquad
\omega_3 (p) = - \frac{1}{2 p} [ h_{11} (p) + h_{33} (p) ], 
\label{Lorentz}  
\end{eqnarray}
where the index $I$ takes the transverse components $I = 1, 2$. The other four components
are expressible by the above six ones. 

In this respect, it is worthwhile to consider the two BRST transformations for asymptotic fields.
The GCT BRST transformation in terms of the components in (\ref{Lorentz}) takes the form\footnote{The field $K_{\mu\nu}$
is not an independent field any longer since its physical contents are included in both $\phi$ and $\psi_{\mu\nu}$.} 
\begin{eqnarray}
&{}& \delta_B^{(1)} \phi (p) = \delta_B^{(1)} h_I (p) = \delta_B^{(1)} \psi_{\mu\nu} (p) 
= \delta_B^{(1)} K_{\mu\nu} (p) = \delta_B^{(1)} \tilde A_\mu (p) = 0, 
\nonumber\\
&{}& \delta_B^{(1)} \omega_\mu (p) = i c_\mu (p), \qquad
\delta_B^{(1)} \bar c_\mu (p) = i b_\mu (p),
\nonumber\\
&{}& \delta_B^{(1)} c^\mu (p) = \delta_B^{(1)} b_\mu (p) = \delta_B^{(1)} \zeta_\mu (p) 
= \delta_B^{(1)} \bar \zeta_\mu (p) = \delta_B^{(1)} \beta_\mu (p) = 0,
\label{GCT-Q_B-Comp}  
\end{eqnarray}
and the ST BRST transformation is given by
\begin{eqnarray}
&{}& \delta_B^{(2)} \phi (p) = \delta_B^{(2)} h_I (p) = \delta_B^{(2)} \psi_{\mu\nu} (p) 
= \delta_B^{(2)} \omega_\mu (p) = 0,
\nonumber\\
&{}& \delta_B^{(2)} K_{\mu\nu} (p) = i ( p_\mu \zeta_\nu (p) + p_\nu \zeta_\mu (p) ), \qquad
\delta_B^{(2)} \tilde A_\mu (p) = \zeta_\mu (p), \qquad
\delta_B^{(2)} \bar \zeta_\mu (p) = i \beta_\mu,
\nonumber\\
&{}& \delta_B^{(2)} \zeta_\mu (p) = \delta_B^{(2)} \beta_\mu (p) 
= \delta_B^{(2)} c^\mu (p) = \delta_B^{(2)} \bar c_\mu (p) = \delta_B^{(2)} b_\mu (p) = 0.
\label{ST-Q_B-Comp}  
\end{eqnarray}
These BRST transformations imply that $\phi (p), h_I (p)$ and $\psi_{\mu\nu} (p)$
could be physical observables while two sets of fields, $\{ \omega_\mu (p), b_\mu (p), 
c_\mu (p), \bar c_\mu (p) \}$ and $\{ \tilde A_\mu (p), \beta_\mu (p), 
\zeta_\mu (p), \bar \zeta_\mu (p) \}$ might belong to BRST quartets, which are dropped
from the physical state by the Kugo-Ojima subsidiary condition (\ref{Phys-Hilbert}) \cite{Kugo-Ojima}.  
However, $b_\mu (p), c_\mu (p)$ and $\bar c_\mu (p)$ are simple pole fields obeying 
$p^2 b_\mu (p) = p^2 c_\mu (p) = p^2 \bar c_\mu (p) = 0$, whereas $\omega_\mu (p)$ is a dipole field satisfying 
$( p^2 )^2 \omega_\mu (p) = 0$, so that a naive Kugo-Ojima's quartet mechanism cannot be applied 
to the GCT BRST quartet, $\{ \omega_\mu (p), b_\mu (p), c_\mu (p), \bar c_\mu (p) \}$ in a straightforward manner. 

To understand the GCT BRST quartet mechanism, let us present 4D CRs in terms of the components in (\ref{Lorentz}). 
From Eqs. (\ref{FT-4D-CR3}), (\ref{FT-4D-CR6}), (\ref{FT-4D-CR8}) and the definition (\ref{Lorentz}), it is straightforward to 
derive the following 4D CRs:
\begin{eqnarray}
&{}& [ h_I (p), h_J^\dagger (q) ] = \frac{\kappa^2}{2} \delta_{IJ} \theta (p^0) \delta(p^2) \delta^4 (p-q).
\label{Lor-4D-CR1}
\\
&{}& [ h_I (p), \omega_\mu^\dagger (q) ] = [ h_I (p), b_\mu^\dagger (q) ]
= [ b_\mu (p), b_\nu^\dagger (q) ] = 0. 
\label{Lor-4D-CR2}
\\
&{}& [ \omega_\mu (p), b_\nu^\dagger (q) ] = i \frac{1}{2} \eta_{\mu\nu} \theta (p^0) 
\delta(p^2) \delta^4 (p-q). 
\label{Lor-4D-CR3}
\\
&{}& \{ c^\mu (p), \bar c_\nu^\dagger (q) \} = - i \delta_\nu^\mu \theta (p^0) 
\delta(p^2) \delta^4 (p-q). 
\label{Lor-4D-CR4}
\end{eqnarray}
In addition to them, we have a little complicated expression for $[ \omega_\mu (p), \omega_\nu^\dagger (q) ]$
because $\omega_\mu (p)$ is a dipole field, but luckily enough this expression is unnecessary
for our purpose \cite{Kugo-Ojima}. 
It is known how to take out a simple pole field from a dipole field, which amounts to
using an operator defined by \cite{Kugo-Ojima} 
\begin{eqnarray}
{\cal D}_p = \frac{1}{2 |\vec{p}|^2} p_0 \frac{\partial}{\partial p_0} + c, 
\label{D-ope}
\end{eqnarray}
where $c$ is a constant. Using this operator, we can define a simple pole field
$\hat h_{\mu\nu} (p)$ from the dipole field $h_{\mu\nu} (p)$, which
obeys $(p^2)^2 h_{\mu\nu} (p) = 0$, as
\begin{eqnarray}
\hat h_{\mu\nu} (p) &\equiv& h_{\mu\nu} (p) - {\cal D}_p ( p^2 h_{\mu\nu} (p) )
\nonumber\\
&=& h_{\mu\nu} (p) - i {\cal D}_p \Biggl\{ 4 \kappa^2 \Biggl[ p_{(\mu} b_{\nu)} (p) 
+ \frac{(\beta_1 + 2 \beta_2) \kappa^2 \gamma^2}{2 \beta_1 (\beta_1 + 4 \beta_2)} 
p_\mu p_\nu p_\rho b^\rho (p) \Biggr] 
\nonumber\\
&+& \frac{2 \beta_2 \kappa^2 \gamma}{\beta_1 (\beta_1 + 4 \beta_2)} 
p_\mu p_\nu p_\rho \beta^\rho (p) \Biggr\},
\label{Simple-field}
\end{eqnarray}
where in the last equality we have used the Fourier transform of the linearized field equation (\ref{Box-h})
(cf. Eq. (\ref{FT-Box-h})).
It is then possible to prove the equation\footnote{A proof is given in Appendix E.} 
\begin{eqnarray}
p^2 \hat h_{\mu\nu} (p) = 0.
\label{Simple-field2}
\end{eqnarray}
Furthermore, in Eq. (\ref{Lorentz}) we redefine $\omega_\mu$ by $\hat \omega_\mu$ as
\begin{eqnarray}
&{}& \hat \omega_0 (p) = \frac{1}{2 p} [ \hat h_{00} (p) - \hat h_{11} (p) ],  \qquad
\hat \omega_I (p) = \frac{1}{p} \hat h_{0I} (p),  
\nonumber\\
&{}& \hat \omega_3 (p) = - \frac{1}{2 p} [ \hat h_{11} (p) + \hat h_{33} (p) ].
\label{hat-omega}  
\end{eqnarray}
The key point is that with this redefinition from $\omega_\mu$ to $\hat \omega_\mu$,
the GCT BRST transformation and the 4D CRs remain unchanged owing to
$\delta_B^{(1)} b_\mu = \delta_B^{(1)} \beta_\mu = 0$ and $[ b_\mu (p), b^\dagger_\nu (q) ] 
= [ b_\mu (p), \beta^\dagger_\nu (q) ] = [ h_I (p), b^\dagger_\mu (q) ] = [ h_I (p), \beta^\dagger_\mu (q) ] = 0$, 
those are,
\begin{eqnarray}
&{}& \delta_B  \hat \omega_\mu (p) = i c_\mu (p),  \quad
[ \hat \omega_\mu (p), b^\dagger_\nu (q) ] = [ \omega_\mu (p), b^\dagger_\nu (q) ],
\nonumber\\
&{}& [ h_I (p), \hat \omega^\dagger_\mu (q) ] = [ h_I (p), \omega^\dagger_\mu (q) ].
\label{point}  
\end{eqnarray}

Now that it turns out that the fields, $\{ h_I, \hat \omega_\mu, b_\mu,
c_\mu, \bar c_\mu \}$ are simple pole fields,\footnote{Without the redefinition,
$h_I (p)$ is already a simple pole field as can be seen in Eq. (\ref{Lor-4D-CR1}).}
we can obtain the standard creation and annihilation operators in the three-dimensional 
Fourier expansion from those in the four-dimensional one through the relation (\ref{3D-4D}). 
As a result, the three-dimensional (anti)commutation relations, which are denoted as 
$[ \Phi (\vec{p}), \Phi^\dagger (\vec{q}) \}$ with
$\Phi (\vec{p}) \equiv \{ h_I (\vec{p}), \hat \omega_\mu (\vec{p}), b_\mu (\vec{p}), 
c_\mu (\vec{p}), \bar c_\mu (\vec{p}) \}$, are given by\footnote{The bracket $[ A, B \}$ 
is the graded commutation relation denoting either commutator or anticommutator, according to 
the Grassmann-even or odd character of $A$ and $B$, i.e., $[ A, B \} = A B - (-)^{|A| |B|} B A$.}  
\begin{eqnarray}
[ \Phi (\vec{p}), \Phi^\dagger (\vec{q}) \} &=&
\left(
\begin{array} {cc|cc|cc}
    &        \frac{\kappa^2}{2} \delta_{IJ}     &     &    &    &    \\ 
\hline
    &        & [ \hat \omega_\mu (\vec{p}), \hat \omega_\nu^\dagger (\vec{q}) ]  & i \frac{1}{2} \eta_{\mu\nu}  \\ 
    &        &  -i \frac{1}{2}  \eta_{\mu\nu}   &  0       &              \\
\hline   
    &        &       &    &  & - i \eta_{\mu\nu} \\  
    &        &      & & -i \eta_{\mu\nu} &            \\
\end{array}
\right) 
\times \delta ( \vec{p} - \vec{q} )_.
\label{G-3D-CRs}  
\end{eqnarray}
The (anti)commuatation relations (\ref{G-3D-CRs}) have in essence the same structure as those 
of the Yang-Mills theory \cite{Kugo-Ojima}. In particular, the positive coefficient $\frac{\kappa^2}{2} \delta_{IJ}$ 
means that $h_I$ have a positive norm. Hence, we find that $h_I$, which corresponds to the transverse graviton, 
is the physical observable of positive norm while a set of fields $\{ \hat \omega_\mu, b_\mu, c_\mu, 
\bar c_\mu \}$ belongs to a BRST quartet. This quartet appears in the physical subspace only 
as zero norm states by the Kugo-Ojima subsidiary conditions (\ref{Phys-Hilbert}).
It is worthwhile to stress that both the massive scalar and the massless graviton of positive 
norm appear in the physical Hilbert space.\footnote{That the massive scalar $\phi$ has a positive norm can be 
understood from Eq. (\ref{FT-4D-CR1}).}

Next, let us consider another BRST quartet, $\{ \tilde A_\mu (p), \beta_\mu (p), \zeta_\mu (p), \bar \zeta_\mu (p) \}$.
All the members in this quartet are dipole ghost fields, so that BRST quartet mechanism works for this quartet
in a direct manner. This BRST quartet is therefore dropped from the physical Hilbert space by the Kugo-Ojima subsidiary
condition (\ref{Phys-Hilbert}) \cite{Kugo-Ojima}.    

Finally, we would like to exhibit that the massive ghost has a negative norm, thereby violating the unitarity of quadratic
gravity. For this aim, let us take a Lorentz frame, $p^\mu = ( p, 0, 0, 0 ) = ( M, 0, 0, 0 )$, and select five independent 
components $\psi_A$ taking $A = 1, 2, \dots, 5$ in $\psi_{\mu\nu}$ as
\begin{eqnarray}
&{}& \psi_1 (p) = \frac{1}{2} ( \psi_{11} (p) - \psi_{22} (p) ),  \qquad
\psi_2 (p) = \frac{\sqrt{3}}{2} ( \psi_{11} (p) + \psi_{22} (p) ),
\nonumber\\
&{}& \psi_3 (p) = \psi_{12} (p),  \qquad
\psi_4 (p) = \psi_{13} (p),  \qquad
\psi_5 (p) = \psi_{23} (p).
\label{Lorentz2}  
\end{eqnarray}
Note that the other five components can be expressed by the above five ones via the equations
$\partial^\mu \psi_{\mu\nu} = \eta^{\mu\nu} \psi_{\mu\nu} = 0$ in Eq. (\ref{Massive-ghost-eq}). 
Under these conditions, Eq. (\ref{FT-4D-CR2}) leads to 
\begin{eqnarray}
[ \psi_A (p), \psi_B^\dagger (q) ] = - \frac{1}{2 \kappa^2 \gamma^2} \delta_{AB} \theta(p^0)
\delta( p^2 + M^2 ) \delta^4 ( p - q ).
\label{Psi-4d-FT}  
\end{eqnarray}
The negative sign on the RHS shows that the massive ghost certainly has a negative norm.

To summarize, we have demonstrated that physical states of quadratic gravity are composed of
the spin $0$ massive scalar, the spin $2$ massless graviton and the spin $2$ massive ghost.
In particular, the massive ghost has negative norm so it breaks the unitarity of the physical S-matrix.

\section{Conclusions}

In this article, we have presented the manifestly covariant quantization of quadratic gravity in the de Donder gauge condition 
for general coordinate invariance on the basis of the BRST transformation. After constructing the BRST-invariant quantum action
and setting up the canonical (anti)commutation relations (CCRs), we have explicitly calculated the important equal-time commutation 
relations (ETCRs). Of course, along almost the similar line all the ETCRs can be in principle evaluated in terms of the CCRs, 
field equations and BRST transformations.

In particlular, we have focused our attention on the ETCRs between the metric tensor and its time derivatives in detail, 
and found that they vanish identically as in Eq. (\ref{general-dot-g-dot-g}), which is one of the most surprising results in this study. 
It is of interest to recall what becomes of the ETCRs in the other gravitational theories. 
In the conventional general relativity, as can be seen in Eq. (\ref{GR-gg-CR}), the ETCR, $[ \dot g_{\rho\sigma}, g_{\mu\nu}^\prime ]$ 
is not vanishing, which reflects the fact that the metric tensor is dynamical and carries physical degrees of
freedom associated with the graviton. In the famous higher-derivative gravities, those are, conformal gravity \cite{Oda-Ohta} 
and $f(R)$ gravity \cite{Oda-f}, it is not vanishing either. As is well known, conformal gravity has $7$ physical degrees of freedom 
which correspond to the massless graviton of $2$ physical degrees and the massive ghost of $5$ degrees whereas $f(R)$ gravity, 
which includes $R^2$ gravity as a special case, possesses $3$ physical degrees of freedom which correspond to the massless graviton 
of $2$ physical degrees and the massive scalar of $1$ degree, which is sometimes called ``scalaron''. From this perspective, 
we can say that the vanishing ETCRs occur only when the massive ghost coexists with the massive scalar, so it might be related to 
the perturbative renormalizability of quadratic gravity. 

It is also valuable to point out that the metric tensor field usually has the dual role as the dynamical field and the geometrical standard, 
which makes it difficult to construct a quantum gravity based on general relativity \cite{Wald}. On the other hand, in quadratic gravity 
under consideration, the dual role is separated out and the dynamical role is transferred to different fields, i.e. $\psi_{\mu\nu}, h_{\mu\nu}$
and $\phi$, which are linear combination of the metric field, auxiliary fields and FP ghosts. This situation might also make it possible to construct 
a renormalizable quantum gravity in case of quadratic gravity.

Finally, we should briefly comment on confinement of the massive ghost, thereby recovering the unitarity of the physical 
S-matrix in quadratic gravity within the framework of canonical formalism. As mentioned at the beginning of Section 6, we have assumed that 
all elementary fields have their own asymptotic fields and there is no bound state. Then, a natural question arises that 
if there is some bound state, how the results obtained in Sections 6 and 7 would be modified, in particular, 
the destiny of the massive ghost violating the unitarity.\footnote{There might be a question about the existence of asymptotic fields. 
However, the assumption called $\it{asymptotic \, completeness}$ is a crucial postulate in quantum field theories 
since it is the only known general principle which uniquely determines the representation space of field operators \cite{N-O-text}, 
so we preserve the assumption of the asymptotic completeness.}  Actually, such a question has been
already proposed in Ref. \cite{Kawasaki} where it is assumed that a bound state with the same mass of the massive ghost is formed 
between the metric tensor field and the FP ghosts, and it is a BRST-partner of the massive ghost. Then, via the BRST quartet 
mechanism, the massive ghost is confined in the zero-norm states as in the FP ghosts, thereby recovering the unitarity of 
the physical S-matrix. It is certain that this idea is very interesting but has some unsolved problems. For instance, it is quite
unclear how the bound state can be made and why the bound state has exactly the same mass as the massive ghost.
In a separate paper, we shall advocate a more sophisticated mechanism for the confinement of the massive ghost
in the future.


\appendix
\addcontentsline{toc}{section}{Appendix~\ref{app:scripts}: Training Scripts}
\section*{Appendix}
\label{app:scripts}
\renewcommand{\theequation}{A.\arabic{equation}}
\setcounter{equation}{0}

\section{Derivation of $[ \dot K_{\rho\sigma}, K_{\mu\nu}^\prime ]$ at the linearized level}

It is difficult to derive a full expression for the ETCR, $[ \dot K_{\rho\sigma}, K_{\mu\nu}^\prime ]$
but it is rather easy to do so at the linearized level.  In this appendix, we present a derivation of the ETCR, 
$[ \dot K_{\rho\sigma}, K_{\mu\nu}^\prime ]$ at the linearized level.

At the linearized level, the conjugate momenta, $\pi_g^{\mu\nu}$ and $\pi_K^{\mu\nu}$ take the form
\begin{eqnarray}
\pi_g^{\mu\nu} &=& - \frac{1}{4 \kappa^2} ( - \eta^{0 \lambda} \eta^{\mu\nu} \eta^{\sigma\tau} 
- \eta^{0 \tau} \eta^{\mu\lambda} \eta^{\nu\sigma} - \eta^{0 \sigma} \eta^{\mu\tau} \eta^{\nu\lambda} 
+ \eta^{0 \lambda} \eta^{\mu\tau} \eta^{\nu\sigma} 
+ \eta^{0 \tau} \eta^{\mu\nu} \eta^{\lambda\sigma}
\nonumber\\
&+& \eta^{0 (\mu} \eta^{\nu)\lambda} \eta^{\sigma\tau} ) \partial_\lambda \varphi_{\sigma\tau}
- \frac{1}{2} ( 2 \eta^{0 (\mu} \eta^{\nu)\rho} -  \eta^{\mu\nu} \eta^{0 \rho} )  b_\rho
\nonumber\\
&-& \gamma \Big( \partial^{(\mu} \bar K^{\nu)0} - \frac{1}{2} \partial^0 \bar K^{\mu\nu} 
- \frac{1}{2} \eta^{\mu\nu} \partial_\rho \bar K^{0\rho} \Big), 
\nonumber\\
\pi_K^{\mu\nu} &=& - \gamma \Big[ - \frac{1}{2} \Big( \eta^{\mu\rho} \eta^{\nu\sigma} 
- \frac{1}{2} \eta^{\mu\nu} \eta^{\rho\sigma} \Big) 
( 2 \partial_{(\rho} \varphi_{\sigma) 0} - \dot \varphi_{\rho\sigma} )
\nonumber\\
&-& \frac{1}{4} ( 2 \eta^{0(\mu} \eta^{\nu)\rho} - \eta^{\mu\nu} \eta^{0\rho} ) \partial_\rho \varphi \Big] 
- \eta^{0(\mu} \beta^{\nu)}, 
\label{L-CCM}  
\end{eqnarray}
where $\bar K^{\mu\nu} \equiv K^{\mu\nu} - \frac{1}{2} \eta^{\mu\nu} K$ and $K \equiv \eta^{\mu\nu} K_{\mu\nu}$.
  
From these canonical conjugate momenta, we can express $\dot K_{ij}$ as
\begin{eqnarray}
\dot K_{ij} = 2 \partial_{(i} K_{j)0} - \frac{2}{\gamma} \Big( \eta_{i \mu} \eta_{j \nu} 
- \frac{1}{2} \eta_{ij} \eta_{\mu\nu} \Big) \Big( \pi_g^{\mu\nu} + \frac{1}{2 \kappa^2 \gamma} \pi_K^{\mu\nu} \Big)
+ \frac{1}{2 \kappa^2 \gamma^2} \eta_{ij} \beta^0. 
\label{L-dot-K}  
\end{eqnarray}
Using this expression, it is easy to derive the ETCR:
\begin{eqnarray}
[ \dot K_{ij}, K_{\mu\nu}^\prime ] = i \frac{1}{2 \kappa^2 \gamma^2} [ \eta_{i \mu} \eta_{j \nu} + \eta_{i \nu} \eta_{j \mu} 
- \eta_{ij} ( \eta_{\mu\nu} + \delta_\mu^0 \delta_\nu^0 ) ] \delta^3, 
\label{dot-Kij-K}  
\end{eqnarray}
where we have utilized $[ \beta_\rho, K_{\mu\nu}^\prime ] = - i ( \delta_\mu^0 \eta_{\rho\nu} + \delta_\nu^0 \eta_{\rho\mu}
+ \delta_\rho^0 \delta_\mu^0 \delta_\nu^0 ) \delta^3$, which is the linearized version of Eq. (\ref{beta-K2}).
Moreover, the linearized gauge condition $\partial_\mu K^{\mu\nu} = 0$ produces
\begin{eqnarray}
[ \dot K_{0\rho}, K_{\mu\nu}^\prime ] = 0. 
\label{dot-K0rho-K}  
\end{eqnarray}
Then, Eqs. (\ref{dot-Kij-K}) and (\ref{dot-K0rho-K}) are summarized to be
\begin{eqnarray}
[ \dot K_{\rho\sigma}, K_{\mu\nu}^\prime ] &=& i \frac{1}{2 \kappa^2 \gamma^2} ( - \eta_{\rho\sigma} \eta_{\mu\nu} 
+ 2 \eta_{\rho(\mu} \eta_{\nu)\sigma} - \delta_\rho^0 \delta_\sigma^0 \eta_{\mu\nu} 
- \delta_\mu^0 \delta_\nu^0 \eta_{\rho\sigma} + 2 \delta_\rho^0 \eta_{\sigma(\mu} \delta_{\nu)}^0 
\nonumber\\
&+& 2 \delta_\sigma^0 \eta_{\rho(\mu} \delta_{\nu)}^0 + \delta_\rho^0 \delta_\sigma^0 \delta_\mu^0 \delta_\nu^0 ) \delta^3. 
\label{L-dot-K-K}  
\end{eqnarray}

\renewcommand{\theequation}{B.\arabic{equation}}
\setcounter{equation}{0}

\section{Useful ETCRs in flat Minkowski background}

In this appendix, we write down useful ETCRs and field equations in a flat Minkowski background, which are extensively 
used in deriving 4D CRs in Section 7. By means of various CCRs and ETCRs shown in this article, the proof of these ETCRs is so straightforward 
that we omit it.

In case of the ETCRs at the linearized level in a flat Minkowski background, one can utilize a general formula:
\begin{eqnarray}
[ \dot A, B^\prime ] = - [ A, \dot B^\prime ], 
\label{P-der}  
\end{eqnarray}
where $A$ and $B$ are generic Heisenberg operators.
Here let us write down useful forms of field equations for $A_\mu, \beta_\mu, \zeta_\mu$ and $\bar \zeta_\mu$:
\begin{eqnarray}
\ddot A_0 &=& \frac{\beta_1}{2 (\beta_1 + \beta_2)} \left( \Delta A_0 + \frac{\beta_1 + 2 \beta_2}{\beta_1} 
\partial_i \dot A_i - \frac{\beta_2}{\beta_1} \dot K \right),
\nonumber\\
\ddot A_i &=& \Delta A_i + \frac{\beta_1 + 2 \beta_2}{\beta_1} ( - \partial_i \dot A_0 + \partial_i \partial_j A_j ) 
- \frac{\beta_2}{\beta_1} \partial_i K,
\nonumber\\
\ddot \beta_0 &=& \frac{1}{2} ( \Delta \beta_0 + \partial_i \dot \beta_i ),
\qquad
\ddot \beta_i = \Delta \beta_i - \partial_i \dot \beta_0 + \partial_i \partial_j \beta_j,
\nonumber\\
\ddot \zeta_0 &=& \frac{1}{2} ( \Delta \zeta_0 + \partial_i \dot \zeta_i ),
\qquad
\ddot \zeta_i = \Delta \zeta_i - \partial_i \dot \zeta_0 + \partial_i \partial_j \zeta_j,
\nonumber\\
\ddot{\bar \zeta}_0 &=& \frac{1}{2} ( \Delta \bar \zeta_0 + \partial_i \dot{\bar \zeta}_i ),
\qquad
\ddot{\bar \zeta_i} = \Delta \bar \zeta_i - \partial_i \dot{\bar \zeta}_0 + \partial_i \partial_j \bar \zeta_j.
\label{Leq-A-beta}  
\end{eqnarray}

 Using these linearized field equations and relevant CCRs and ETCRs, we can derive the following ETCRs:
\begin{eqnarray}
&{}& [ \dot A_\mu, A_\nu^\prime ] = i \frac{1}{4 \beta_1} \left[ \eta_{\mu\nu} + \frac{\beta_1 + 2 \beta_2}{2 (\beta_1 + \beta_2)} 
\delta_\mu^0 \delta_\nu^0 \right] \delta^3,  \qquad
[ \dot A_\mu, \dot A_\nu^\prime ] = - i \frac{\beta_1 + 2 \beta_2}{4 \beta_1 (\beta_1 + \beta_2)} \delta_{(\mu}^0 \delta_{\nu)}^i \partial_i \delta^3,
\nonumber\\
&{}& [ \dot A_\mu, \ddot A_\nu^\prime ] = i \Biggl\{ \frac{1}{4 \beta_1} \eta_{\mu\nu} \Delta
+ \delta_\mu^0 \delta_\nu^0 \left[ \frac{\beta_1 + 2 \beta_2}{4 \beta_1 (\beta_1 + \beta_2)} \Delta
+ \frac{3 \beta_2^2}{8 (\beta_1 + \beta_2)^2 \kappa^2 \gamma^2} \right]
\nonumber\\
&{}& + \frac{\beta_1 + 2 \beta_2}{8 \beta_1 (\beta_1 + \beta_2)} \delta_\mu^i \delta_\nu^j 
\partial_i \partial_j \Biggr\} \delta^3,
\nonumber\\
&{}&  [ \varphi_{\mu\nu}, \dot A_\rho^\prime ] = [ K_{\mu\nu}, \dot A_\rho^\prime ] = 0,
\nonumber\\
&{}&  [ \varphi_{\mu\nu}, \ddot A_\rho^\prime ] = - i \frac{\beta_2}{2 (\beta_1 + \beta_2) \gamma} \delta_\rho^0
( \eta_{\mu\nu} - 2 \delta_\mu^0 \delta_\nu^0 ) \delta^3,
\nonumber\\
&{}&  [ K_{\mu\nu}, \ddot A_\rho^\prime ] = - i \frac{\beta_2}{4 (\beta_1 + \beta_2) \kappa^2 \gamma^2} \delta_\rho^0
( \eta_{\mu\nu} + \delta_\mu^0 \delta_\nu^0 ) \delta^3,
\nonumber\\
&{}&  [ A_\mu, \dot \beta_\nu^\prime ] = - i \left( \eta_{\mu\nu} + \frac{1}{2} \delta_\mu^0 \delta_\nu^0 \right) \delta^3,
\qquad [ A_\mu, \ddot \beta_\nu^\prime ] = i \delta_{(\mu}^0 \delta_{\nu)}^i \partial_i \delta^3,
\nonumber\\
&{}&  [ K_{\mu\nu}, \beta_\rho^\prime ] = i ( \delta_\mu^0 \eta_{\rho\nu} + \delta_\nu^0 \eta_{\rho\mu}
+ \delta_\mu^0 \delta_\nu^0 \delta_\rho^0 ) \delta^3, 
\nonumber\\
&{}& [ K_{\mu\nu}, \dot \beta_\rho^\prime ] = - i ( \delta_\mu^0 \delta_\nu^0 \delta_\rho^i 
+ \delta_\mu^0 \delta_\nu^i \delta_\rho^0 + \delta_\mu^i \delta_\nu^0 \delta_\rho^0 
+ 2 \eta_{\rho(\mu} \delta_{\nu)}^i ) \partial_i \delta^3,
\nonumber\\
&{}& [ K_{\mu\nu}, \ddot \beta_\rho^\prime ] = i [ 2 ( \delta_{(\mu}^0 \eta_{\nu)\rho} 
+ \delta_\mu^0 \delta_\nu^0 \delta_\rho^0 ) \Delta + ( \delta_\mu^0 \delta_\nu^i \delta_\rho^j
+ \delta_\mu^i \delta_\nu^0 \delta_\rho^j + \delta_\mu^i \delta_\nu^j \delta_\rho^0 ) \partial_i \partial_j ] \delta^3.  
\label{L-ETCRs1}  
\end{eqnarray}
 
The other useful ETCRs are summarized as 
\begin{eqnarray}
&{}& \{ \bar \zeta_\mu, \dot \zeta_\nu^\prime \} = \left( \eta_{\mu\nu} 
+ \frac{1}{2} \delta_\mu^0 \delta_\nu^0 \right) \delta^3,
\qquad
\{ \bar \zeta_\mu, \ddot \zeta_\nu^\prime \} = - \{ \zeta_\mu, \ddot{\bar\zeta}_\nu^\prime \}
= - \delta_{(\mu}^0 \delta_{\nu)}^i \partial_i \delta^3,
\nonumber\\
&{}& \{ c^\mu, \dot{\bar c}^\prime_\nu \} = - \{ \bar c_\nu, \dot c^{\prime \mu} \} = \delta_\mu^\nu \delta^3,
\nonumber\\
&{}& [ \varphi_{\mu\nu}, b_\rho^\prime ] = 2 i \delta_{(\mu}^0 \eta_{\nu)\rho} \delta^3,
\qquad
[ \varphi_{\mu\nu}, \dot b_\rho^\prime ] = - 2 i \delta_{(\mu}^i \eta_{\nu)\rho} \partial_i \delta^3,
\nonumber\\
&{}& [ \varphi_{\mu\nu}, \dot K_{\rho\sigma}^\prime ] = i \frac{1}{\gamma} [ \eta_{\mu\nu} \eta_{\rho\sigma}
- 2 \eta_{\mu(\rho} \eta_{\sigma)\nu} + \eta_{\mu\nu} \delta_\rho^0 \delta_\sigma^0 
- 2 ( \eta_{\mu(\rho} \delta_{\sigma)}^0 \delta_\nu^0 + \eta_{\nu(\rho} \delta_{\sigma)}^0 \delta_\mu^0 )
- 2 \delta_\mu^0 \delta_\nu^0 \delta_\rho^0 \delta_\sigma^0 ] \delta^3,
\nonumber\\
&{}& [ \varphi_{\mu\nu}, \dot K^\prime ] = i \frac{1}{\gamma} ( \eta_{\mu\nu} - 2 \delta_\mu^0 \delta_\nu^0 )
\delta^3,
\qquad  [ \varphi_{\mu\nu}, \phi^\prime ] = i \frac{4 \kappa^2 \gamma}{\beta_1 + 4 \beta_2} 
\delta_{(\mu}^0 \delta_{\nu)}^i \partial_i \delta^3,
\nonumber\\
&{}&
[ \varphi_{\mu\nu}, \dot \phi^\prime ] = i \frac{\beta_1}{(\beta_1 + \beta_2) \gamma} ( \eta_{\mu\nu} 
- 2 \delta_\mu^0 \delta_\nu^0 ) \delta^3 - i \frac{2 \kappa^2 \gamma}{\beta_1 + 4 \beta_2} 
( \delta_\mu^0 \delta_\nu^0 \Delta + \delta_\mu^i \delta_\nu^j \partial_i \partial_j ) \delta^3,
\nonumber\\
&{}&
[ A_\mu, \phi^\prime ] = i \frac{3 \beta_2}{4 (\beta_1 + \beta_2)(\beta_1 + 4 \beta_2)} 
\delta_\mu^0 \delta^3,
\qquad
[ A_\mu, \dot \phi^\prime ] = - i \frac{3 \beta_2}{4 (\beta_1 + \beta_2)(\beta_1 + 4 \beta_2)} 
\delta_\mu^i \partial_i \delta^3,
\nonumber\\
&{}&
[ K_{\mu\nu}, \phi^\prime ] = - i \frac{1}{\beta_1 + 4 \beta_2} \delta_{(\mu}^0 \delta_{\nu)}^i
\partial_i \delta^3,
\nonumber\\
&{}&
[ K_{\mu\nu}, \dot \phi^\prime ] = i \frac{\beta_1}{2 (\beta_1 + \beta_2) \kappa^2 \gamma^2} 
( \eta_{\mu\nu} + \delta_\mu^0 \delta_\nu^0 ) \delta^3
+ i \frac{1}{2 (\beta_1 + 4 \beta_2)} ( \delta_\mu^0 \delta_\nu^0 \Delta 
+ \delta_\mu^i \delta_\nu^j \partial_i \partial_j ) \delta^3.
\label{L-ETCRs2}  
\end{eqnarray}

\renewcommand{\theequation}{C.\arabic{equation}}
\setcounter{equation}{0}

\section{A derivation of $[ h_{\mu\nu} (x), h_{\rho\sigma} (y) ]$}

Since the derivation of this 4D CR is most complicated among all the 4D CRs, we wish to present its 
derivation in an explicit manner.
 
Since $h_{\mu\nu}$ is a massless dipole ghost field, we can express it as in Eq. (\ref{E-h-exp}).
By means of this expression, the 4D CR, $[ h_{\mu\nu} (x), h_{\rho\sigma} (y) ]$ takes the form:
\begin{eqnarray}
&{}& [ h_{\mu\nu} (x), h_{\rho\sigma} (y) ] = - \int d^3 z \Biggl( D(y-z) \overleftrightarrow{\partial}_0^z 
[ h_{\mu\nu} (x), h_{\rho\sigma} (z) ] + E(y-z) \overleftrightarrow{\partial}_0^z 
[ h_{\mu\nu} (x), \Box h_{\rho\sigma} (z) ] \Biggr)
\nonumber\\
&{}& = - \int d^3 z \Biggl( D(y-z) [ h_{\mu\nu} (x), \dot h_{\rho\sigma} (z) ] - \partial_0^z D(y-z) 
[ h_{\mu\nu} (x), h_{\rho\sigma} (z) ]  
\nonumber\\
&{}& + E(y-z) [ h_{\mu\nu} (x), \Box \dot h_{\rho\sigma} (z) ] - \partial_0^z E(y-z) 
\left. [ h_{\mu\nu} (x), \Box h_{\rho\sigma} (z) ] \Biggr) \right|_{z^0 = x^0}
\nonumber\\
&{}& \equiv I + J \equiv ( I_1 + I_2 ) + ( J_1 + J_2 ).
\label{hh-4D-CR}  
\end{eqnarray}
where we will use Eq. (\ref{Box-h}) for $\Box h_{\mu\nu}$ in $J$.

After a little tedious but straightforward calculation, we can get the result for $I_1, I_2$ and $I$:
\begin{eqnarray}
&{}& I_1 \equiv - \int d^3 z \, D(y-z) \left. [ h_{\mu\nu} (x), \dot h_{\rho\sigma} (z) ] \right|_{z^0 = x^0}
\nonumber\\
&{}& = - 2 i \kappa^2 [ \eta_{\mu\nu} \eta_{\rho\sigma} - 2 \eta_{\mu(\rho} \eta_{\sigma)\nu}
- 2 ( \eta_{\rho(\mu} \delta_{\nu)}^0 \delta_\sigma^0 + \eta_{\sigma(\mu} \delta_{\nu)}^0 \delta_\rho^0 ) ] D(x-y) 
+ i \frac{2 (\beta_1 + 2 \beta_2) \kappa^4 \gamma^2}{\beta_1 (\beta_1 + 4 \beta_2)}
\nonumber\\
&{}& \times [ ( \eta_{\mu\nu} - 2 \delta_\mu^0 \delta_\nu^0 )( \delta_\rho^0 \delta_\sigma^0 \partial_0^2 
+ \delta_\rho^i \delta_\sigma^j \partial_i \partial_j )
+ ( \eta_{\rho\sigma} - 2 \delta_\rho^0 \delta_\sigma^0 )( \delta_\mu^0 \delta_\nu^0 \partial_0^2 
+ \delta_\mu^i \delta_\nu^j \partial_i \partial_j ) ] D(x-y)   
\nonumber\\
&{}& - i \frac{16 (\beta_1 + 2 \beta_2) \kappa^4 \gamma^2}{\beta_1 (\beta_1 + 4 \beta_2)} 
\delta_{(\mu}^0 \delta_{\nu)}^i \delta_{(\rho}^0 \delta_{\sigma)}^j \partial_i \partial_j D(x-y)
- i \frac{\kappa^4 \gamma^2}{\beta_1} [ \eta_{\mu\rho} ( \delta_\nu^0 \delta_\sigma^0 \partial_0^2 
+ \delta_\nu^i \delta_\sigma^j \partial_i \partial_j ) 
\nonumber\\
&{}& + \eta_{\mu\sigma} ( \delta_\nu^0 \delta_\rho^0 \partial_0^2 + \delta_\nu^i \delta_\rho^j \partial_i \partial_j )
+ \eta_{\nu\rho} ( \delta_\mu^0 \delta_\sigma^0 \partial_0^2 + \delta_\mu^i \delta_\sigma^j \partial_i \partial_j )
+ \eta_{\nu\sigma} ( \delta_\mu^0 \delta_\rho^0 \partial_0^2 + \delta_\mu^i \delta_\rho^j \partial_i \partial_j ) ] D(x-y)
\nonumber\\
&{}& - i \frac{2 (\beta_1 + 2 \beta_2)(\beta_1 - 2 \beta_2) \kappa^6 \gamma^4}{\beta_1^2 (\beta_1 + 4 \beta_2)^2}
[ ( \delta_\mu^0 \delta_\nu^0 \partial_0^2 + \delta_\mu^i \delta_\nu^j \partial_i \partial_j )
( \delta_\rho^0 \delta_\sigma^0 \partial_0^2 + \delta_\rho^k \delta_\sigma^l \partial_k \partial_l )
\nonumber\\
&{}& + 4 \delta_{(\mu}^0 \delta_{\nu)}^i \delta_{(\rho}^0 \delta_{\sigma)}^j \partial_0^2 \partial_i \partial_j ] D(x-y).
\label{I-1}  
\\
&{}& I_2 \equiv \int d^3 z \, \partial_0^z D(y-z) \Big. [ h_{\mu\nu} (x), h_{\rho\sigma} (z) ] \Big|_{z^0 = x^0}
\nonumber\\
&{}& = i \frac{4 (\beta_1 + 2 \beta_2) \kappa^4 \gamma^2}{\beta_1 (\beta_1 + 4 \beta_2)} 
[ ( \eta_{\mu\nu} - 2 \delta_\mu^0 \delta_\nu^0 ) \delta_{(\rho}^0 \delta_{\sigma)}^i
+ ( \eta_{\rho\sigma} - 2 \delta_\rho^0 \delta_\sigma^0 ) \delta_{(\mu}^0 \delta_{\nu)}^i ] 
\partial_0 \partial_i D(x-y)   
\nonumber\\
&{}& - i \frac{4 (\beta_1 + 2 \beta_2)(\beta_1 - 2 \beta_2) \kappa^6 \gamma^4}{\beta_1^2 (\beta_1 + 4 \beta_2)^2} 
[ ( \delta_\mu^0 \delta_\nu^0 \partial_0^2 + \delta_\mu^j \delta_\nu^k \partial_j \partial_k ) \delta_{(\rho}^0 \delta_{\sigma)}^i
+ ( \delta_\rho^0 \delta_\sigma^0 \partial_0^2 + \delta_\rho^j \delta_\sigma^k \partial_j \partial_k ) \delta_{(\mu}^0 \delta_{\nu)}^i ]
\nonumber\\
&{}& \times \partial_0 \partial_i D(x-y)
- i \frac{2 \kappa^4 \gamma^2}{\beta_1} ( \eta_{\mu\rho} \delta_{(\nu}^0 \delta_{\sigma)}^i 
+ \eta_{\mu\sigma} \delta_{(\nu}^0 \delta_{\rho)}^i + \eta_{\nu\rho} \delta_{(\mu}^0 \delta_{\sigma)}^i 
+ \eta_{\nu\sigma} \delta_{(\mu}^0 \delta_{\rho)}^i ) 
\nonumber\\
&{}& \times \partial_0 \partial_i D(x-y).
\label{I-2}  
\\
&{}& I \equiv I_1 + I_2 
\nonumber\\
&{}& = - 2 i \kappa^2 [ \eta_{\mu\nu} \eta_{\rho\sigma} - 2 \eta_{\mu(\rho} \eta_{\sigma)\nu}
- 2 ( \eta_{\rho(\mu} \delta_{\nu)}^0 \delta_\sigma^0 + \eta_{\sigma(\mu} \delta_{\nu)}^0 \delta_\rho^0 ) ] D(x-y) 
+ i \frac{2 (\beta_1 + 2 \beta_2) \kappa^4 \gamma^2}{\beta_1 (\beta_1 + 4 \beta_2)}
\nonumber\\
&{}& \times [ ( \eta_{\mu\nu} - 2 \delta_\mu^0 \delta_\nu^0 ) \partial_\rho \partial_\sigma 
+ ( \eta_{\rho\sigma} - 2 \delta_\rho^0 \delta_\sigma^0 ) \partial_\mu \partial_\nu 
- 8 \delta_{(\mu}^0 \delta_{\nu)}^i \delta_{(\rho}^0 \delta_{\sigma)}^j \partial_i \partial_j ] D(x-y)   
\nonumber\\
&{}& - i \frac{\kappa^4 \gamma^2}{\beta_1} ( \eta_{\mu\rho} \partial_\nu \partial_\sigma 
+ \eta_{\mu\sigma} \partial_\nu \partial_\rho + \eta_{\nu\rho} \partial_\mu \partial_\sigma
+ \eta_{\nu\sigma} \partial_\mu \partial_\rho ) D(x-y)
\nonumber\\
&{}& - i \frac{2 (\beta_1 + 2 \beta_2)(\beta_1 - 2 \beta_2) \kappa^6 \gamma^4}{\beta_1^2 (\beta_1 + 4 \beta_2)^2}
\partial_\mu \partial_\nu \partial_\rho \partial_\sigma D(x-y).
\label{I}  
\end{eqnarray}
 
Next, let us move on to the evaluation of the $J$ part. We find that calculating $J$ part is easier rather than doing 
the above $I$ part since only the terms involving $b_\mu$ field in (\ref{Box-h}) contribute to $[ h_{\mu\nu} (x), 
\Box h_{\rho\sigma} (z) ]$ and $[ h_{\mu\nu} (x), \Box \dot h_{\rho\sigma} (z) ]$ owing to Eq. (\ref{4D-CR10}). 

Here we would like to calculate $J_2$ in an explicit manner. Let us start with the definition of $J_2$:
\begin{eqnarray}
&{}& J_2 \equiv \int d^3 z \, \partial_0^z E(y-z) \Big. [ h_{\mu\nu} (x), \Box h_{\rho\sigma} (z) ] \Big|_{z^0 = x^0}
\nonumber\\
&{}& = - 4 \kappa^2 \int d^3 z \, \partial_0^z E(y-z) \Big. [ h_{\mu\nu} (x), \partial_{(\rho} b_{\sigma)} (z) 
- \frac{(\beta_1 + 2 \beta_2) \kappa^2 \gamma^2}{2 \beta_1 (\beta_1 + 4 \beta_2)} \partial_\rho
\partial_\sigma \partial_\tau b^\tau (z) ] \Big|_{z^0 = x^0},
\label{J2-1}  
\end{eqnarray}
where we have inserted Eq. (\ref{Box-h}) and utilized the 4D CR, $[ h_{\mu\nu} (x), \beta_\rho (y) ] = 0$ as mentioned above. 

From the 4D CR, $[ h_{\mu\nu} (x), b_\rho (y) ] = - i ( \eta_{\mu\rho} \partial_\nu + \eta_{\nu\rho} \partial_\mu ) D(x-z)$, 
it is easy to derive the following ETCRs:
\begin{eqnarray}
&{}& \Big. [ h_{\mu\nu} (x), \partial_{(\rho} b_{\sigma)} (z) ] \Big|_{z^0 = x^0} = - i [ \eta_{\mu\rho} \delta_{(\nu}^0 \delta_{\sigma)}^i 
+ \eta_{\mu\sigma} \delta_{(\nu}^0 \delta_{\rho)}^i + \eta_{\nu\rho} \delta_{(\mu}^0 \delta_{\sigma)}^i 
+ \eta_{\nu\sigma} \delta_{(\mu}^0 \delta_{\rho)}^i ] \partial_i \delta^3,
\nonumber\\
&{}& \Big. [ h_{\mu\nu} (x), \partial_\rho \partial_\sigma \partial_\tau b^\tau (z) ] \Big|_{z^0 = x^0} 
= - 4 i [ ( \delta_\mu^0 \delta_\nu^0 \Delta + \delta_\mu^j \delta_\nu^k \partial_j \partial_k ) \delta_{(\rho}^0 \delta_{\sigma)}^i
\nonumber\\
&{}& + ( \delta_\rho^0 \delta_\sigma^0 \Delta + \delta_\rho^j \delta_\sigma^k \partial_j \partial_k ) 
\delta_{(\mu}^0 \delta_{\nu)}^i ] \partial_i \delta^3,
\label{J2-ETCRs}  
\end{eqnarray}
where we have used $\Box D(x) = 0, D(0, \vec{x}) = 0, \partial_0 D(0, \vec{x}) = - \delta^3(x)$.
With the help of this equation, $J_2$ can be cast to the form:
\begin{eqnarray}
&{}& J_2 = i \frac{8 (\beta_1 + 2 \beta_2) \kappa^4 \gamma^2}{\beta_1 (\beta_1 + 4 \beta_2)} 
( \delta_\mu^0 \delta_\nu^0 \delta_{(\rho}^0 \delta_{\sigma)}^i + \delta_\rho^0 \delta_\sigma^0 \delta_{(\mu}^0 \delta_{\nu)}^i ) 
\partial_0 \partial_i D(x-y)   
\nonumber\\
&{}& - 4 i \kappa^2 ( \eta_{\mu\rho} \delta_{(\nu}^0 \delta_{\sigma)}^i + \eta_{\mu\sigma} \delta_{(\nu}^0 \delta_{\rho)}^i 
+ \eta_{\nu\rho} \delta_{(\mu}^0 \delta_{\sigma)}^i + \eta_{\nu\sigma} \delta_{(\mu}^0 \delta_{\rho)}^i ) 
\partial_0 \partial_i E(x-y)
\nonumber\\
&{}& + i \frac{8 (\beta_1 + 2 \beta_2) \kappa^4 \gamma^2}{\beta_1 (\beta_1 + 4 \beta_2)} 
[ ( \delta_\mu^0 \delta_\nu^0 \partial_0^2 + \delta_\mu^j \delta_\nu^k \partial_j \partial_k ) \delta_{(\rho}^0 \delta_{\sigma)}^i
+ ( \delta_\rho^0 \delta_\sigma^0 \partial_0^2 + \delta_\rho^j \delta_\sigma^k \partial_j \partial_k ) \delta_{(\mu}^0 \delta_{\nu)}^i ]
\nonumber\\
&{}& \times \partial_0 \partial_i E(x-y),
\label{J2-2}  
\end{eqnarray}
where $\Box E(x) = D(x)$ was used. 

In a similar way, $J_1$ can be calculated to be
\begin{eqnarray}
&{}& J_1 \equiv - \int d^3 z \, E(y-z) \Big. [ h_{\mu\nu} (x), \Box \dot h_{\rho\sigma} (z) ] \Big|_{z^0 = x^0}
\nonumber\\
&{}& = - 2 i \kappa^2 ( \eta_{\mu\rho} \delta_\nu^0 \delta_\sigma^0 + \eta_{\mu\sigma} \delta_\nu^0 \delta_\rho^0
+ \eta_{\nu\rho} \delta_\mu^0 \delta_\sigma^0 + \eta_{\nu\sigma} \delta_\mu^0 \delta_\rho^0 ) D(x-y)
\nonumber\\
&{}& + i \frac{4 (\beta_1 + 2 \beta_2) \kappa^4 \gamma^2}{\beta_1 (\beta_1 + 4 \beta_2)}
[ 2 \delta_\mu^0 \delta_\nu^0 \delta_\rho^0 \delta_\sigma^0 \partial_0^2 
+ ( \delta_\mu^0 \delta_\nu^0 \delta_\rho^i \delta_\sigma^j 
+ \delta_\mu^i \delta_\nu^j \delta_\rho^0 \delta_\sigma^0 ) \partial_i \partial_j 
+ 4 \delta_{(\mu}^0 \delta_{\nu)}^i \delta_{(\rho}^0 \delta_{\sigma)}^j \partial_i \partial_j ] D(x-y)
\nonumber\\
&{}&
- 2 i \kappa^2 [ \eta_{\mu\rho} ( \delta_\nu^0 \delta_\sigma^0 \partial_0^2 + \delta_\nu^i \delta_\sigma^j \partial_i \partial_j ) 
+ \eta_{\mu\sigma} ( \delta_\nu^0 \delta_\rho^0 \partial_0^2 + \delta_\nu^i \delta_\rho^j \partial_i \partial_j )
+ \eta_{\nu\rho} ( \delta_\mu^0 \delta_\sigma^0 \partial_0^2 + \delta_\mu^i \delta_\sigma^j \partial_i \partial_j )
\nonumber\\
&{}& + \eta_{\nu\sigma} ( \delta_\mu^0 \delta_\rho^0 \partial_0^2 + \delta_\mu^i \delta_\rho^j \partial_i \partial_j ) ] E(x-y)
+ i \frac{4 (\beta_1 + 2 \beta_2) \kappa^4 \gamma^2}{\beta_1 (\beta_1 + 4 \beta_2)}
[ ( \delta_\mu^0 \delta_\nu^0 \partial_0^2 + \delta_\mu^i \delta_\nu^j \partial_i \partial_j )
\nonumber\\
&{}& \times ( \delta_\rho^0 \delta_\sigma^0 \partial_0^2 + \delta_\rho^k \delta_\sigma^l \partial_k \partial_l )
+ 4 \delta_{(\mu}^0 \delta_{\nu)}^i \delta_{(\rho}^0 \delta_{\sigma)}^j \partial_0^2 \partial_i \partial_j ] E(x-y).
\label{J1}  
\end{eqnarray}

We can therefore get the result of $J = J_1 + J_2$:
\begin{eqnarray}
&{}& J \equiv J_1 + J_2 
\nonumber\\
&{}& = - 4 i \kappa^2 ( \eta_{\rho(\mu} \delta_{\nu)}^0 \delta_\sigma^0 + \eta_{\sigma(\mu} \delta_{\nu)}^0 \delta_\rho^0 ) D(x-y) 
+ i \frac{4 (\beta_1 + 2 \beta_2) \kappa^4 \gamma^2}{\beta_1 (\beta_1 + 4 \beta_2)}
( \delta_\mu^0 \delta_\nu^0 \partial_\rho \partial_\sigma + \delta_\rho^0 \delta_\sigma^0 \partial_\mu \partial_\nu 
\nonumber\\
&{}& + 4 \delta_{(\mu}^0 \delta_{\nu)}^i \delta_{(\rho}^0 \delta_{\sigma)}^j \partial_i \partial_j ) D(x-y)   
- 4 i \kappa^2 ( \eta_{\mu(\rho} \partial_{\sigma)} \partial_\nu  
+ \eta_{\nu(\rho} \partial_{\sigma)} \partial_\mu ) E(x-y)
\nonumber\\
&{}& + i \frac{4 (\beta_1 + 2 \beta_2) \kappa^4 \gamma^2}{\beta_1 (\beta_1 + 4 \beta_2)}
\partial_\mu \partial_\nu \partial_\rho \partial_\sigma E(x-y).
\label{J}  
\end{eqnarray}
Finally, adding $I$ and $J$ leads to $[ h_{\mu\nu} (x), h_{\rho\sigma} (y) ]$ in Eq. (\ref{4D-CR5}).

\renewcommand{\theequation}{D.\arabic{equation}}
\setcounter{equation}{0}

\section{Derivation of Fourier transform}

In this appendix, we present a derivation of Eq. (\ref{FT-4D-CR4}). The Fourier transform 
of the other 4D CRs is similarly calculated.

From the formula (\ref{Phi-2}), the dipole ghost field $h_{\mu\nu}$ can be expressible as
\begin{eqnarray}
h_{\mu\nu} (p) = \frac{i}{2 (2 \pi)^{\frac{3}{2}}} \theta (p^0) \int d^3 z \, e^{-ipz}
\overleftrightarrow{\partial}_0^z [ \delta(p^2) h_{\mu\nu} (z) + \delta^\prime(p^2) \Box h_{\mu\nu} (z) ].
\label{FT-h-DP}  
\end{eqnarray}
We can also express $\tilde A_\mu (p)$ in a similar manner since  $\tilde A_\mu (p)$ is a dipole 
ghost field. Thus, we have the following commutator:
\begin{eqnarray}
&{}& [ h_{\mu\nu} (p), \tilde A_\rho^\dagger (q) ] = \frac{1}{4 (2 \pi)^3} \theta (p^0) \theta (q^0) \int d^3 y \, d^3 z \, e^{-ipy}
\overleftrightarrow{\partial}_0^y e^{iqz} \overleftrightarrow{\partial}_0^z [ \delta(p^2) h_{\mu\nu} (y) 
\nonumber\\
&{}& + \delta^\prime(p^2) \Box h_{\mu\nu} (y), \delta(q^2) \tilde A_\rho (z) + \delta^\prime(q^2) \Box \tilde A_\rho (z) ].
\nonumber\\
&{}& = \frac{1}{4 (2 \pi)^3} \theta (p^0) \theta (q^0) \int d^3 y \, d^3 z \, e^{-ipy}
\overleftrightarrow{\partial}_0^y e^{iqz} \overleftrightarrow{\partial}_0^z \Biggl\{ \delta(p^2) \delta(q^2) 
\Biggl[ i \frac{\beta_2}{2 (\beta_1 + \beta_2) \gamma m^2} 
\nonumber\\
&{}& \times \Biggl( \eta_{\mu\nu} + \frac{2}{m^2} \partial_\mu \partial_\nu \Biggr) \partial_\rho D(y-z) 
+ i \frac{\beta_2}{(\beta_1 + \beta_2) \gamma m^2} \partial_\mu \partial_\nu \partial_\rho E(y-z) \Biggr]
\nonumber\\
&{}& + [ \delta(p^2) \delta^\prime(q^2) + \delta^\prime(p^2) \delta(q^2) ] \, i \frac{\beta_2}{(\beta_1 + \beta_2) \gamma m^2} 
\partial_\mu \partial_\nu \partial_\rho D(y-z) \Biggr\}
\nonumber\\
&{}& \equiv A_1 + A_2 + A_3, 
\label{FT-h-A}  
\end{eqnarray}
where we have used $\Box E(x) = D(x), \Box^2 E(x) = \Box D(x) = 0$ and Eq. (\ref{4D-CR6}).

Let us evaluate each term $A_i ( i = 1, 2, 3 )$ in order.
\begin{eqnarray}
&{}& A_1 \equiv \frac{1}{4 (2 \pi)^3} \theta (p^0) \theta (q^0) \int d^3 y \, d^3 z \, e^{-ipy}
\overleftrightarrow{\partial}_0^y e^{iqz} \overleftrightarrow{\partial}_0^z \delta(p^2) \delta(q^2) 
\nonumber\\
&{}& \times i \frac{\beta_2}{2 (\beta_1 + \beta_2) \gamma m^2} 
\Biggl( \eta_{\mu\nu} + \frac{2}{m^2} \partial_\mu \partial_\nu \Biggr) 
\partial_\rho D(y-z)
\nonumber\\
&{}& = i \frac{\beta_2}{8 (\beta_1 + \beta_2) \gamma m^2} \theta (p^0) \theta (q^0) \delta(p^2) \delta(q^2)
\int d^4 k \, ( k^0 + p^0 ) ( k^0 + q^0 ) \, e^{i [ (p^0 - k^0) y^0 + (k^0 - q^0) z^0 ]}
\nonumber\\
&{}& \times \Biggl( \eta_{\mu\nu} - \frac{2}{m^2} k_\mu k_\nu \Biggr) k_\rho \epsilon(k^0) \delta(k^2) 
\delta^3 (p - k) \delta^3 (k - q), 
\label{Ev-A1}  
\end{eqnarray}
where we have used the definition of the invariant delta function $D(x)$ in Eq. (\ref{D-function}),
and performed the integral $\int d^3 y \, d^3 z$.  Here the useful mathematical formulae are
\begin{eqnarray}
\theta (q^0) \delta(q^2) &=& \frac{1}{2 q^0} \delta ( q^0 - |\vec{q}| ), 
\nonumber\\
\epsilon(k^0) \delta(k^2) &\equiv& [ \theta(k^0) - \theta(-k^0) ] \delta (k^2) 
\nonumber\\
 &=& \frac{1}{2 |\vec{k}|} [ \delta ( k^0 - |\vec{k}| ) - \delta ( k^0 + |\vec{k}| ) ].
\label{theta-epsilon}  
\end{eqnarray}
Then, we find that a contribution from $k^0 = - |\vec{k}|$ vanishes in $A_1$ owing to the presence of the factor
$( k^0 + p^0 ) ( k^0 + q^0 )$. Thus, the integration over $k_\mu$ produces  
\begin{eqnarray}
A_1 = i \frac{\beta_2}{8 (\beta_1 + \beta_2) \gamma m^2} 
\theta (p^0) \delta(p^2) \Biggl( \eta_{\mu\nu} - \frac{2}{m^2} p_\mu p_\nu \Biggr) 
p_\rho \delta^4 (p - q). 
\label{Ev-A1-end}  
\end{eqnarray}

In a perfectly similar way, one can evaluate the remaining $A_2, A_3$. In the process of the calculation,
it is useful to recall the following formulae:
\begin{eqnarray}
\theta (q^0) \delta^\prime(q^2) &=& - \frac{1}{4} \Biggl[ \frac{1}{|\vec{q}|^3} \delta ( q^0 - |\vec{q}| )
+ \frac{1}{|\vec{q}|^2} \delta^\prime( q^0 - |\vec{q}| ) \Biggr], 
\nonumber\\
\epsilon(k^0) \delta^\prime(k^2) &\equiv& [ \theta(k^0) - \theta(-k^0) ] \delta^\prime(k^2)
\nonumber\\
&=& - \frac{1}{4} \Biggl\{ \frac{1}{|\vec{k}|^3} [ \delta ( k^0 - |\vec{k}| ) - \delta ( k^0 + |\vec{k}| ) ] 
+ \frac{1}{|\vec{k}|^2} [ \delta^\prime ( k^0 - |\vec{k}| ) 
\nonumber\\
&+& \delta^\prime ( k^0 + |\vec{k}| ) ] \Biggr\}.
\label{theta-epsilon2}  
\end{eqnarray}
The result is of the form:
\begin{eqnarray}
A_2 &\equiv& \frac{1}{4 (2 \pi)^3} \theta (p^0) \theta (q^0) \int d^3 y \, d^3 z \, e^{-ipy}
\overleftrightarrow{\partial}_0^y e^{iqz} \overleftrightarrow{\partial}_0^z \delta(p^2) \delta(q^2) 
i \frac{\beta_2}{(\beta_1 + \beta_2) \gamma m^2} \partial_\mu \partial_\nu \partial_\rho E(y-z)
\nonumber\\
&=& - i \frac{\beta_2}{4 (\beta_1 + \beta_2) \gamma m^2} 
\theta (p^0) \delta^\prime(p^2) p_\mu p_\nu p_\rho \delta^4 (p - q), 
\nonumber\\
A_3 &\equiv& \frac{1}{4 (2 \pi)^3} \theta (p^0) \theta (q^0) \int d^3 y \, d^3 z \, e^{-ipy}
\overleftrightarrow{\partial}_0^y e^{iqz} \overleftrightarrow{\partial}_0^z  
[ \delta(p^2) \delta^\prime(q^2) + \delta^\prime(p^2) \delta(q^2) ] 
\nonumber\\
&\times& i \frac{\beta_2}{(\beta_1 + \beta_2) \gamma m^2} \partial_\mu \partial_\nu \partial_\rho D(y-z)
\nonumber\\
&=& - i \frac{\beta_2}{2 (\beta_1 + \beta_2) \gamma m^2} 
\theta (p^0) \delta^\prime(p^2) p_\mu p_\nu p_\rho \delta^4 (p - q).
\label{Ev-A2&3}  
\end{eqnarray}
Finally, adding all the $A_i$, we can arrive at Eq. (\ref{FT-4D-CR4}).

\renewcommand{\theequation}{E.\arabic{equation}}
\setcounter{equation}{0}

\section{A proof of Eq. (\ref{Simple-field2})}

We will give a proof of Eq. (\ref{Simple-field2}) in this appendix. The Fourier transform of Eq. (\ref{Box-h})
is of form:
\begin{eqnarray}
p^2 h_{\mu\nu} (p) = 4 i \kappa^2 \Biggl[ p_{(\mu} b_{\nu)} (p)
+ \frac{(\beta_1 + 2 \beta_2) \kappa^2 \gamma^2}{2 \beta_1 (\beta_1 + 4 \beta_2)}
p_\mu p_\nu p_\rho b^\rho (p) \Biggr]
+ i \frac{2 \beta_2 \kappa^2 \gamma}{\beta_1 (\beta_1 + 4 \beta_2)}
p_\mu p_\nu p_\rho \beta^\rho (p). 
\label{FT-Box-h}  
\end{eqnarray}
In order to evaluate ${\cal D}_p ( p^2 h_{\mu\nu} (p) )$, let us calculate the following
quantities:
\begin{eqnarray}
{\cal D}_p ( p_{(\mu} b_{\nu)} ) 
&=& \frac{1}{2 | \vec{p} |^2} p_0 \Biggl( \delta_{(\mu}^0 b_{\nu)} + p_{(\mu} \frac{ \partial b_{\nu)}}{\partial p_0} \Biggr)
+ c p_{(\mu} b_{\nu)},
\nonumber\\
{\cal D}_p ( p_\mu p_\nu p_\rho b^\rho ) 
&=& \frac{1}{2 | \vec{p} |^2} p_0 \Biggl( \delta_\mu^0 p_\nu p_\rho b^\rho + \delta_\nu^0 p_\mu p_\rho b^\rho 
+ \delta_\rho^0 p_\mu p_\nu b^\rho + p_\mu p_\nu p_\rho \frac{\partial b^\rho}{\partial p_0} \Biggr)
+ c p_\mu p_\nu p_\rho b^\rho,
\nonumber\\
{\cal D}_p ( p_\mu p_\nu p_\rho \beta^\rho ) 
&=& \frac{1}{2 | \vec{p} |^2} p_0 \Biggl( \delta_\mu^0 p_\nu p_\rho \beta^\rho + \delta_\nu^0 p_\mu p_\rho \beta^\rho 
+ \delta_\rho^0 p_\mu p_\nu \beta^\rho + p_\mu p_\nu p_\rho \frac{\partial \beta^\rho}{\partial p_0} \Biggr)
\nonumber\\
&+& c p_\mu p_\nu p_\rho \beta^\rho.
\label{Dp-op}
\end{eqnarray}

Using these equations and field equations $p^2 b_\mu (p) = p^2 p_\rho \beta^\rho (p) = 0$, we can reduce 
$p^2 \hat h_{\mu\nu} (p)$ to the form:
\begin{eqnarray}
p^2 \hat h_{\mu\nu} (p) &=& p^2 h_{\mu\nu} (p) - p^2 {\cal D}_p ( p^2 h_{\mu\nu} (p) )
\nonumber\\
&=& p^2 h_{\mu\nu} (p) - \frac{i}{2 | \vec{p} |^2} p_0 p^2 \Biggl\{ 4 \kappa^2 
\Biggl[ p_{(\mu} \frac{ \partial b_{\nu)}}{\partial p_0}
+ \frac{(\beta_1 + 2 \beta_2) \kappa^2 \gamma^2}{2 \beta_1 (\beta_1 + 4 \beta_2)} 
p_\mu p_\nu p_\rho \frac{\partial b^\rho}{\partial p_0} \Biggr] 
\nonumber\\
&+& \frac{2 \beta_2 \kappa^2 \gamma}{\beta_1 (\beta_1 + 4 \beta_2)} 
\Biggl( p_\mu p_\nu \beta^0 + p_\mu p_\nu p_\rho \frac{\partial \beta^\rho}{\partial p_0} \Biggr) \Biggr\}.
\label{p-hat-h}
\end{eqnarray}
To simplify this expression further, let us take the differentiation of the field equations with respect to $p_0$
whose result is given by 
\begin{eqnarray}
p^2 \frac{\partial b_\mu}{\partial p_0} = 2 p_0 b_\mu, \qquad
p^2 p_\rho \frac{\partial \beta^\rho}{\partial p_0} = 2 p_0 p_\rho \beta^\rho - p^2 \beta^0.
\label{diff-p-f.e.}
\end{eqnarray}
Substituting Eq. (\ref{diff-p-f.e.}) into the RHS on Eq. (\ref{p-hat-h}), and using the on-shell
relation, $p_0^2 = - p^2 + | \vec{p} |^2$, the field equations again and Eq. (\ref{FT-Box-h}), 
we find that $p^2 \hat h_{\mu\nu} (p) = 0$ as expected.



\begin{thebibliography}{99}

\bibitem{Luca}
L. Buoninfante, {``Strict Renormalizability as a Paradigm for Fundamental Physics", 
arXiv:2504.05900 [hep-th] and references therein.}    

\bibitem{Stelle1}
K. S. Stelle, {``Renormalization of Higher Derivative Quantum Gravity'',
Phys. Rev. {\bf D 16} (1977) 953.}

\bibitem{Starobinsky}
A. A. Starobinsky, {``A New Type of Isotropic Cosmological Models without Singularity'',
Phys. Lett. {\bf B 91} (1980) 99.}

\bibitem{Anselmi}
D. Anselmi, {``On the Quantum Field Theory of the Gravitational Interactions", JHEP {\bf 06} 
(2017) 086.}    

\bibitem{Salvio}
A. Salvio, {``Quadratic Gravity'', Front. in Phys. {\bf 6} (2018) 77.}

\bibitem{Strumia}
A. Strumia, {``Interpretation of Quantum Mechanics with Indefinite Norm'', MDPI Physics {\bf 1} (2019) 17.}

\bibitem{Donoghue}
J. E. Donoghue and G. Menezes, {``On Quadratic Gravity'', Nuovo. Cim. {\bf C 45} (2022) 26.}

\bibitem{Kubo-Kugo}
J. Kubo and T. Kugo, {"Unitarity Violation in Field Theories of Lee-Wick's Complex Ghost", PTEP {\bf 2023} 
(2023) 123B02.}

\bibitem{Holdom}
B. Holdom, {``Making Sense of Ghosts'', Nucl. Phys. {\bf B 1008} (2024) 116696.}

\bibitem{Luca2}
L. Buoninfante, {``Remarks on Ghost Resonances", 
JHEP {\bf 02} (2025) 186.}    

\bibitem{Kugo-Ojima}
T. Kugo and I. Ojima, {"Local Covariant Operator Formalism of Nonabelian Gauge Theories
and Quark Confinement Problem", Prog. Theor. Phys. Suppl. {\bf 66} (1979) 1.}

\bibitem{Nakanishi}
N. Nakanishi, {"Indefinite Metric Quantum Field Theory of General Gravity", 
Prog. Theor. Phys. {\bf 59} (1978) 972.}

\bibitem{N-O-text}
N. Nakanishi and I. Ojima, {"Covariant Operator Formalism of Gauge Theories and Quantum Gravity", 
World Scientific Publishing, 1990 and references therein.}

\bibitem{Kimura1}
S. Kawasaki, T. Kimura and K. Kitago, {"Canonical Quantum Theory of Gravitational Field with Higher Derivatives", 
Prog. Theor. Phys. {\bf 66} (1981) 2085.}

\bibitem{Kimura2}
S. Kawasaki and T. Kimura, {"Canonical Quantum Theory of Gravitational Field with Higher Derivatives. II", 
Prog. Theor. Phys. {\bf 68} (1982) 1749.}

\bibitem{Kimura3}
S. Kawasaki and T. Kimura, {"Canonical Quantum Theory of Gravitational Field with Higher Derivatives. III", 
Prog. Theor. Phys. {\bf 69} (1983) 1015.}

\bibitem{Oda-Ohta}
I. Oda and M. Ohta, {``Quantum Conformal Gravity", JHEP {\bf 02} (2024) 213.}    

\bibitem{Oda-Conf}
I. Oda, {``Conformal Symmetry in Quantum Gravity", Eur. Phys. J.  {\bf C 84} (2024) 887.}    

\bibitem{Oda-f}
I. Oda, {``BRST Formalism of $f(R)$ Gravity", arXiv:2410.20270 [hep-th].}    

\bibitem{Oda-Q}
I. Oda, {``Quantum Scale Invariant Gravity in de Donder Gauge'', Phys. Rev. {\bf D 105} (2022) 066001.}   

\bibitem{Oda-W}
I. Oda, {``Quantum Theory of Weyl Invariant Scalar-tensor Gravity'', Phys. Rev. {\bf D 105} (2022) 120618.}   

\bibitem{Oda-Saake}
I. Oda and P. Saake, {"BRST Formalism of Weyl Conformal Gravity", 
Phys. Rev. {\bf D 106} (2022) 106007.}    

\bibitem{Oda-Corfu}
I. Oda, {``BRST formalism of Weyl Invariant Gravity and Confinement of Massive Tensor Ghost'', 
PoS CORFU2023 (2024) 158.}   

\bibitem{MTW}
C. W. Misner, K. S. Thorne and J. A. Wheeler, {``Gravitation", W H Freeman and Co (Sd), 1973.}

\bibitem{Banks}
T. Banks and N. Seiberg,  {``Symmetries and Strings in Field Theory and Gravity'', Phys. Rev.
{\bf D 83} (2011) 084019.}

\bibitem{Peskin}
M. E. Peskin, {``Introduction to String and Superstring Theory II, in From the Planck Scale to 
the Weak Scale", TASI 1986, ed. by H. E. Haber, World Scientific (1987) 277.}

\bibitem{Wald}
R. M. Wald, {``General Relativity'', The University of Chicago Press, 1984.}

\bibitem{Buchbinder}
I. L. Buchbinder and I. L. Shapiro, {``Introduction to Quantum Field Theory with Applications to
Quantum Gravity", Oxford University Press, 2021.}

\bibitem{Kawasaki}
S. Kawasaki and K. Kimura, {"A Possible Mechanism of Ghost Confinement in a Renormalizable 
Quantum Gravity", Prog. Theor. Phys. {\bf 65} (1981) 1767.}



\end{thebibliography}
\end{document}